\documentclass[preprint, sort&compress]{elsarticle}
\usepackage{graphicx}
\usepackage{graphics}
\usepackage{epsfig}
\usepackage{amsmath}
\usepackage{amsfonts}
\usepackage{amssymb}
\usepackage{dcolumn}
\usepackage{bm}
\usepackage{longtable}
\usepackage{multirow}

\begin{document}

\title{Non-equilibrium magnetic interactions in strongly correlated systems}

\author[RU]{A. Secchi}
\ead{a.secchi@science.ru.nl}

\author[UH]{S. Brener}

\author[UH]{A. I. Lichtenstein}

\author[RU]{M. I. Katsnelson}

\address[RU]{Institute for Molecules and Materials, Radboud University Nijmegen, 6525 AJ Nijmegen, The Netherlands}
\address[UH]{Institut f\"ur Theoretische Physik, Universitat Hamburg, Jungiusstra{\ss}e 9, D-20355 Hamburg, Germany}

\date{\today}

\begin{abstract}
We formulate a low-energy theory for the magnetic interactions between electrons in the multi-band Hubbard model under non-equilibrium conditions determined by an external time-dependent electric field which simulates laser-induced spin dynamics. We derive expressions for dynamical exchange parameters in terms of non-equilibrium electronic Green functions and self-energies, which can be computed, e.g., with the methods of time-dependent dynamical mean-field theory. Moreover, we find that a correct description of the system requires, in addition to exchange, a new kind of magnetic interaction, that we name \emph{twist exchange}, which formally resembles Dzyaloshinskii-Moriya coupling, but is not due to spin-orbit, and is actually due to an effective three-spin interaction. Our theory allows the evaluation of the related time-dependent parameters as well. 
\end{abstract}


\maketitle

\section{Introduction}

Manipulation of the magnetic order of materials via electromagnetic fields on increasingly fast time scales is one of the most intriguing issues of modern magnetism \cite{Kirilyuk10}. While there seems to be a limit (of the order of $\approx 2$ ps) on the time scale attainable in magnetization switching via magnetic fields \cite{Tudosa04}, pioneering experimental works dating back to the 1990's \cite{Beaurepaire96, Hohlfeld97, Scholl97, Gudde99} demonstrated that ultrashort optical laser pulses trigger magnetic phenomena on sub-picosecond time scales. These early works opened the field of ultrafast spin dynamics, which has been characterized by a flourishing of experimental efforts in the last 15 years \cite{Kirilyuk10}, including inverse Faraday effect \cite{Kimel05}, all-optical helicity-dependent switching of the magnetization \cite{Stanciu07}, complete write-read events requiring a time of only 30 ps \cite{Vahaplar09}, reversal of lattice magnetization in ferrimagnets via a transient ferromagnetic state \cite{Radu11}. Besides their obvious practical role in the design of a new generation of ultra-fast memories, these researches have important implications for a fundamental understanding of the interaction between light and matter on short time scales.

From the point of view of theory, the study of magnetization in a realistic solid-state system is a challenging problem. In equilibrium, magnetic interactions in magnetic metals and semiconductors are known to be non-Heisenberg \cite{NagaevBook, Auslender82, Irkhin85, Turzhevskii90}, that is, the lengths of magnetic moments and values of exchange parameters depend on the magnetic configuration for which they are calculated. The accurate calculation of these quantities must then follow from an ab-initio formulation, and represents a complicated many-body problem. In the case of equilibrium, the expressions for computing exchange parameters have been given years ago, either within the multiple-scattering formalism in density functional theory \cite{Lichtenstein87} or, more recently, in terms of electronic Green functions and self-energies within the Matsubara scheme, for a multiband Hubbard model \cite{Katsnelson00, Katsnelson02, Katsnelson10}. These formulas are commonly used now for a very broad circle of magnetic materials, see, e.g., Refs. \cite{Szunyogh98, Boukhvalov04, Wan06, Sato10, Simon11, Delczeg12}.

One could think of using these first-principle methods to calculate equilibrium exchange parameters for further use within a classical Heisenberg model to simulate spin dynamics. However, this approach is not expected to allow for a satisfactory description of ultrafast magnetism. In fact, the time scale of the laser pulses is typically faster than the typical scale of exchange interactions ($\approx 10 \div 100$ fs). This implies that magnetic interactions cannot be treated adiabatically, i.e., the relevant parameters (such as exchange) depend on time. Ab-initio spin dynamics \cite{Antropov95, Antropov96}, also, is based on \emph{equilibrium} electronic structure.

Within the LDA++ approach \cite{Lichtenstein98}, the first-principles electronic structure is mapped to the \emph{multiband Hubbard model}. A multiband Hubbard model, with realistic tight-binding and interaction parameters, is likely to be general and flexible enough to describe many strongly correlated systems relevant for ultrafast magnetism. In order to include a time-dependent optical excitation, we allow the hopping parameters to depend on time. Then, the main approximation we can take advantage of consists in the fact that spin dynamics is known to be much slower than electron dynamics in relevant systems \cite{Beaurepaire96}. This means that an effective atomistic model can be derived, with time-dependent parameters accounting for the magnetic interactions mediated by the fast electronic dynamics, which can be computed from first principles. Technically, we need to separate the spin degrees of freedom from the electronic ones, and derive an effective action for the spin variables, after integrating on the electronic variables.

The needed formalism is that of non-equilibrium Green functions, a very general tecnique which resulted from the contributions of several authors (for reviews and lectures, see Refs.\cite{KadanoffBaymBook, Rammer86, Wagner91, Leeuwen06, KamenevBook}). One typically makes a distinction between the equilibrium formalism, due to Matsubara \cite{Matsubara55}, and the non-equilibrium formalism developed mainly by Schwinger \cite{Schwinger61} and Keldysh \cite{Keldysh64}, which neglects initial (equilibrium) correlations. The Schwinger-Keldysh formulation has been applied to the study of spin dynamics, e.g. for a single spin in a Josephson junction \cite{Zhu04}, or in a junction between ferromagnets \cite{Fransson08, Fransson10}, or combined with the mean field approximation for the treatment of magnetic interaction \cite{Bhattacharjee12}. However, a first-principle study of an extended and strongly correlated system out of equilibrium, such as a fermionic multi-band Hubbard model, has not been attempted yet, and in this paper we present a general framework for this relevant condensed-matter model. Moreover, since we cannot realistically neglect the initial correlations, we need to use the Kadanoff-Baym formalism \cite{KadanoffBaymBook, Wagner91, Leeuwen06}, which unifies the approaches of Schwinger, Keldysh and Matsubara.

The advantage of this approach is that it does not need any assumption on the time dependence of the external field, so there is no restriction on time scales, which allows to study the role of non-adiabatic and non-Heisenberg effects in magnetization dynamics. It is also suitable to make a first-principle formulation of quantum noise, whose time scale may be comparable to that of the ultrafast pulse, invalidating the Landau-Lifschitz equations, and whose treatment has up to now relied on phenomenological parameters \cite{Atxitia09}.

Besides being an initial step towards a realistic first-principle model of spin dynamics in condensed-matter systems, our work is immediately relevant to the following problem. Let us consider a magnetically ordered system with negligible spin-orbit interactions and anisotropy. Applying a purely electric field does not generate spin dynamics, in the sense that it cannot flip the spins of individual electrons, because the electric field does not couple directly with spins, and there is no spin-orbit. Nevertheless, a time-dependent electric field modulates the possibility of electronic hopping between different lattice sites (and orbitals), which in turn affects the strength of magnetic interactions, such as exchange. Therefore, if the electron spins are not all parallel in equilibrium, and the electronic dynamics is fast enough, then one can assume that the magnetic structure, at each time, evolves in the way that minimizes the \emph{spin action}, with instantaneous parameters characterizing the magnetic interactions that depend on the time-dependent electronic configuration and hopping. The only constraint on the minimization is that the total numbers of electrons with spin up and down along a chosen direction must remain constant. This would be a first direct application of the results of this contribution.

This Article is organized as follows: in Sections 2 - 9 we detail the method which we use to derive the effective action for the spins, in Sections 10 - 13 we present the results, and Section 14 contains our conclusions. In particular, Section 2 contains a short review of the Kadanoff-Baym formalism, intended for non-expert readers. In Section 3 we present the fermionic multiband Hubbard model, and in Section 4 we show how to transform from the fermionic representation to a bosonic one, where the bosons are related to the directions of the spin axes. In Section 5 we discuss how to integrate away the fermionic degrees of freedom from the total action, obtaining an effective bosonic action for the low-energy excitations on the top of the ground state, which is shown in Section 6. In Section 7 we introduce the Kadanoff-Baym equations of motion, which allow to express the magnetic interactions appearing in the bosonic action in terms of fermionic Green functions and self-energies, as we do in Section 8; in Section 9 we show a way to slightly simplify the action. After this introductory part, in Section 10 we start to present the results of the derivation, showing that the out-of-equilibrium multiband Hubbard model, close to the equilibrium ground state, can be mapped into a rather complicated effective model that includes two kinds of time-dependent magnetic interactions. In Section 11 we show that our formalism correctly recovers the known results \cite{Katsnelson00} in the equilibrium case, while in Section 12 we discuss the general non-equilibrium case, together with some relevant approximations which can simplify the calculations. In Section 13 we give the expression of the dynamical spin stiffness, which is an interesting observable in view of comparison with experiments. In Section 14 we summarize our results and discuss possible further developments. Appendices A - G include mathematical details.

\section{Kadanoff-Baym approach to non-equilibrium systems}

For the sake of completeness, we here review briefly the main concepts of the Kadanoff-Baym approach to the study of non-equilibrium systems \cite{KadanoffBaymBook, Rammer86, Wagner91}, which combines the approaches of Matsubara \cite{Matsubara55}, Schwinger \cite{Schwinger61} and Keldysh \cite{Keldysh64}. 

\subsection{Kadanoff-Baym time contour}

\begin{figure}
\centering
\includegraphics[scale = 0.35]{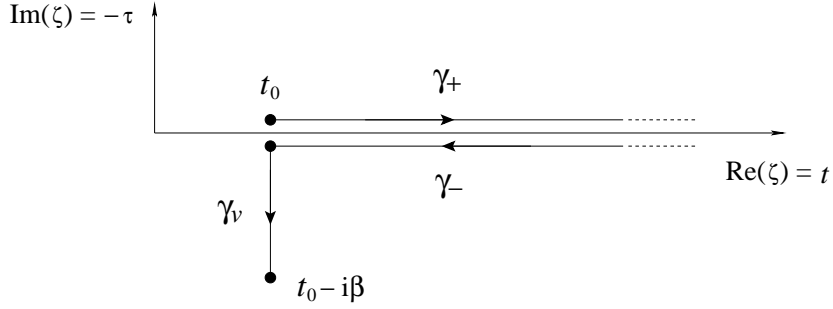}
\caption{Schematic representation of the Kadanoff-Baym contour. Branches $\gamma_+$ and $\gamma_-$ are displaced for graphical convenience, but they both coincide with a portion of the real time axis, extending from $t_0$ to $\infty$.}
\label{fig KB}
\end{figure}

Given a time-dependent Hamiltonian $\hat{H}(t)$, the equation of motion for the density operator is $\text{d} \hat{\rho}(t) / \text{d}t = - \text{i} \left[ \hat{H}(t), \hat{\rho}(t) \right]$, which can be solved formally as $\hat{\rho}(t) = \hat{U}(t, t_0) \hat{\rho}(t_0) \hat{U}(t_0, t)$, where $\hat{\rho}(t_0)$ is the (supposedly known) density operator at a reference time $t_0$, and the evolution operator is
\begin{align}
\hat{U}(t, t') = \Theta(t - t') \mathcal{T} \text{exp}\left( - \text{i} \int_{t'}^{t} \text{d} t_1 \hat{H}(t_1)  \right) + \Theta(t' - t) \widetilde{\mathcal{T}} \text{exp}\left( - \text{i} \int_{t'}^{t} \text{d} t_1 \hat{H}(t_1)  \right) 
\end{align}
for $t \ne t'$, while $\hat{U}(t, t) = 1$; the symbol $\Theta(t)$ denotes the step function [$\Theta(x > 0) = 1, \Theta(x < 0) = 0$]. The expectation value for the observable $\hat{O}$ at time $t$ is:
\begin{align}
O(t) \equiv \frac{\text{Tr} \left[ \hat{O} \hat{\rho}(t) \right]}{\text{Tr} \left[\hat{\rho}(t) \right]}  =  \frac{\text{Tr} \left[ \hat{\rho}(t_0) \hat{U}(t_0, t) \hat{O}  \hat{U}(t, t_0)  \right]}{\text{Tr} \left[\hat{\rho}(t_0) \right]},
\label{exp value O}
\end{align}
where the trace is evaluated over the complete many-body Hilbert space, and in the last passage the cyclic property of the trace and the identity $\hat{U}(t, t') \hat{U}(t', t) = \hat{U}(t, t) = \hat{1}$ have been used. We choose the reference time $t_0$ in such a way that, for $t < t_0$, the Hamiltonian is independent of time and the system is in equilibrium. Therefore, we can use for $\hat{\rho}(t_0)$ the grand-canonical equilibrium expression,
\begin{align}
\hat{\rho}(t_0) = \frac{\text{e}^{- \beta \left( \hat{H}_0 - \mu \hat{N} \right)} }{\text{Tr} \left[ \text{e}^{- \beta \left( \hat{H}_0 - \mu \hat{N} \right)} \right]},
\end{align} 
where $\hat{H}_0 = \hat{H}(t \le t_0)$, $\hat{N}$ is the number-of-particle operator and $\mu$ is the chemical potential. We assume $\left[ \hat{H}_0, \hat{N} \right] = 0$. We extend the time domain to the complex plane, defining the complex time variable $\zeta \equiv t - \text{i} \tau$, with the understanding that $\hat{H}(\zeta) \equiv \hat{H}(t)$ depends only on the real part of time, and we define the evolution operator in imaginary time
\begin{align}
\text{e}^{- \beta \left( \hat{H}_0 - \mu \hat{N} \right)} \equiv \hat{U}_v(t_0 - \text{i} \beta, t_0).
\end{align}
Thus, we can write Eq.\eqref{exp value O} as:
\begin{align}
O(t) &  = \frac{\text{Tr} \left[ \hat{U}_v(t_0 - \text{i} \beta, t_0) \hat{U}(t_0,  \infty) \hat{U}( \infty, t) \hat{O}  \hat{U}(t, t_0)  \right]}{\text{Tr} \left[  \text{e}^{- \beta \left( \hat{H}_0 - \mu \hat{N} \right)   } \right]} \nonumber \\
&  = \frac{\text{Tr} \left[ \hat{U}_v(t_0 - \text{i} \beta, t_0) \hat{U}(t_0, t) \hat{O}  \hat{U}(t, \infty)  \hat{U}(\infty, t_0) \right]}{\text{Tr} \left[  \text{e}^{- \beta \left( \hat{H}_0 - \mu \hat{N} \right)   } \right]},
\label{exp value operator}
\end{align}
which shows that the evaluation of $O(t)$ requires that we let the system evolve along three time domains in the complex plane: a \emph{forward} branch $\gamma_+ \equiv (t_0, \infty)$, a \emph{backward} branch $\gamma_- \equiv (\infty, t_0)$, and a segment on the imaginary (vertical) axis of time $\gamma_v \equiv  (t_0, t_0 - \text{i} \beta)$, which we call the \emph{vertical} branch (see Fig.\ref{fig KB}). It must be noted that the time value which we have labelled as $\infty$ may actually be chosen as a completely arbitrary (finite) value. The total domain over which the system evolves is the \emph{Kadanoff-Baym contour}
\begin{align}
\Gamma \equiv  \gamma_+ \cup \gamma_- \cup \gamma_v \equiv (t_0, \infty) \cup (\infty, t_0) \cup (t_0, t_0 - \text{i} \beta).
\end{align}
We define the total evolution operator on $\Gamma$ as:
\begin{align}
\hat{U}_{\Gamma} \equiv \hat{U}_v(t_0 - \text{i} \beta, t_0) \hat{U}(t_0, \infty) \hat{U}(\infty, t_0).
\label{U Gamma} 
\end{align}
From Eq.\eqref{exp value operator}, we see that the computation of $O(t)$ is realized by opening the contour $\Gamma$ at the instant $t$ either on the branch $\gamma_+$ or on the branch $\gamma_-$, inserting there the Schr\"odinger-represented operator $\hat{O}$ in $\hat{U}_{\Gamma}$, and evaluating the trace of the resulting operator. The inclusion of the branch $\gamma_v$ is required to treat systems where the inital correlations are not negligible, as it is the case in typical solid-state systems. In the cases where the initial correlations can be neglected, the initial density matrix is of the single-particle kind and there is no need to express it in a contour formulation: in such conditions, one can restrict the contour to $\gamma_+ \cup \gamma_-$, possibly with $t_0 \rightarrow - \infty$, which is the Schwinger-Keldysh contour. In equilibrium conditions, on the other hand, the Hamiltonian is time-independent and the contour is restricted to $\gamma_v$, which is the Matsubara contour. The Kadanoff-Baym formulation, therefore, unifies and generalizes the other approaches, and allows to treat the most general case of a system in non-equilibrium with initial correlations.

\subsection{Path integral formulation}

The Kadanoff-Baym partition function is defined as
\begin{align}
\mathcal{Z} \equiv \text{Tr}\left[ \hat{U}_{\Gamma}\right] \Big/ \text{Tr} \left[  \text{e}^{- \beta \left( \hat{H}_0 - \mu \hat{N} \right)   } \right],
\label{partition}
\end{align}
where $\hat{U}_{\Gamma}$ is given in Eq.\eqref{U Gamma}. Since $ \hat{U}(t_0, \infty) \hat{U}(\infty, t_0) = \hat{1}$, it follows that $\mathcal{Z} = 1$. Despite this apparent triviality, expressing $\mathcal{Z}$ by means of path integrals allows to derive a non-equilibrium action, from which it is possible to extract physical information \cite{KamenevBook}. To do this, we start by denoting the Hamiltonian as $\hat{H}\left[ \hat{\phi}^{\dagger}, \hat{\phi}; t\right]$ for $t > t_0$, where $\hat{\phi}^{\dagger}$ and $\hat{\phi}$ represent the sets of fermionic creation and annihilation operators, respectively. Instead, for $t \le t_0$ and on the branch $\gamma_v$ the Hamiltonian is constant and we denote it as $\hat{H}_0\left[ \hat{\phi}^{\dagger}, \hat{\phi}\right]$. We parameterize the branch $\gamma_v$ by means of the real variable
\begin{align}
\tau = - \text{Im}(\zeta),
\end{align} 
which is equal to $0$ for $\zeta = t_0$ and to $\beta$ for $\zeta = t_0 - \text{i}\beta$. On the $\gamma_v$ contour, $\zeta = t_0 + \text{i} \text{ Im}(\zeta) = t_0 - \text{i} \tau$. Then, standard manipulations lead to the expression
\begin{align}
\mathcal{Z} = \int \mathcal{D} \left[\bar{\phi}, \phi \right] \text{e}^{\text{i} S\left[\bar{\phi}, \phi \right]},
\label{Z initial}
\end{align}
where the effective action $S\left[\bar{\phi}, \phi \right]$, written in terms of the Grassmann variables $\left(\bar{\phi},  \phi \right)$ relative to the $\left(\hat{\phi}^{\dagger}, \hat{\phi}\right)$ operators, is:
\begin{align}
S\left[\bar{\phi}, \phi \right] = & \int_{t_0 + \epsilon}^{\infty} \text{d}t  \Big\{ \text{i} \, \bar{\phi}_+(t) \cdot  \dot{\phi}_+(t - \epsilon)  - H\left[\bar{\phi}_+(t), \phi_+(t - \epsilon); t  \right] \nonumber \\
& \quad \quad -  \text{i}  \, \bar{\phi}_-(t - \epsilon) \cdot \dot{\phi}_-(t - \epsilon)  + H\left[\bar{\phi}_-(t - \epsilon), \phi_-(t); t \right] \Big\} \nonumber \\
& + \text{i} \int_{\varepsilon}^{\beta} \text{d}\tau \Big\{ \bar{\phi}_v(\tau) \cdot \dot{\phi}_v(\tau - \varepsilon) + K\left[ \bar{\phi}_v(\tau), \phi_v(\tau - \varepsilon)\right] \Big\}  ,
\label{action S}
\end{align}
where $K \equiv H_{0} - \mu N$, and which requires some explanations. First, Eq.\eqref{action S} is written, for convenience, in terms of real time variables $t$ and $\tau$, instead of contour variables. Therefore, since each value of $t$ corresponds to two distinct points on the Kadanoff-Baym contour (one on $\gamma_+$ and one on $\gamma_-$), the time-dependent Grassmann fields must be specified by the index $\pm$ if their argument is on the real-time axis, while we use the label $v$ for the fields with the argument on the branch $\gamma_v$. Then, we have introduced two infinitesimally small positive quantities, $\epsilon$ and $\varepsilon$, in order to emphasize a subtle technical point, namely the fact that any product of $\hat{\phi}^{\dagger}$ and $\hat{\phi}$ fermionic operators appearing in the Hamiltonian transforms in the path integral formulation into a product of $\bar{\phi}$ and $\phi$ Grassmann fields in which the fields $\phi$ are evaluated at an instant occurring infinitesimally before (in the contour sense) the instant when the $\bar{\phi}$ fields are evaluated \cite{KamenevBook}. While this aspect is often neglected, on the basis that the fields are assumed to be continuous functions, for our purposes it will be important to keep explicitly track of this discrete structure because we will have to deal with discontinuous functions, such as the correlators (Green functions) originating from the Grassmann numbers $(\bar{\phi}, \phi)$. In these cases, we will need to consider carefully the direction along which the independent variables approach the discontinuity point, which will be possible in our formulation. At the appropriate stage of the calculations, we will send $\epsilon \rightarrow 0^+$ and $\varepsilon \rightarrow 0^+$. On the other hand, whenever $\epsilon$ or $\varepsilon$ appears inside the argument of a continuous function, we are allowed to send it to $0$ immediately, as we already did implicitly in the case of the time-dependent external field. Finally, derivatives of fields are always meant to be taken from the right side, i.e., $\dot{\phi}(t) = \lim_{\epsilon \rightarrow 0^{+}} \left[ \phi(t + \epsilon) - \phi(t) \right] / \epsilon$.

\section{Multi-band Hubbard model}

In order to model an electronic system driven out of equilibrium by a time-dependent external field (e.g. a laser pulse), we must consider a Hamiltonian of the form
\begin{align}
\hat{H}(t) \equiv \hat{H}_T(t) + \hat{H}_V,
\end{align}
where $\hat{H}_T(t)$ is the single-particle Hamiltonian, including the time-dependent field, and $\hat{H}_V$ is the (time-independent) interaction potential between the electrons. We will treat a multi-band Hubbard model \cite{Lichtenstein98, Feng12}, therefore the electronic single-particle states are identified by three labels: the site index $i$, the orbital index $\lambda$ and the spin index $\sigma$. We ignore spin-orbit coupling and assume that the external field is diagonal in spin indices (we are therefore excluding magnetic fields, but including purely electric fields which are relevant for modelling all-optical experiments). The single-particle Hamiltonian is then given by
\begin{align}
\hat{H}_T(t) & \equiv \sum_{i_a \lambda_a} \sum_{i_b \lambda_b} T_{i_a \lambda_a, i_b \lambda_b}(t) \sum_{\sigma} \hat{\phi}^{\dagger}_{i_a \lambda_a \sigma} \hat{\phi}_{i_b \lambda_b \sigma}    \equiv \sum_a \sum_b  T_{a b}(t) \sum_{\sigma} \hat{\phi}^{\dagger}_{a \sigma} \hat{\phi}_{b \sigma} \nonumber \\
&  = \sum_a \sum_b T_{a b}(t) \hat{\phi}^{\dagger}_{a} \cdot \hat{\phi}_{b}
\end{align}
where we have grouped the site and orbital indexes according to $a \equiv (i_a, \lambda_a)$ and $b \equiv (i_b, \lambda_b)$, and we have defined the spinor fermionic operators
\begin{align}
\hat{\phi}^{\dagger}_a = \left( \begin{matrix} \hat{\phi}^{\dagger}_{a \uparrow} & \hat{\phi}^{\dagger}_{a \downarrow} \end{matrix} \right), \quad \quad
\hat{\phi}_{b} = \left( \begin{matrix} \hat{\phi}_{b \uparrow} \\ \hat{\phi}_{b \downarrow} \end{matrix} \right) .
\end{align}
The matrix element $T_{ab}(t) = T^*_{ba}(t)$ of the single-particle Hamiltonian is written as
\begin{align}
T_{ab}(t) = T_{ab} + f_{ab}(t),
\end{align}
where $T_{ab}$ is the time-independent hopping parameter due to electronic structure, and $f_{ab}(t)$ is the time-dependent matrix element of the perturbing field. We denote as $t_0$ the time at which the external field is switched on: $f_{a b}(t) = 0$ for $t \le t_0$. The interaction potential generating $\hat{H}_V$ is assumed to be on-site, i.e., 
\begin{align}
\hat{H}_V \equiv \frac{1}{2} \sum_{i} \sum_{\lambda_1 \lambda_2 \lambda_3 \lambda_4} \sum_{\sigma \sigma'} V_{\lambda_1 \lambda_2 \lambda_3 \lambda_4} \hat{\phi}^{\dagger}_{i \lambda_1 \sigma} \hat{\phi}^{\dagger}_{i \lambda_2 \sigma'} \hat{\phi}_{i \lambda_3 \sigma'} \hat{\phi}_{i \lambda_4 \sigma}.
\end{align}

Including magnetic fields and spin-orbit coupling complicates the analysis significantly, and will be left to future work. However, this model already allows to describe some interesting magnetic phenomena. For example, consider a system which contains both spin-up and spin-down electrons. The arrangement of the spins within each lattice site depends initially on the equilibrium magnetic interactions, mainly exchange. Applying a time dependent electrostatic field on a portion of the sample, as we shall show, may change the strength of the magnetic interactions. If in a certain region of the sample the coupling switches, e.g., from antiferromagnetic to ferromagnetic, then this might generate a re-arrangement of the total spin in each lattice site as a \emph{purely electronic} phenomenon, due to the inter-site hopping. Hence, even without individual spin rotations, domains with an ordering which is different from the initial one may originate as a consequence of electronic transfer between atomic sites.

\section{Rotation of the spin quantization axes}

\subsection{Holstein-Primakoff bosons}

For each site, we define a rotation matrix, acting in the space of spinor fermionic operators, as:
\begin{align}
R_i(z) \equiv  \left( \begin{matrix} \sqrt{1 - \left| \xi_i(z) \right|^2} &  \xi_i^*(z) \\
                                           -\xi_i(z) & \sqrt{1 - \left| \xi_i(z) \right|^2} \end{matrix} \right),  
\label{R matrix}
\end{align}
where $z$ is a contour variable which parameterizes the Kadanoff-Baym contour. In Eq.\eqref{R matrix} we have introduced the boson fields
\begin{align}
\xi_i(z) \equiv - \text{e}^{\text{i}\varphi_i(z)} \sin\left[ \theta_i(z) / 2 \right] ,
\label{def xi} 
\end{align}
with $\theta_i \in \left[ 0, \pi \right[$ , $\varphi_i \in \left[ 0, 2 \pi \right[ $ being the polar angles that determine the spin axis on site $i$ at time $z$; it holds that $R_i^{\dagger}(z) \cdot R_i(z) = 1$.

We transform the Grassmann variables appearing in the action, Eq.\eqref{action S}, according to
\begin{align}
& \bar{\phi}_{a \pm}(t) = \bar{\psi}_{a \pm}(t) \cdot R^{\dagger}_{a \pm}(t), \quad \quad \phi_{a \pm}(t) =  R_{a \pm}(t) \cdot \psi_{a \pm}(t), \nonumber \\
& \bar{\phi}_{a v}(\tau) = \bar{\psi}_{a v}(\tau) \cdot R^{\dagger}_{a v}(\tau), \quad \quad \phi_{a v}(\tau) =  R_{a v}(\tau) \cdot \psi_{a v}(\tau).
\label{spin rotation}
\end{align}
To understand the meaning of the rotation that we have just introduced, the following considerations are in order. The local vector spin operator for site-orbital $a$, in the laboratory reference frame, is
\begin{align}
\hat{\boldsymbol{\sigma}}_a = \hat{\phi}^{\dagger}_a \cdot \boldsymbol{\sigma} \cdot  \hat{\phi}_a,
\label{spin operator}
\end{align}
where $\boldsymbol{\sigma}$ is the vector of Pauli matrices. The expectation value of this operator on the state $ \hat{\phi}^{\dagger}_{a \sigma} \left| 0 \right>$ is 
\begin{align}
\left< 0 \left| \hat{\phi}_{a \sigma}     \hat{\boldsymbol{\sigma}}_a        \hat{\phi}^{\dagger}_{a \sigma} \right| 0 \right> = \sigma \boldsymbol{u}_z ,
\end{align}
where $\sigma \in \lbrace \uparrow, \downarrow \rbrace \equiv \lbrace + , - \rbrace$. Instead, the expectation value of the spin operator on the state $ \hat{\psi}^{\dagger}_{a \sigma} \left| 0 \right>$ is 
\begin{align}
\left< 0 \left| \hat{\psi}_{a \sigma}     \hat{\boldsymbol{\sigma}}_a        \hat{\psi}^{\dagger}_{a \sigma} \right| 0 \right>  =   \sigma \boldsymbol{e}_a,
\label{exp value spin psi up}
\end{align}
where we have used Eqs.\eqref{def xi} and \eqref{spin rotation}, and the unit vector $\boldsymbol{e}_a$ is given by
\begin{align}
\boldsymbol{e}_a \equiv \boldsymbol{u}_x \sin(\theta_a) \cos(\varphi_a) + \boldsymbol{u}_y \sin(\theta_a) \sin(\varphi_a) + \boldsymbol{u}_z \cos(\theta_a).
\label{unit vector spin}
\end{align} 
Therefore, $\boldsymbol{e}_a$ (which depends on time) has the meaning of the unit spin vector on site-orbital $a$, measured in the laboratory reference frame, if the site $a$ hosts a $\psi_{\uparrow}$ electron \cite{Schulz90}. Expression \eqref{unit vector spin} can also be written as
\begin{align}
\boldsymbol{e}_a \equiv  \sqrt{1 - \left| \xi_a \right|^2} \Big[ -   \left( \xi_a + \xi^*_a \right) \boldsymbol{u}_x  +  \text{i}  \left( \xi_a - \xi^*_a \right) \boldsymbol{u}_y \Big] + \boldsymbol{u}_z \left( 1 - 2 \left| \xi_a \right|^2 \right),
\end{align} 
which shows that $\xi$ and $\xi^*$ are, effectively, nothing else than Holstein-Primakoff bosons \cite{SpinWavesBook, AuerbachBook, YosidaBook}.

\subsection{Transformed action}

We assume the rotation matrices $R(t)$ to be differentiable functions of $t$ over the interval $] t_0 + \epsilon, \infty[$, and the rotation matrices $R(\tau)$ to be differentiable functions of $\tau$ over the interval $]  \varepsilon, \beta[$. So, in the evaluation of Eq.\eqref{action S} we can already apply the replacements $R(t - \epsilon) \rightarrow R(t)$, $R(\tau - \varepsilon) \rightarrow R(\tau)$. Instead, we need to keep finite $\epsilon$ and $\varepsilon$ in the arguments of the fermionic fields $(\bar{\psi}, \psi)$.  

Since we are considering a on-site interaction, $H_V$ is rotationally invariant, hence it is unaffacted by the transformation defined in Eq.\eqref{spin rotation}, which means that $H_V[\bar{\phi}, \phi]  = H_V[\bar{\psi}, {\psi}]$. On the other hand, the single-particle Hamiltonian acquires a dependence on the bosonic fields, $H_T[\bar{\phi}, \phi; t]  = H_T[\bar{\psi}, {\psi}, \xi^*, \xi; t]$.

The transformed action, depending on the fermionic (Grassmann) $\psi$ fields and on the bosonic (complex) $\xi$ fields, is written as:
\begin{align}
S\left[\bar{\phi}, \phi \right] \equiv S\left[\bar{\psi}, {\psi}, \xi^{*}, \xi\right] = S\left[\bar{\psi}, \psi \right]+ S'\left[\bar{\psi}, {\psi}, \xi^{*}, \xi\right],
\end{align}
with $S\left[\bar{\psi}, \psi \right]$ denoting the original expression of the action, with the $\left( \bar{\phi}, \phi \right)$ fermions replaced by the $\left( \bar{\psi}, \psi \right)$ fermions, and
\begin{align}
S' & \left[\bar{\psi}, {\psi}, \xi^{*}, \xi\right]\equiv  \int_{\varepsilon}^{\beta} \text{d} \tau  \bar{\psi}_{v}(\tau) \cdot  \Delta_{v}(\tau) \cdot \psi_{v}(\tau - \varepsilon) \nonumber \\
&  + \int_{t_0 + \epsilon}^{\infty} \text{d}t  \Big[ \bar{\psi}_{+}(t) \cdot \Delta_{+}(t) \cdot \psi_{+}(t - \epsilon) - \bar{\psi}_{-}(t - \epsilon) \cdot  \Delta_{-}(t)  \cdot \psi_{-}(t) \Big] ,
\end{align}
where the quantities $ \Delta_{v}$ and $\Delta_{\pm}$ are matrices in both the spaces of Hubbard indexes and spin indexes, and they depend on the fields $( \xi_v, \xi^*_v )$ and $( \xi_{\pm}, \xi^*_{\pm} )$, respectively. Expressing explicitly their Hubbard-space structure,  
\begin{align}
& \Delta_{ab, \pm}(t)  =  \text{i}  R^{\dagger}_{a \pm}(t) \dot {R}_{a \pm}(t) \delta_{a b} -  T_{a b}(t) \left[ R^{\dagger}_{a \pm}(t) R_{b \pm}(t) - 1 \right] , \nonumber\\
& \Delta_{ab, v}(\tau) =  \text{i} R^{\dagger}_{a v}(\tau) \dot {R}_{a v}(\tau) \delta_{a b} +  \text{i} T_{a b} \left[ R^{\dagger}_{a v}(\tau) R_{b v}(\tau) - 1 \right] .
\end{align}
These are now matrices in spin space, which include diagonal and non-diagonal components. We separate the corresponding parts of the action:
\begin{align}
S' = S^{(1)} + S^{(2)}, 
\end{align}
with
\begin{align}
& S^{(1)} \equiv  \sum_{a} \sum_b \sum_{\sigma} \Bigg\{ \int_{\varepsilon}^{\beta} \text{d} \tau  \bar{\psi}^{\sigma}_{a v}(\tau) \, \Delta^{\sigma \bar{\sigma}}_{a b, v}(\tau) \, \psi^{\bar{\sigma}}_{b v}(\tau - \varepsilon)  \nonumber \\
& \quad \quad \quad + \int_{t_0 + \epsilon}^{\infty} \text{d}t  \Big[ \bar{\psi}^{\sigma}_{a +}(t) \, \Delta^{\sigma \bar{\sigma}}_{ab, +}(t) \, \psi^{\bar{\sigma}}_{b +}(t - \epsilon) - \bar{\psi}^{\sigma}_{a -}(t - \epsilon) \, \Delta^{\sigma \bar{\sigma}}_{ab, -}(t) \, \psi^{\bar{\sigma}}_{b -}(t) \Big] \Bigg\} , \nonumber \\
& S^{(2)} \equiv  \sum_{a} \sum_b \sum_{\sigma} \Bigg\{ \int_{\varepsilon}^{\beta} \text{d} \tau  \bar{\psi}^{\sigma}_{a v}(\tau) \, \Delta^{\sigma \sigma}_{a b, v}(\tau) \, \psi^{\sigma}_{b v}(\tau - \varepsilon) \nonumber \\
& \quad \quad \quad + \int_{t_0 + \epsilon}^{\infty} \text{d}t   \Big[ \bar{\psi}^{\sigma}_{a +}(t) \, \Delta^{\sigma \sigma}_{ab, +}(t) \, \psi^{\sigma}_{b +}(t - \epsilon) - \bar{\psi}^{\sigma}_{a -}(t - \epsilon) \, \Delta^{\sigma \sigma}_{ab, -}(t) \, \psi^{\sigma}_{b -}(t) \Big] \Bigg\},
\end{align}
where we have introduced the notation $\bar{\sigma} \equiv - \sigma$.

\subsection{Partition function}

The partition function is written in terms of the rotated fields as:
\begin{align}
\mathcal{Z} = \int \mathcal{D}\left[  \bar{\psi}, \psi \right] \text{e}^{\text{i} S \left[ \bar{\psi}, \psi\right]}  \int \mathcal{D}\left[  \theta, \varphi \right] \text{e}^{\text{i} S' \left[ \bar{\psi}, \psi, \xi^*(\theta, \varphi), \xi(\theta, \varphi) \right]} ,
\label{Z with bosons}
\end{align}
where the symbol 
\begin{align}
\int \mathcal{D}\left[  \theta, \varphi \right] = \prod_a  \left[ \frac{1}{4 \pi} \int_{0}^{\pi} \text{d} \theta_a \sin(\theta_a) \int_{0}^{2 \pi} \text{d} \varphi_a \right]
\end{align}
means integration over all possible orientations $(\theta_a, \varphi_a)$ at all lattice sites. The dependence of the fields $(\xi_a^*, \xi_a)$ on the angles $(\theta_a, \varphi_a)$ is given by Eq.\eqref{def xi}. It can be shown that integrating over the angles $(\theta_a, \varphi_a)$ ensures that the partition function as written in Eq.\eqref{Z with bosons} is equal to Eq.\eqref{Z initial}, i.e., Eq.\eqref{Z with bosons} is an identity. Since $\varphi$ is the phase of the complex number $-\xi$, while $\sin(\theta / 2)$ is its modulus, we see that
\begin{align}
& \int_{0}^{\pi} \text{d} \theta \sin(\theta) \int_{0}^{2 \pi} \text{d} \varphi \, f \left[ - \sin(\theta / 2) \text{e}^{\text{i} \varphi}, - \sin(\theta / 2) \text{e}^{-\text{i} \varphi} \right] \nonumber \\
& = 4 \iint_{C_1} \text{d} \text{Re}(\xi)  \; \text{d} \text{Im}(\xi) \; f \left[ \xi, \xi^* \right],
\end{align}
where the integration domain $C_1$ is the circle of radius equal to 1 in the complex plane, centered on $0$, i.e., $\left| \xi \right|^2 < 1$, which is described by $\varphi \in \left[0, 2 \pi \right[$ and $\theta \in \left[ 0, \pi\right[$. We therefore change the path variables from $(\theta, \varphi)$ to $(\xi, \xi^*)$, by introducing the notation
\begin{align}
\int \mathcal{D}\left[  \theta, \varphi \right] = \prod_a  \left[ \frac{1}{\pi} \iint_{C_1} \text{d} \text{Re}(\xi_a)  \; \text{d} \text{Im}(\xi_a)  \right] \equiv \int \mathcal{D}\left[  \xi^*, \xi \right],
\end{align}
so that
\begin{align}
\mathcal{Z} \equiv \int \mathcal{D}\left[  \bar{\psi}, \psi \right] \text{e}^{\text{i} S \left[ \bar{\psi}, \psi\right]}  \int \mathcal{D}\left[  \xi^*, \xi \right] \text{e}^{\text{i} S' \left[ \bar{\psi}, \psi, \xi^*, \xi \right]} .
\end{align}

\subsection{Low-energy theory}

The introduction of the Holstein-Primakoff bosons allows to decouple the dynamics of electronic and spin degrees of freedom. Now suppose that the equilibrium spin configuration is collinear (ferro-, antiferro-, or ferri- magnetic), with $\boldsymbol{u}_z$ being the initial direction of atomic spin alignment. Our goal is to derive a theory for the low-energy excitations on the top of the equilibrium configuration. Such excitations corrispond to small deviations of the atomic spins from the direction of $\boldsymbol{u}_z$, i.e., we can assume that the polar angles $\theta_i$ be small. We can therefore approximate
\begin{align}
R_i \approx \left( \begin{matrix} 1 - \frac{\left|\xi_i\right|^2}{2} &  \xi_i^* \\
                                           -\xi_i & 1 - \frac{\left|\xi_i\right|^2}{2} \end{matrix} \right).  
\end{align}
Now the unitarity of the transformation \eqref{spin rotation} holds but for corrections of the order of $\left| \xi_i\right|^4$. Under this approximation, 
\begin{align}
R^{\dagger}_{a s} \dot {R}_{as} \approx 
\left( \begin{matrix}  \frac{\dot{\xi}_{a s} \xi_{a s}^* - \xi_{a s} \dot{\xi}_{a s}^*}{2} &   \dot{\xi}_{a s}^* \\ -\dot{\xi}_{a s} & \frac{\xi_{a s} \dot{\xi}_{a s}^* - \dot{\xi}_{a s} \xi_{a s}^*}{2} \end{matrix} \right),
\end{align}
\begin{align}
R^{\dagger}_{a s} R_{b s} - 1 \approx 
\left( \begin{matrix} \xi^*_{a s} \xi_{b s} - \frac{\left| \xi_{a s} \right|^2 + \left| \xi_{b s} \right|^2}{2}   &  \xi_{b s}^* - \xi_{a s}^* \\ \xi_{a s} - \xi_{b s}   &  \xi_{a s} \xi^*_{b s} - \frac{\left| \xi_{a s} \right|^2 + \left| \xi_{b s} \right|^2}{2} \end{matrix} \right) .
\end{align}
This procedure is justified if we limit ourselves to the description of small rotations of the spins with respect to the direction of the initial quantization axis.

For future convenience, we define permutation operators $P_{a b}$, which substitute $a$ with $b$ and $b$ with $a$ in the expressions they act upon, where $a$ and $b$ are Hubbard indexes. Moreover, in order to adopt a compact notation, we will sometimes put arrows $\overrightarrow{ }, \overleftarrow{ }$ over the permutation and derivative operators to indicate the direction along which they act. The $\Delta$ matrices are written as:
\begin{align}
& \Delta^{\uparrow \downarrow}_{ab, \pm}(t) = \left[ \delta_{a b} \text{i} \frac{\overrightarrow{\partial}}{\partial t} + T_{a b}(t) \left( 1 - \overrightarrow{P}_{a b} \right) \right] \xi^*_{a, \pm}(t) , \nonumber \\
& \Delta^{\downarrow \uparrow}_{ab, \pm}(t) = - \left[ \delta_{a b} \text{i} \frac{\overrightarrow{\partial}}{\partial t} + T_{a b}(t) \left( 1 - \overrightarrow{P}_{a b} \right) \right] \xi_{a, \pm}(t) , 
\end{align}
\begin{align}
& \Delta^{\uparrow \downarrow}_{ab, v}(\tau) = \left[ \delta_{a b} \text{i} \frac{\overrightarrow{\partial}}{\partial \tau} - \text{i} T_{a b} \left( 1 - \overrightarrow{P}_{a b} \right) \right] \xi^*_{a v}(\tau) , \nonumber \\
& \Delta^{\downarrow \uparrow}_{ab, v}(\tau) = - \left[ \delta_{a b} \text{i} \frac{\overrightarrow{\partial}}{\partial \tau} - \text{i} T_{a b} \left( 1 - \overrightarrow{P}_{a b} \right) \right] \xi_{a v}(\tau) ;
\end{align}
\begin{align}
& \Delta^{\uparrow \uparrow}_{ab, \pm}(t) = \delta_{a b}  \xi_{a \pm}(t) \frac{\text{i}}{2}  \left( \frac{\overleftarrow\partial}{\partial t} - \frac{\overrightarrow\partial}{\partial t}\right) \xi^*_{a \pm}(t) \nonumber \\
& \quad \quad \quad \quad + T_{a b}(t) \left[ \frac{\left| \xi_{a \pm}(t) \right|^2 + \left| \xi_{b \pm}(t) \right|^2}{2} - \xi_{a \pm}^*(t) \xi_{b \pm}(t)\right],\nonumber \\
& \Delta^{\downarrow \downarrow}_{ab, \pm}(t) =  \delta_{a b}  \xi^*_{a \pm}(t) \frac{\text{i}}{2}  \left( \frac{\overleftarrow\partial}{\partial t} - \frac{\overrightarrow\partial}{\partial t}\right) \xi_{a \pm}(t) \nonumber \\
& \quad \quad \quad \quad + T_{a b}(t) \left[ \frac{\left| \xi_{a \pm}(t) \right|^2 + \left| \xi_{b \pm}(t) \right|^2}{2} - \xi_{a \pm}(t) \xi_{b \pm}^*(t)\right], 
\end{align}
\begin{align}
& \Delta^{\uparrow \uparrow}_{ab, v}(\tau) = \delta_{a b}  \xi_{a v}(\tau) \frac{\text{i}}{2}  \left( \frac{\overleftarrow\partial}{\partial \tau} - \frac{\overrightarrow\partial}{\partial \tau}\right) \xi^*_{a v}(\tau) \nonumber \\
& \quad \quad \quad \quad - \text{i} T_{a b} \left[ \frac{\left| \xi_{a v}(\tau) \right|^2 + \left| \xi_{b v}(\tau) \right|^2}{2} - \xi_{a v}^*(\tau) \xi_{b v}(\tau)\right],\nonumber \\
& \Delta^{\downarrow \downarrow}_{ab, v}(\tau) =  \delta_{a b}  \xi^*_{a v}(\tau) \frac{\text{i}}{2}  \left( \frac{\overleftarrow\partial}{\partial \tau} - \frac{\overrightarrow\partial}{\partial \tau}\right) \xi_{a v}(\tau) \nonumber \\
& \quad \quad \quad \quad -\text{i} T_{a b} \left[ \frac{\left| \xi_{a v}(\tau) \right|^2 + \left| \xi_{b v}(\tau) \right|^2}{2} - \xi_{a v}(\tau) \xi_{b v}^*(\tau)\right].
\end{align} 
It is seen that the $\Delta_{a b}^{\sigma \bar{\sigma}}$'s are linear, while the $\Delta_{a b}^{\sigma \sigma}$'s are quadratic, in the bosonic fields $(\xi^*, \xi)$.

To be consistent with the small $\theta$'s approximation, we must expand the partition function in series of the bosonic fields and retain only up to quadratic terms. We obtain 
\begin{align}
\mathcal{Z} & = \int \mathcal{D}\left[\bar{\psi}, \psi, \xi^*, \xi \right] \text{e}^{\text{i} \left \{ S\left[\bar{\psi}, \psi \right]+ S'\left[\bar{\psi}, {\psi}, \xi^{*}, \xi\right] \right\} } \nonumber \\
& \approx \int \mathcal{D}\left[\bar{\psi}, \psi, \xi^*, \xi \right] \text{e}^{\text{i} S\left[\bar{\psi}, \psi \right]} \left\{ 1 + \text{i} S'\left[\bar{\psi}, {\psi}, \xi^{*}, \xi\right] - \frac{1}{2} \left( S^{(1)}\left[\bar{\psi}, {\psi}, \xi^{*}, \xi\right] \right)^2  \right\} \nonumber \\
& \equiv \mathcal{Z}^{(0)} + \mathcal{Z}^{(1)} + \mathcal{Z}^{(2)},
\end{align}
where
\begin{align}
& \mathcal{Z}^{(0)} \equiv \int \mathcal{D}\left[\bar{\psi}, \psi, \xi^*, \xi \right] \text{e}^{\text{i} S\left[\bar{\psi}, \psi \right]} = \int \mathcal{D}\left[\xi^*, \xi \right] , \nonumber \\
& \mathcal{Z}^{(1)} \equiv \int \mathcal{D}\left[\bar{\psi}, \psi, \xi^*, \xi \right] \text{e}^{\text{i} S\left[\bar{\psi}, \psi \right]} \left\{   \text{i} S^{(1)}\left[\bar{\psi}, {\psi}, \xi^{*}, \xi\right]  - \frac{1}{2} \left( S^{(1)}\left[\bar{\psi}, {\psi}, \xi^{*}, \xi\right] \right)^2  \right\} , \nonumber \\
& \mathcal{Z}^{(2)} \equiv \int \mathcal{D}\left[\bar{\psi}, \psi, \xi^*, \xi \right] \text{e}^{\text{i} S\left[\bar{\psi}, \psi \right]} \text{i} S^{(2)}\left[\bar{\psi}, {\psi}, \xi^{*}, \xi\right].
\label{parts of Z}
\end{align}

\section{Fermionic correlators}

In the next sections, we will integrate out the fermionic fields, by exploiting the expression for the single-particle Green function (for $s \in \lbrace +, -, v \rbrace$, and $z = t$ or $z = \tau$, accordingly),
\begin{align}
\int \mathcal{D}\left[\bar{\psi}, \psi \right] \text{e}^{\text{i} S\left[\bar{\psi}, \psi \right]} \text{i} \bar{\psi}^{\sigma'}_{a s'}(z') \psi^{\sigma}_{b s}(z) \equiv  G_{b a, s s'}^{\sigma \sigma'}(z, z').
\end{align}
Since the Hamiltonian is spin-independent (i.e., the hopping is diagonal in spin space), the Green functions with $\sigma \ne \sigma'$ are zero in our system. Therefore, 
\begin{align}
&  \int \mathcal{D}\left[\bar{\psi}, \psi, \xi^*, \xi \right] \text{e}^{\text{i} S\left[\bar{\psi}, \psi \right]}    \text{i} S^{(1)}\left[\bar{\psi}, {\psi}, \xi^{*}, \xi\right]  = 0 \nonumber \\
& \Rightarrow \mathcal{Z}^{(1)} = \int \mathcal{D}\left[\bar{\psi}, \psi, \xi^*, \xi \right] \text{e}^{\text{i} S\left[\bar{\psi}, \psi \right]} \left\{   - \frac{1}{2} \left( S^{(1)}\left[\bar{\psi}, {\psi}, \xi^{*}, \xi\right] \right)^2  \right\} .
\label{simpl Z(1)}
\end{align}
As a consequence, the action will contain no terms linear in the bosonic fields, which is due to the fact that we are not including spin-orbit coupling. Therefore, our model will not include Dzyaloshinskii-Moriya interactions \cite{Katsnelson10}. Since we will need only spin-diagonal single-particle Green functions, we will label them with just one spin index. We also need the expression for the following two-particle Green function,
\begin{align}
- \int \mathcal{D}\left[\bar{\psi}, \psi \right] \text{e}^{\text{i} S\left[\bar{\psi}, \psi \right]} \bar{\psi}^{\sigma}_{a s}(z) \psi^{\bar{\sigma}}_{b s}(z) \bar{\psi}^{\bar{\sigma}}_{a' s'}(z') \psi^{\sigma}_{b' s'}(z')  =  \chi^{\bar{\sigma} \sigma \sigma \bar{\sigma}}_{b a b' a', s s s' s'}(z, z, z', z').
\end{align}
We adopt the approximation
\begin{align}
\chi^{\bar{\sigma} \sigma \sigma \bar{\sigma}}_{b a b' a', s s s' s'}(z, z, z', z') \approx - G^{\bar{\sigma}}_{b a', s s'}(z, z') G^{\sigma}_{b' a, s' s}(z', z),
\label{novertex}
\end{align}
which corresponds to neglecting the vertex in the two-particle Dyson equation, and where we have already taken into account the fact that correlators $G^{\sigma \bar{\sigma}}$ are zero in our system. It must be noted that Eq.\eqref{novertex} is the only approximation that we adopt on the many-body level. For the equilibrium case, the same approximation was used in Ref.\cite{Katsnelson02} and it was shown that it leads to the standard expression for exchange parameters \cite{Lichtenstein87, Katsnelson00}. 

All quantities will then be written in terms of single-particle Green functions. The correlators can be classified according to the positions of their time arguments on the Kadanoff-Baym contour, and put in correspondence with non-equilibrium Green functions written in terms of the field operators, as follows:
\begin{align}
& G^{\sigma}_{b a, + +}(t_1, t_2) \equiv G^{\mathbb{T} \sigma}_{b a}(t_1, t_2) = -\text{i} \left<  \mathcal{T}_t \left[ \hat{\psi}_b^{\sigma}(t_1) \hat{\psi}_a^{\dagger \sigma}(t_2) \right] \right> , \nonumber \\
& G^{\sigma}_{b a, - -}(t_1, t_2) \equiv G^{\widetilde{\mathbb{T}} \sigma}_{b a}(t_1, t_2) = -\text{i} \left<  \widetilde{\mathcal{T}}_t \left[ \hat{\psi}_b^{\sigma}(t_1) \hat{\psi}_a^{\dagger \sigma}(t_2) \right] \right>, \nonumber \\
& G^{\sigma}_{b a, + -}(t_1, t_2) \equiv G^{< \sigma}_{b a}(t_1, t_2) = \text{i} \left<   \hat{\psi}_a^{\dagger \sigma}(t_2)  \hat{\psi}_b^{\sigma}(t_1)   \right>, \nonumber \\
& G^{\sigma}_{b a, - +}(t_1, t_2) \equiv G^{> \sigma}_{b a}(t_1, t_2) = - \text{i} \left<  \hat{\psi}_b^{\sigma}(t_1) \hat{\psi}_a^{\dagger \sigma}(t_2)     \right>, \nonumber \\
& G^{\sigma}_{b a, v v}(\tau_1, \tau_2) \equiv G^{\mathbb{M} \sigma}_{b a}(\tau_1 - \tau_2) = -\text{i} \left<  \mathcal{T}_{\tau} \left[ \hat{\psi}_b^{\sigma}(\tau_1) \hat{\psi}_a^{\dagger \sigma}(\tau_2) \right] \right> , \nonumber \\
& G^{\sigma}_{b a, \pm v}(t, \tau) \equiv G^{\urcorner \sigma}_{b a}(t, \tau) = \text{i} \left< \hat{\psi}^{\dagger \sigma}_a(\tau) \hat{\psi}^{\sigma}_b(t) \right>, \nonumber \\
& G^{\sigma}_{b a, v \pm}(\tau, t) \equiv G^{\ulcorner \sigma}_{b a}(\tau, t) = - \text{i} \left< \hat{\psi}^{\sigma}_b(\tau) \hat{\psi}^{\dagger \sigma}_a(t) \right>,
\label{correlators}
\end{align}  
where we have used the notation 
\begin{align}
\left< \hat{O}(z_1, z_2) \right> \equiv   \text{Tr}\left\{ \text{e}^{-\beta \left( \hat{H}_0 - \mu \hat{N} \right) }  \hat{O}(z_1, z_2) \right\} \Big/ \text{Tr}\left\{ \text{e}^{-\beta \left( \hat{H}_0 - \mu \hat{N} \right) }\right\},
\end{align}
and
\begin{align}
& \hat{\psi}(t) \equiv \hat{U}(t_0, t) \hat{\psi} \hat{U}(t, t_0), \nonumber \\
& \hat{\psi}(\tau) \equiv \hat{U}_v(t_0, t_0 - \text{i} \tau) \hat{\psi} \hat{U}_v(t_0 - \text{i} \tau, t_0) = \text{e}^{\tau \left( \hat{H}_0 - \mu \hat{N} \right)} \hat{\psi} \text{e}^{- \tau \left( \hat{H}_0 - \mu \hat{N} \right)},
\end{align}
where the identity $\hat{U}_v(t_0 , t_0 - \text{i} \tau) = \text{e}^{\tau \left( \hat{H}_0 - \mu \hat{N} \right)}$ follows from the fact that the Hamiltonian is constant (equal to $\hat{H}_0$) on the vertical branch of the Kadanoff-Baym contour. The correlators $G^{< \sigma}_{b a}(t_1, t_2)$ and $G^{> \sigma}_{b a}(t_1, t_2)$ are continuous functions of their time arguments $(t_1, t_2)$, while $G^{\mathbb{T} \sigma}_{b a}(t_1, t_2)$ and $G^{\widetilde{\mathbb{T}} \sigma}_{b a}(t_1, t_2)$ are discontinuous across the line $t_1 = t_2$ if $b = a$. In fact, from the definitions of the Green functions related to the real-time branches, it follows that, for $t \ne t'$,
\begin{align}
& G^{\mathbb{T}}_{b a}(t, t') = \Theta(t - t') G^>_{b a}(t, t') + \Theta(t' - t) G^<_{b a}(t, t'), \nonumber \\
& G^{\widetilde{\mathbb{T}}}_{b a}(t, t') = \Theta(t' - t) G^>_{b a}(t, t') + \Theta(t - t') G^<_{b a}(t, t'),
\label{discontinuity relations}
\end{align}
where $\Theta(t - t')$ is the step function. Therefore,
\begin{align}
& \lim_{\epsilon \rightarrow 0^+} G^{\mathbb{T}}_{b a}(t, t + \epsilon) = \lim_{\epsilon \rightarrow 0^+} G^{\mathbb{T}}_{b a}(t - \epsilon, t) = G^<_{b a}(t, t), \nonumber \\
& \lim_{\epsilon \rightarrow 0^+} G^{\mathbb{T}}_{b a}(t, t - \epsilon) = \lim_{\epsilon \rightarrow 0^+} G^{\mathbb{T}}_{b a}(t + \epsilon, t) = G^>_{b a}(t, t),
\label{T disc}
\end{align}
and
\begin{align}
& \lim_{\epsilon \rightarrow 0^+} G^{\widetilde{\mathbb{T}}}_{b a}(t, t + \epsilon) = \lim_{\epsilon \rightarrow 0^+} G^{\widetilde{\mathbb{T}}}_{b a}(t - \epsilon, t) = G^>_{b a}(t, t), \nonumber \\
& \lim_{\epsilon \rightarrow 0^+} G^{\widetilde{\mathbb{T}}}_{b a}(t, t - \epsilon) = \lim_{\epsilon \rightarrow 0^+} G^{\widetilde{\mathbb{T}}}_{b a}(t + \epsilon, t) = G^<_{b a}(t, t).
\label{anti-T disc}
\end{align}
The discontinuity is:
\begin{align}
G^<_{b a}(t, t) - G^>_{b a}(t, t) = \text{i} \delta_{a b}.
\label{discontinuity > <}
\end{align}
We introduce the notation
\begin{align}
\rho_{b a}(t) \equiv - \text{i} \, G^{<}_{b a}(t, t) = \left<   \hat{\psi}_a^{\dagger}(t) \, \hat{\psi}_b(t)   \right> = \left<  \hat{U}(t_0, t) \, \hat{\psi}_a^{\dagger} \, \hat{\psi}_b  \hat{U}(t, t_0)   \right> 
\label{pair correlation function}
\end{align}
to denote the time-dependent pair correlation function; in the case $a = b$ we will drop one of the identical subscripts and denote as $\rho_a(t)$ the occupation number on site-orbital $a$ at time $t$. From the definitions, Eqs.\eqref{correlators}, it is seen that the following property holds:
\begin{align}
\left[ G^{\lessgtr}_{a b}(t, t')   \right]^* = - G^{\lessgtr}_{b a}(t', t) ,
\label{cc G mu}
\end{align}

Since the Hamiltonian is constant on the vertical branch of the Kadanoff-Baym contour, the Matsubara Green function depends on $(\tau, \tau')$ only via the difference $(\tau - \tau')$, as indicated in Eqs.\eqref{correlators}. Therefore, the Matsubara Green function $G^{\mathbb{M}}(\tau)$ depends on one time argument only, $\tau \in (-\beta, \beta)$. It is convenient to put
\begin{align}
G^{\mathbb{M}}_{b a}( \tau - \tau' )  \equiv \Theta(\tau - \tau') G^{\mathbb{D}}_{b a}(\tau - \tau') + \Theta(\tau' - \tau) G^{\mathbb{U}}_{b a}(\tau - \tau'), 
\label{formally Matsubara}
\end{align}
which is consistent with the definition of $G^{\mathbb{M}}$ given in Eqs.\eqref{correlators}. One can immediately see that
\begin{align}
&  \lim_{\varepsilon \rightarrow 0^+} G_{b a}^{\mathbb{M}}(- \varepsilon) = G^{\mathbb{U}}_{b a}(0) = \text{i} \left<  \hat{\psi}^{\dagger}_a \hat{\psi}_b \right> = G_{b a}^<(t_0, t_0) \equiv \text{i} \rho_{b a}, \nonumber \\
&  \lim_{\varepsilon \rightarrow 0^+} G_{b a}^{\mathbb{M}}(\varepsilon) = G^{\mathbb{D}}_{b a}(0) = - \text{i} \left<  \hat{\psi}_b \hat{\psi}^{\dagger}_a  \right> = G_{b a}^>(t_0, t_0) \equiv \text{i} ( \rho_{b a} - \delta_{a b} ),
\label{boundary 0 Matsubara}
\end{align}
where we have introduced the equilibrium pair correlation function, i.e., the pair correlation function at $t = t_0$, $\rho_{b a} \equiv - \text{i} G_{b a}^<(t_0, t_0)$. Therefore,
\begin{align}
\lim_{\varepsilon \rightarrow 0^+} \left[ G_{b a}^{\mathbb{M}}(- \varepsilon) - G_{b a}^{\mathbb{M}}(\varepsilon) \right]   = \text{i} \delta_{a b}.
\end{align}
Values of $G^{\mathbb{M}}(\tau)$ at the borders $\pm \beta$ and close to $0$ are related:
\begin{align}
G^{\mathbb{M}}(\beta) = - G^{\mathbb{M}}(-\varepsilon), \quad \quad \quad G^{\mathbb{M}}(-\beta) = - G^{\mathbb{M}}(\varepsilon),
\end{align}
where $\varepsilon \rightarrow 0^+$.

Finally, the functions $G^{\urcorner}$ and $G^{\ulcorner}$ are continuous, and they satisfy the following boundary conditions:
\begin{align}
&  G^{\urcorner}_{b a}(t_0, 0) = \text{i} \left< \hat{\psi}^{\dagger}_a \hat{\psi}_b \right> = \text{i} \rho_{b a}, \nonumber \\
&  G^{\ulcorner}_{b a}(0, t_0) = - \text{i} \left< \hat{\psi}_b \hat{\psi}^{\dagger}_a \right> = \text{i} ( \rho_{b a} - \delta_{a b} )  .
\end{align}

In the following, we will express the partition function in terms of an effective action involving the $(\xi, \xi^*)$ fields only, which will be achieved by integrating the fermionic fields out, by means of Eqs.\eqref{correlators}. We will apply the Keldysh rotation to the bosonic fields living on the Schwinger-Keldysh contour \cite{KamenevBook},
\begin{align}
\left( \begin{matrix} \xi_{a +}(t) \\ \xi_{a -}(t)\end{matrix}\right) = \frac{1}{\sqrt{2}} \left( \begin{matrix} 1 & 1 \\ 1 & -1 \end{matrix}\right) \left( \begin{matrix} \xi_{a C}(t) \\ \xi_{a Q}(t)\end{matrix}\right) .
\label{Keldysh rotation}
\end{align}

\subsection{Evaluation of $\mathcal{Z}^{(1)}$}

We now calculate $\mathcal{Z}^{(1)}$ explicitly. From Eq.\eqref{simpl Z(1)}, we see that this quantity requires the evaluation of double time integrals only. Since the domain where the Green functions $G(z, z')$ is discontinuous is one-dimensional (the line $z =z'$), so it has zero measure with respect to the integration domain, we can already take the limits $\epsilon \rightarrow 0^+$ and $\varepsilon \rightarrow 0^+$. This gives:
\begin{align}
& \mathcal{Z}^{(1)}  = - \int \mathcal{D}\left[\xi^*, \xi \right] \sum_{a, a'} \sum_{b, b'} \Bigg\{  \int_{t_0}^{\infty} \text{d}t  \int_{t_0}^{\infty} \text{d}t'  \left( \begin{matrix} \Delta^{\uparrow \downarrow}_{a b, +}(t), &  \Delta^{\uparrow \downarrow}_{a b, -}(t) \end{matrix} \right) \nonumber \\
& \quad \quad \quad \cdot \left( \begin{matrix} G_{b a'}^{\mathbb{T} \downarrow}(t, t') G_{b' a}^{\mathbb{T} \uparrow}(t', t) & 
                     -G_{b a'}^{< \downarrow}(t, t') G_{b' a}^{> \uparrow}(t', t) \\
                     -G_{b a'}^{> \downarrow}(t, t') G_{b' a}^{< \uparrow}(t', t) &
                      G_{b a'}^{\widetilde{\mathbb{T}} \downarrow}(t, t') G_{b' a}^{\widetilde{\mathbb{T}} \uparrow}(t', t)
\end{matrix} \right)
\left( \begin{matrix} \Delta^{\downarrow \uparrow}_{a' b', +}(t') \\  \Delta^{\downarrow \uparrow}_{a' b', -}(t') \end{matrix} \right) \nonumber \\
& + \int_{0}^{\beta} \text{d} \tau \int_{0}^{\beta} \text{d} \tau' \Delta^{\uparrow \downarrow}_{a b, v}(\tau) \, \Delta^{\downarrow \uparrow}_{a' b', v}(\tau') \, G^{\mathbb{M} \downarrow}_{b a'}(\tau - \tau') \, G^{\mathbb{M} \uparrow}_{b' a}(\tau' - \tau) \nonumber \\
&  + \int_{t_0}^{\infty}  \text{d}t \int_{0}^{\beta}  \text{d} \tau \, \sum_{\sigma}  \Big[ \Delta^{\sigma \bar{\sigma}}_{a b, +}(t) - \Delta^{\sigma \bar{\sigma}}_{a b, -}(t)\Big] \, \Delta^{\bar{\sigma} \sigma}_{a' b', v}(\tau) \, G^{\urcorner \bar{\sigma}}_{b a'}(t , \tau) \, G^{\ulcorner \sigma}_{b' a}(\tau , t)  \Bigg\}.
\label{Z(1) first passage}
\end{align}

We now perform the Keldysh rotation. Since the $\Delta_{\pm}$'s are linear functionals of the $\xi_{\pm}$'s, they transform under the Keldysh rotation exactly as the fields themselves, i.e., 
\begin{align}
& \left( \begin{matrix} \Delta^{\uparrow \downarrow}_{a b, +}(t), &  \Delta^{\uparrow \downarrow}_{a b, -}(t) \end{matrix} \right) = 
\left( \begin{matrix} \Delta^{\uparrow \downarrow}_{a b, C}(t), &  \Delta^{\uparrow \downarrow}_{a b, Q}(t) \end{matrix} \right) \frac{1}{\sqrt{2}} \left( \begin{matrix} 1 & 1 \\ 1 & -1 \end{matrix}\right), \nonumber \\
& \left( \begin{matrix} \Delta^{\downarrow \uparrow}_{a' b', +}(t') \\  \Delta^{\downarrow \uparrow}_{a' b', -}(t') \end{matrix} \right) =
\frac{1}{\sqrt{2}} \left( \begin{matrix} 1 & 1 \\ 1 & -1 \end{matrix}\right) 
\left( \begin{matrix} \Delta^{\downarrow \uparrow}_{a' b', C}(t') \\  \Delta^{\downarrow \uparrow}_{a' b', Q}(t') \end{matrix} \right).
\end{align}
Therefore, the first term in Eq.\eqref{Z(1) first passage} produces a coupling between the $\xi_C$ and the $\xi_Q$ fields via the matrix
\begin{align}
& \frac{1}{2} \left( \begin{matrix} 1 & 1 \\ 1 & -1 \end{matrix}\right) \left( \begin{matrix} 
                      G_{b a'}^{\mathbb{T} \downarrow}(t, t') G_{b' a}^{\mathbb{T} \uparrow}(t', t) & 
                     -G_{b a'}^{< \downarrow}(t, t') G_{b' a}^{> \uparrow}(t', t) \\
                     -G_{b a'}^{> \downarrow}(t, t') G_{b' a}^{< \uparrow}(t', t) &
                      G_{b a'}^{\widetilde{\mathbb{T}} \downarrow}(t, t') G_{b' a}^{\widetilde{\mathbb{T}} \uparrow}(t', t)
\end{matrix} \right)  \left( \begin{matrix} 1 & 1 \\ 1 & -1 \end{matrix}\right) \nonumber \\
& \equiv \left( \begin{matrix} 0 & \chi^A_{b' a b a'}(t, t') \\ \chi^R_{b' a b a'}(t, t') & \chi^K_{b' a b a'}(t, t') \end{matrix} \right),
\end{align}
whose non-zero elements are:
\begin{align}
& \chi^K_{b' a b a'}(t, t') \equiv G_{b a'}^{< \downarrow}(t, t') \, G_{b' a}^{> \uparrow}(t', t) + G_{b a'}^{> \downarrow}(t, t') \, G_{b' a}^{< \uparrow}(t', t) , \nonumber \\
& \chi^A_{b' a b a'}(t, t') \equiv \Theta(t' - t) \left[ G_{b a'}^{< \downarrow}(t, t') \, G_{b' a}^{> \uparrow}(t', t) - G_{b a'}^{> \downarrow}(t, t') \, G_{b' a}^{< \uparrow}(t', t) \right] , \nonumber \\
& \chi^R_{b' a b a'}(t, t') \equiv - \Theta(t - t') \left[ G_{b a'}^{< \downarrow}(t, t') \, G_{b' a}^{> \uparrow}(t', t) - G_{b a'}^{> \downarrow}(t, t') \, G_{b' a}^{< \uparrow}(t', t) \right] .
\label{RAK functions}
\end{align}
The third term in Eq.\eqref{Z(1) first passage} produces a coupling between the $\xi_Q$ and the $\xi_v$ fields, since $\Delta^{\sigma \bar{\sigma}}_{a b, +}(t) - \Delta^{\sigma \bar{\sigma}}_{a b, -}(t) = \sqrt{2} \Delta^{\sigma \bar{\sigma}}_{a b, Q}(t)$. Thus, $\mathcal{Z}^{(1)}$ gives no coupling at all between the $\xi_C$ and the $\xi_v$ fields. Combining the three terms, the resulting expression is
\begin{align}
\mathcal{Z}^{(1)} & = -   \int \mathcal{D}\left[\xi^*, \xi \right] \sum_{a, a'} \sum_{b, b'} \Bigg\{ \int_{t_0}^{\infty} \text{d}t  \int_{t_0}^{\infty} \text{d}t'  \left( \begin{matrix} \Delta^{\uparrow \downarrow}_{a b, C}(t), &  \Delta^{\uparrow \downarrow}_{a b, Q}(t) \end{matrix} \right) \nonumber \\
& \quad \quad \quad \cdot \left( \begin{matrix} 0 & \chi^A_{b' a b a'}(t, t') \\ \chi^R_{b' a b a'}(t, t') & \chi^K_{b' a b a'}(t, t') \end{matrix} \right)
\left( \begin{matrix} \Delta^{\downarrow \uparrow}_{a' b', C}(t') \\  \Delta^{\downarrow \uparrow}_{a' b', Q}(t') \end{matrix} \right)  \nonumber \\
& +   \int_{0}^{\beta} \text{d} \tau \int_{0}^{\beta} \text{d} \tau' \Delta^{\uparrow \downarrow}_{a b, v}(\tau) \, G^{\mathbb{M} \downarrow}_{b a'}(\tau - \tau') \, \Delta^{\downarrow \uparrow}_{a' b', v}(\tau')  \, G^{\mathbb{M} \uparrow}_{b' a}(\tau' - \tau) \nonumber \\
&  + \sqrt{2} \int_{t_0}^{\infty} \text{d}t \int_0^{\beta} \text{d} \tau \sum_{\sigma = \uparrow, \downarrow}  \Delta^{\sigma \bar{\sigma}}_{a b, Q}(t) \, G^{\urcorner \bar{\sigma}}_{b a'}(t, \tau) \, \Delta^{\bar{\sigma} \sigma}_{a' b', v}(\tau)  \, G^{\ulcorner \sigma}_{b' a}(\tau, t)  \Bigg\}.
\label{Z(1) second passage}
\end{align}

\subsection{Evaluation of $\mathcal{Z}^{(2)}$}

We now calculate $\mathcal{Z}^{(2)}$ explicitly:
\begin{align}
\mathcal{Z}^{(2)} = & \int  \mathcal{D}\left[\xi^*, \xi \right] \sum_{a} \sum_b \sum_{\sigma} \Bigg\{ G_{b a}^{\mathbb{M} \sigma}(- \varepsilon) \int_{\varepsilon}^{\beta}  \text{d} \tau   \, \Delta^{\sigma \sigma}_{a b, v}(\tau)  \nonumber \\
&  + \int_{t_0 + \epsilon}^{\infty} \! \text{d}t \,  \left[ G^{\mathbb{T} \sigma}_{b a}(t - \epsilon, t) \, \Delta^{\sigma \sigma}_{a b, +}(t) - G^{\widetilde{\mathbb{T}} \sigma}_{b a}(t, t - \epsilon) \, \Delta^{\sigma \sigma}_{a b, -}(t) \right]   \Bigg\},
\label{Z(2) first passage}
\end{align}
where we have used the identity $G_{b a, vv}^{\sigma}(\tau - \varepsilon, \tau) \equiv G_{b a}^{\mathbb{M} \sigma}(- \varepsilon)$. From Eq.\eqref{formally Matsubara}, it follows that 
\begin{align}
\lim_{\varepsilon \rightarrow 0^+} G_{b a}^{\mathbb{M} \sigma}(- \varepsilon) = G^{\mathbb{U} \sigma}_{b a}(0).
\end{align}
From Eqs.\eqref{T disc} and \eqref{anti-T disc} we see that
\begin{align}
\lim_{\epsilon \rightarrow 0^+} G^{\mathbb{T} \sigma}_{b a}(t - \epsilon, t) = \lim_{\epsilon \rightarrow 0^+} G^{\widetilde{\mathbb{T}} \sigma}_{b a}(t, t - \epsilon) = \text{i} \left< \hat{U}(t_0, t)\hat{\psi}_a^{\dagger \sigma}  \hat{\psi}_b^{\sigma} \hat{U}(t, t_0)\right> = G^{< \sigma}_{b a}(t, t).
\end{align}
These substitutions remove the discontinuous correlators and we are thus allowed to put $\epsilon = 0$ and $\varepsilon = 0$ in the integration limits in Eq.\eqref{Z(2) first passage}. This gives:
\begin{align}
\mathcal{Z}^{(2)} = & \int \mathcal{D}\left[\xi^*, \xi \right] \sum_{a} \sum_b \sum_{\sigma} \Bigg\{ \int_{t_0}^{\infty} \text{d}t \, G_{b a}^{< \sigma}(t, t)  \, \left[ \Delta^{\sigma \sigma}_{a b, +}(t) - \Delta^{\sigma \sigma}_{a b, -}(t) \right]  \nonumber \\
& \quad \quad + G^{\mathbb{U} \sigma}_{b a}(0) \int_{0}^{\beta} \text{d} \tau \, \Delta^{\sigma \sigma}_{a b, v}(\tau) \Bigg\}.
\label{Z(2) second passage}
\end{align}
We now perform the Keldysh rotation, defined in Eq.\eqref{Keldysh rotation}. The $\Delta^{\sigma \sigma}$'s are not linear in the $\xi$ fields, differently from the $\Delta^{\sigma \bar{\sigma}}$'s, so they do not transform linearly. The result for $\mathcal{Z}^{(2)}$ is:
\begin{align}
\mathcal{Z}^{(2)} & =  \int \mathcal{D}\left[\xi^*, \xi \right]  \sum_{a} \sum_b \sum_{\sigma} \Bigg\{ \int_{t_0 }^{\infty} \text{d}t \, G_{b a}^{< \sigma}(t, t) \nonumber \\
& \quad \quad \cdot \sum_{\kappa = C, Q}  \Bigg[ \delta_{a b} \sigma   \frac{\text{i}}{2}  \left( \dot{\xi}_{a \kappa}(t) \xi^*_{a \bar{\kappa}}(t) - \xi_{a \kappa}(t) \dot{\xi}^*_{a \bar{\kappa}}(t) \right)  \nonumber \\
&  \quad \quad + T_{a b}(t) \frac{1 + \overrightarrow{P}_{a b}}{2} \xi_{a \kappa}(t)  \xi^*_{a \bar{\kappa}}(t)
- T_{a b}(t) \left( \delta_{\sigma \uparrow} \overrightarrow{P}_{a b} + \delta_{\sigma \downarrow} \right)  \xi_{a \kappa}(t) \xi^*_{b \bar{\kappa}}(t) \Bigg] \nonumber \\
& \quad + G^{\mathbb{U} \sigma}_{b a}(0) \int_{0}^{\beta} \text{d} \tau \, \Delta^{\sigma \sigma}_{a b, v}(\tau) \Bigg\},
\end{align}
where we have defined the Keldysh index $\kappa \in \lbrace C, Q \rbrace$, and with $\bar{\kappa}$ we denote the complementary of $\kappa$.

\section{Effective action}

The three parts of the total partition function can be written, respectively, as
\begin{align}
\mathcal{Z}^{(0)} = \! \int \mathcal{D}\!\left[\xi^*, \xi \right], \quad \mathcal{Z}^{(1)} \equiv \! \int \mathcal{D}\!\left[\xi^*, \xi \right] \text{i} S_1\left[\xi^*, \xi \right], \quad \mathcal{Z}^{(2)} \equiv \! \int \mathcal{D}\!\left[\xi^*, \xi \right]  \text{i} S_2\left[\xi^*, \xi \right], 
\end{align}
where $S_1$ and $S_2$ are effective actions which depend functionally on quadratic combinations of the fields $(\xi^*, \xi)$. We can put
\begin{align}
\mathcal{Z} =   \int \mathcal{D}\left[\xi^*, \xi \right] \Big( 1 + \text{i} S_1\left[\xi^*, \xi \right] + \text{i} S_2\left[\xi^*, \xi \right] \Big) \approx   \int \mathcal{D}\left[\xi^*, \xi \right] \text{e}^{\text{i} S\left[\xi^*, \xi \right] },
\end{align}
where the effective action is $S\left[\xi^*, \xi \right] \equiv S_1\left[\xi^*, \xi \right] + S_2\left[\xi^*, \xi \right]$. This total action can be written as the sum of several terms, all quadratic in the Holstein-Primakoff fields, which we classify according to which fields they couple. We label as $S_{x y}$ the action term coupling the fields $\xi^*_x$ and $\xi_y$, with $(x, y) \in \lbrace C, Q, v \rbrace$. $S_1$ provides terms $S_{Q Q}, S_{C Q}, S_{Q C}, S_{v v}, S_{Q v}, S_{v Q}$, while $S_2$ provides terms $S_{C Q}, S_{Q C}, S_{v v}$. The total action is written as:
\begin{align}
S = S_{Q Q} + S_{C Q} + S_{Q C} + S_{v v} + S_{Q v} + S_{v Q}.
\end{align}
The individual terms, including contributions from both $S_1$ and $S_2$, can be presented as follows:
 \begin{align}
& \text{i} S_{Q Q} = \int_{t_0}^{\infty} \text{d}t \int_{t_0}^{\infty} \text{d}t' \,  \xi_{Q}^*(t) \cdot  \hat{A}(t, t') \cdot \xi_{Q}(t'), \nonumber \\
& \text{i} S_{C Q} = \int_{t_0}^{\infty} \text{d}t \left\{ \xi^*_{C}(t) \cdot \hat{B}^{(1)}(t) \cdot \xi_{Q}(t) + \int_{t_0}^{\infty} \text{d}t' \, \xi^*_{C}(t) \cdot \hat{B}^{(2)}(t, t') \cdot \xi_{Q}(t') \right\}, \nonumber \\
& \text{i} S_{Q C} = \int_{t_0}^{\infty} \text{d}t \left\{ \xi^*_{Q}(t) \cdot \hat{C}^{(1)}(t) \cdot \xi_{C}(t) + \int_{t_0}^{\infty} \text{d}t' \, \xi^*_{Q}(t) \cdot \hat{C}^{(2)}(t, t') \cdot \xi_{C}(t') \right\}, \nonumber \\
& \text{i} S_{v v} = \int_{0}^{\beta} \text{d}\tau \left\{ \xi^*_{v}(\tau) \cdot \hat{D}^{(1)}(\tau) \cdot \xi_{v}(\tau) + \int_{0}^{\beta} \text{d}\tau' \, \xi^*_{v}(\tau) \cdot \hat{D}^{(2)}(\tau, \tau') \cdot \xi_{v}(\tau') \right\}, \nonumber \\
& \text{i} S_{Q v} = \int_{t_0}^{\infty} \text{d}t \int_{0}^{\beta} \text{d} \tau \, \xi_{Q}^*(t) \cdot \hat{E}(t, \tau) \cdot \xi_{v}(\tau), \nonumber \\
& \text{i} S_{v Q} = \int_{t_0}^{\infty} \text{d}t \int_{0}^{\beta} \text{d} \tau \, \xi_{v}^*(\tau) \cdot \hat{F}(\tau, t) \cdot \xi_{Q}(t) ,
\label{action individual terms}
\end{align}
where all quantities are either vectors or matrices in Hubbard space, and we have introduced several kernel operators coupling the bosonic fields, which contain time derivatives acting either on the fields at their right side or at their left side. Such operators can be expressed conveniently by using the property that the arrow over a permutation operator can be flipped under the sign of summation over both its indexes, i.e.,
\begin{align}
\sum_{a} \sum_b f_{a b} \overleftarrow{P}_{a b} g_{a b} = \sum_{a} \sum_b f_{b a}  g_{a b} = \sum_{a} \sum_b f_{a b}  g_{b a} = \sum_{a} \sum_b f_{a b} \overrightarrow{P}_{a b} g_{a b};
\end{align}
this property allows to write the individual action terms in Eq.\eqref{action individual terms} as matrix products in Hubbard space. To simplify the notation, for each spin-dependent function $f^{\sigma}$ we define the \emph{charge} combination, $f^{\text{C}} \equiv \left( f^{\uparrow} + f^{\downarrow} \right) / 2$, as well as the \emph{spin} combination, $f^{\text{S}} \equiv \left( f^{\uparrow} - f^{\downarrow} \right) / 2$. We further use Eqs.\eqref{pair correlation function} and \eqref{boundary 0 Matsubara}. The expressions of the kernel operators are then given by:
\begin{align}
\hat{A}_{a a'}(t, t') =  \sum_b \sum_{b'} & \left[ \delta_{ab} \text{i} \overleftarrow{ \frac{\partial}{\partial t} } +  \left( 1 - \overrightarrow{P}_{ab} \right) T_{ab}(t) \right] \chi^K_{b' a b a'}(t, t') \nonumber \\
& \times \left[ \delta_{a' b'} \text{i} \overrightarrow{ \frac{\partial}{\partial t'} } + T_{a' b'}(t') \left( 1 - \overleftarrow{P}_{a' b'} \right) \right],
\label{operator A}
\end{align}
\begin{align}
\hat{B}^{(1)}_{a a'}(t) \equiv & \, \delta_{a a'} \Bigg\{  \frac{\overleftarrow{\partial}}{\partial t} \rho^{\text{S}}_{a}(t) - \rho^{\text{S}}_{a}(t) \frac{\overrightarrow{\partial}}{\partial t}  + \text{i} \sum_b \Big[ \rho^{\text{C}}_{a b}(t) \, T_{b a}(t) +   T_{a b}(t) \, \rho^{\text{C}}_{b a}(t) \Big]  \Bigg\}  \nonumber \\
& - \text{i} \left[ \rho^{\uparrow}_{a' a}(t) \, T_{a a'}(t) + \rho^{\downarrow}_{a a'}(t) \, T_{a' a}(t) \right],
\label{operator B1}
\end{align}
\begin{align}
\hat{B}^{(2)}_{a a'}(t, t') \equiv \sum_b \sum_{b'} &  \left[ \delta_{a b} \text{i} \frac{\overleftarrow{\partial}}{\partial t} + \left( 1 - \overrightarrow{P}_{a b} \right) \, T_{a b}(t) \right] \chi^{A}_{b' a b a'}(t, t') \nonumber \\ 
& \times \left[ \delta_{a' b'} \text{i} \frac{\overrightarrow{\partial}}{\partial t'} + T_{a' b'}(t') \left( 1 - \overleftarrow{P}_{a' b'} \right)  \right],
\label{operator B2}
\end{align}
\begin{align}
\hat{C}^{(1)}_{a a'}(t) = \hat{B}^{(1)}_{a a'}(t),
\label{operator C1}
\end{align}
\begin{align}
\hat{C}^{(2)}_{a a'}(t, t') \equiv \sum_b \sum_{b'} & \left[ \delta_{a b} \text{i} \frac{\overleftarrow{\partial}}{\partial t} + \left( 1 - \overrightarrow{P}_{a b} \right) \, T_{a b}(t) \right] \chi^{R}_{b' a b a'}(t, t') \nonumber \\
& \times \left[ \delta_{a' b'} \text{i} \frac{\overrightarrow{\partial}}{\partial t'} + T_{a' b'}(t') \left( 1 - \overleftarrow{P}_{a' b'} \right)  \right],
\label{operator C2}
\end{align}
\begin{align}
\hat{D}^{(1)}_{a a'}(\tau) = & \, \delta_{a a'}  \Bigg[ \rho^{\text{S}}_{a}  \left( \frac{\overleftarrow{\partial}}{\partial \tau} - \frac{\overrightarrow{\partial}}{\partial \tau}\right) +  \sum_b \Big( \rho^{\text{C}}_{a b} \, T_{b a} + T_{a b} \, \rho^{\text{C}}_{b a}  \Big) \Bigg] \nonumber \\
&  - \left( T_{a a'} \, \rho^{\uparrow}_{a' a}  + \rho^{\downarrow}_{a a'} \, T_{a' a} \right) ,
\label{operator D1}
\end{align}
\begin{align}
\hat{D}^{(2)}_{a a'}(\tau, \tau') \equiv \sum_b \sum_{b'} & \left[ \delta_{a b} \text{i} \frac{\overleftarrow{\partial}}{\partial \tau} - \left( 1 - \overrightarrow{P}_{a b} \right) \text{i} T_{a b} \right]   G^{\mathbb{M} \downarrow}_{b a'}(\tau - \tau') \, G^{\mathbb{M} \uparrow}_{b' a}(\tau' - \tau)   \nonumber \\
& \times \left[ \delta_{a' b'} \text{i} \frac{\overrightarrow{\partial}}{\partial \tau'} - \text{i} T_{a' b'}\left( 1 - \overleftarrow{P}_{a' b'} \right)  \right],
\label{operator D2}
\end{align}
\begin{align}
\hat{E}_{a a'}(t, \tau) = \sqrt{2} \sum_b \sum_{b'} & \left[ \delta_{ab} \text{i} \overleftarrow{ \frac{\partial}{\partial t} } +  \left( 1 - \overrightarrow{P}_{ab} \right) T_{ab}(t) \right] G^{\urcorner \downarrow}_{b a'}(t, \tau) \, G^{\ulcorner \uparrow}_{b' a}(\tau, t) \nonumber \\
& \times \left[ \delta_{a' b'} \text{i} \overrightarrow{ \frac{\partial}{\partial \tau} } - \text{i} T_{a' b'} \left( 1 - \overleftarrow{P}_{a' b'} \right) \right],
\label{operator E}
\end{align}
\begin{align}
\hat{F}_{a' a}(\tau, t) = \sqrt{2} \sum_b \sum_{b'} & \left[ \delta_{a' b'} \text{i} \overleftarrow{ \frac{\partial}{\partial \tau} } - \left( 1 - \overrightarrow{P}_{a' b'} \right) \text{i} T_{a' b'} \right] G^{\urcorner \uparrow}_{b a'}(t, \tau) \, G^{\ulcorner \downarrow}_{b' a}(\tau, t) \nonumber \\
& \times \left[ \delta_{a b} \text{i} \overrightarrow{ \frac{\partial}{\partial t} } + T_{a b}(t) \left( 1 - \overleftarrow{P}_{a b} \right) \right].
\label{operator F}
\end{align}
In all of the above expressions, the action of permutation operators extends out of the square brackets (so, one first has to eliminate the square brackets by doing the products, and then applies the permutation operators in the appropriate directions).

\section{Kadanoff-Baym equations}

\subsection{Green functions and self-energies} 

We now use the Kadanoff-Baym (KB) equations of motion (EsOM) for the Green functions \cite{Leeuwen06} to simplify the expression of the action. We need the following relations and definitions:
\begin{align}
& \overrightarrow{G}^{-1}(t)  \cdot G^{\lessgtr}(t, t') = \left[ \, \Sigma^R \cdot G^{\lessgtr}  +  \Sigma^{\lessgtr} \cdot G^{A}  +  \Sigma^{\urcorner} \star G^{\ulcorner} \, \right](t, t') \equiv W^{\lessgtr}(t, t'), \nonumber \\
&  G^{\lessgtr}(t, t') \cdot \overleftarrow{G}^{-1}(t') = \left[ \, G^{\lessgtr} \cdot \Sigma^A  +  G^{R} \cdot \Sigma^{\lessgtr}    +  G^{\urcorner} \star \Sigma^{\ulcorner} \, \right](t, t') \equiv X^{\lessgtr}(t, t'), \nonumber \\
& \overrightarrow{G}^{-1}(t) \cdot G^{\urcorner}(t, \tau) = \left[ \, \Sigma^R \cdot G^{\urcorner}  +  \Sigma^{\urcorner} \star G^{\mathbb{M}} \, \right](t, \tau) \equiv Y^{\urcorner}(t, \tau) , \nonumber \\
&  G^{\ulcorner}(\tau, t) \cdot \overleftarrow{G}^{-1}(t) = \left[ \, G^{\ulcorner} \cdot \Sigma^{A}  +  G^{\mathbb{M}} \star \Sigma^{\ulcorner} \, \right](\tau, t) \equiv Y^{\ulcorner}(\tau, t), \nonumber \\
& G^{\urcorner}(t, \tau) \cdot \overleftarrow{g}^{-1}(\tau) = \left[ G^{R} \cdot \Sigma^{\urcorner} + G^{\urcorner} \star \Sigma^{\mathbb{M}}\right](t, \tau) \equiv Z^{\urcorner}(t, \tau), \nonumber \\
& \overrightarrow{g}^{-1}(\tau) \cdot G^{\ulcorner}(\tau, t) = \left[ \Sigma^{\ulcorner} \cdot G^{A} + \Sigma^{\mathbb{M}} \star G^{\ulcorner}\right](\tau, t) \equiv Z^{\ulcorner}(\tau, t), \nonumber \\
& \overrightarrow{g}^{-1}(\tau) \!\cdot \! G^{\mathbb{M}}(\tau - \tau') = \text{i} \delta( \tau - \tau')  + \left[ \Sigma^{\mathbb{M}} \star G^{\mathbb{M}} \right](\tau, \tau') \equiv \text{i} \delta( \tau - \tau')  + I^{\mathbb{M}}(\tau, \tau')  , \nonumber \\\
& G^{\mathbb{M}}(\tau - \tau') \! \cdot \! \overleftarrow{g}^{-1}(\tau') = \text{i} \delta( \tau - \tau')  + \left[ G^{\mathbb{M}} \star \Sigma^{\mathbb{M}} \right](\tau, \tau') \equiv \text{i} \delta( \tau - \tau')  + J^{\mathbb{M}}(\tau, \tau'),
\label{KB equations}
\end{align} 
where all the quantities are matrices in the Hubbard space, and the matrix elements of the operators $\overrightarrow{G}^{-1}(t)$, $\overleftarrow{G}^{-1}(t)$, $\overrightarrow{g}^{-1}(\tau)$ and $\overleftarrow{g}^{-1}(\tau)$ are:
\begin{align}
& \overrightarrow{G}_{ab}^{-1}(t) \equiv \delta_{a b} \text{i}\frac{\overrightarrow{\partial}}{\partial t} - T_{a b}(t), \nonumber \\
& \overleftarrow{G}_{ab}^{-1}(t) \equiv - \delta_{a b} \text{i} \frac{\overleftarrow{\partial}}{\partial t} - T_{a b}(t), \nonumber \\
& \overrightarrow{g}^{-1}_{a b}(\tau) \equiv - \frac{\overrightarrow{\partial}}{\partial \tau} \delta_{a b} - ( T_{a b} - \mu \delta_{a b} ), \nonumber \\
& \overleftarrow{g}^{-1}_{a b}(\tau) \equiv  \frac{\overleftarrow{\partial}}{\partial \tau} \delta_{a b} - ( T_{a b} - \mu \delta_{a b} ) .
\end{align}
In Eqs.\eqref{KB equations} the non-equilibrium self-energies $\left(\Sigma^{>}, \Sigma^{<}, \Sigma^{\urcorner}, \Sigma^{\ulcorner} \right)$ and the equilibrium (Matsubara) self-energy $\Sigma^{\mathbb{M}}$ have been introduced, as well as the retarded ($R$) and advanced ($A$) combinations of Green functions and self-energies:
\begin{align}
& G^{R}(t , t') \equiv  \Theta(t - t') \left[G^{>}(t , t') - G^{<}(t , t') \right], \nonumber \\
& G^{A}(t , t') \equiv - \Theta(t' - t) \left[G^{>}(t , t') - G^{<}(t , t') \right], \nonumber \\
& \Sigma^{R}(t , t') \equiv  \delta(t - t') \overline{\Sigma}(t) + \Theta(t - t') \left[\Sigma^{>}(t , t') - \Sigma^{<}(t , t') \right], \nonumber \\
& \Sigma^{A}(t , t') \equiv  \delta(t - t') \overline{\Sigma}(t) - \Theta(t' - t) \left[\Sigma^{>}(t , t') - \Sigma^{<}(t , t') \right], 
\label{ret adv G}
\end{align}
where $\overline{\Sigma}(t)$ is the Hartree-Fock self-energy \cite{Leeuwen06}. We have also used the following notations:
\begin{align}
& \left[ f \cdot g \right](z, z')  \equiv  \int_{t_0}^{\infty} \text{d} t \, f(z, t) \cdot g(t, z'), \nonumber \\
& \left[ f \star g \right](z, z')  \equiv - \text{i} \int_{0}^{\beta} \text{d} \tau  f(z, \tau) \cdot g(\tau, z').
\end{align}
From Eqs.\eqref{KB equations} and Eq.\eqref{cc G mu}, it can be proved that
\begin{align}
\Big[ W^{\lessgtr}_{a a'}(t, t') \Big]^* = - X^{\lessgtr}_{a' a}(t', t).
\label{symmetry X W}
\end{align}

\subsection{Symmetry breaking}

Despite the fact that the Hamiltonian is invariant under uniform rotation of all spins around a same arbitrary axis, the ground state exhibits a spontaneous symmetry breaking (SSB). The aim of our treatment is to treat low-energy excitations on the top of the magnetically ordered ground state. The direction of the preferred quantization axis of the ground state, $\boldsymbol{u}_z$, may be selected by microscopic magnetic fields, which are not included explicitly in our Hamiltonian. To implement the SSB, we assume that Green functions are spin-dependent [for simplicity, the spin index was not written explicitly in Eqs.\eqref{KB equations}], with $G^{\uparrow} \ne G^{\downarrow}$, where the source of the spin dependence is included in the self-energies, $\Sigma^{\uparrow} \ne \Sigma^{\downarrow}$. In reality, this corresponds to spin-polarized self-consistent solutions of density functional \cite{Lichtenstein87} or dynamical mean-field theory \cite{Katsnelson00} equations.

\subsection{Properties of the Matsubara functions}

From Eqs.\eqref{KB equations}, using Eqs.\eqref{formally Matsubara} and \eqref{boundary 0 Matsubara}, one obtains the equations for the continuous functions $G^{\mathbb{U}}$ and $G^{\mathbb{D}}$, valid for $\tau \ne \tau'$:
\begin{align}
& \Theta(\tau - \tau') \, \overrightarrow{g}^{-1}(\tau) \cdot G^{\mathbb{D}}(\tau - \tau') + \Theta(\tau' - \tau) \, \overrightarrow{g}^{-1}(\tau) \cdot G^{\mathbb{U}}(\tau - \tau') =  I^{\mathbb{M}}(\tau, \tau'), \nonumber \\
& G^{\mathbb{D}}(\tau - \tau') \cdot \overleftarrow{g}^{-1}(\tau') \, \Theta(\tau - \tau') +  G^{\mathbb{U}}(\tau - \tau') \cdot \overleftarrow{g}^{-1}(\tau') \, \Theta(\tau' - \tau)  = J^{\mathbb{M}}(\tau, \tau').
\label{Matsubara final}
\end{align}
Moreover, from the definition of the Matsubara Green function, the following boundary conditions follow:
\begin{align}
& G^{\mathbb{M}}(\tau + \beta) = -  G^{\mathbb{M}}(\tau) \quad \quad \quad \text{for } -\beta < \tau < 0, \nonumber \\
& G^{\mathbb{M}}(\tau - \beta) = -  G^{\mathbb{M}}(\tau) \quad \quad \quad \text{for } 0 < \tau < \beta,
\end{align}
which can be written compactly as
\begin{align}
G^{\mathbb{M}}[\tau - \text{sign}(\tau) \beta] = -  G^{\mathbb{M}}(\tau) \quad \quad \quad \text{for } -\beta < \tau < \beta \wedge \tau \ne 0. 
\label{Matsubara boundary}
\end{align}
Equation \eqref{Matsubara boundary} implies that we can put: 
\begin{align}
& G^{\mathbb{M}}(\tau) \equiv  \sum_{n = -\infty}^{+ \infty} \text{e}^{\text{i} \omega_n \tau} G^{\mathbb{M}}(\omega_{n}), \nonumber \\
& G^{\mathbb{M}}(\omega_{n}) = \frac{1}{\beta} \int_{0}^{\beta} \text{d} \tau \text{e}^{- \text{i} \omega_n \tau}  G^{\mathbb{M}}(\tau),
\label{Fourier Matsubara}
\end{align}
where the quantities
\begin{align}
\omega_n \equiv \frac{(2 n + 1) \pi}{\beta}
\end{align}
are the fermionic Matsubara frequencies. They are such that 
\begin{align}
\int_{0}^{\beta} \text{d} \tau \text{e}^{\text{i} (\omega_m - \omega_n) \tau} = \beta \delta_{n m}.
\label{delta Matsubara}
\end{align}
Since the same boundary conditions as in Eqs.\eqref{Matsubara boundary} apply to Matsubara self-energies, we can analogously put
\begin{align}
& \Sigma^{\mathbb{M}}(\tau) \equiv  \sum_{n = -\infty}^{+ \infty} \text{e}^{\text{i} \omega_n \tau} \Sigma^{\mathbb{M}}(\omega_{n}), \nonumber \\
& \Sigma^{\mathbb{M}}(\omega_{n}) = \frac{1}{\beta} \int_{0}^{\beta} \text{d} \tau \text{e}^{- \text{i} \omega_n \tau}  \Sigma^{\mathbb{M}}(\tau).
\end{align}
This implies that
\begin{align}
I^{\mathbb{M}}(\tau, \tau') &\equiv - \text{i} \int_{0}^{\beta} \text{d} \tau_1   \Sigma^{\mathbb{M}}(\tau - \tau_1) \cdot G^{\mathbb{M}}(\tau_1 - \tau') \nonumber \\
& = - \text{i} \beta  \sum_{n = -\infty}^{+ \infty} \text{e}^{\text{i} \omega_n (\tau - \tau') } \Sigma^{\mathbb{M}}(\omega_{n}) \cdot G^{\mathbb{M}}(\omega_{n}) ,
\label{I Fourier}
\end{align}
where we have used Eq.\eqref{delta Matsubara}. Analogously,
\begin{align}
J^{\mathbb{M}}(\tau, \tau') & \equiv - \text{i}  \int_{0}^{\beta} \text{d} \tau_1  G^{\mathbb{M}}(\tau - \tau_1) \cdot \Sigma^{\mathbb{M}}(\tau_1 - \tau')  \nonumber \\
& = - \text{i} \beta  \sum_{n = -\infty}^{+ \infty} \text{e}^{\text{i} \omega_n (\tau - \tau') } G^{\mathbb{M}}(\omega_{n}) \cdot \Sigma^{\mathbb{M}}(\omega_{n}) .
\label{J Fourier}
\end{align}
Therefore, the quantities $I^{\mathbb{M}}(\tau, \tau') \equiv I^{\mathbb{M}}(\tau - \tau')$ and $J^{\mathbb{M}}(\tau, \tau') \equiv J^{\mathbb{M}}(\tau - \tau')$ depend on $\tau$ and $\tau'$ only via the difference $(\tau - \tau')$, exactly as the Matsubara Green functions. In the following, we will employ any of the two notations, according to convenience.

It is now useful to derive some symmetry properties of the Matsubara Green functions and of their self-energies. From the definition, see Eqs.\eqref{correlators}, it can be seen that
\begin{align}
G^{\mathbb{M}}_{a' a}(\tau) = - \left[ G^{\mathbb{M}}_{a a'}(\tau) \right]^*,
\label{symmetry Matsubara tau}
\end{align}
which implies [see Eq.\eqref{Fourier Matsubara}]
\begin{align}
G^{\mathbb{M}}_{a' a}(\omega_n) = - \left[ G^{\mathbb{M}}_{a a'}( - \omega_n) \right]^* .
\label{symmetry Matsubara omega}
\end{align}
Then, we consider the two equations of motion for the Matsubara Green functions, see among Eqs.\eqref{KB equations}, and compare the complex conjugate of the first one with the second one, with the help of Eq.\eqref{symmetry Matsubara tau} and the fact that $\partial f(\tau - \tau') / \partial \tau = - \partial f(\tau - \tau') / \partial \tau'$. Doing this, we obtain the identity
\begin{align}
- \Big[ I_{a a'}^{\mathbb{M}}(\tau - \tau') \Big]^* \equiv J_{a' a}^{\mathbb{M}}(\tau - \tau')   ,
\end{align}
which, written in frequency space, gives 
\begin{align}
\Sigma^{\mathbb{M}}_{b a}(\omega_n) = - \left[ \Sigma^{\mathbb{M}}_{a b} (-\omega_n) \right]^*.
\label{symmetry Sigma omega}
\end{align}
This implies that, in systems where $\Sigma^{\mathbb{M}}_{a b} = \Sigma^{\mathbb{M}}_{b a}$, $\Sigma^{\mathbb{M}}_{a b}(\tau)$ is purely imaginary:
\begin{align}
\Sigma^{\mathbb{M}}_{a b}(\tau) & = \! \left[ \sum_{\omega_n > 0} + \sum_{\omega_n < 0} \right] \text{e}^{\text{i} \omega_n \tau} \Sigma^{\mathbb{M}}_{a b}(\omega_n) = \! \! \sum_{\omega_n > 0} \Big[ \text{e}^{\text{i} \omega_n \tau} \Sigma^{\mathbb{M}}_{a b}(\omega_n) + \text{e}^{- \text{i} \omega_n \tau} \Sigma^{\mathbb{M}}_{a b}(- \omega_n) \Big] \nonumber \\
& = \sum_{\omega_n > 0} \Bigg\{ \text{e}^{\text{i} \omega_n \tau} \Sigma^{\mathbb{M}}_{a b}(\omega_n) - \text{e}^{- \text{i} \omega_n \tau} \left[ \Sigma^{\mathbb{M}}_{b a}(\omega_n) \right]^* \Bigg\}  \nonumber \\
& \stackrel{\text{if } \Sigma^{\mathbb{M}}_{a b} = \Sigma^{\mathbb{M}}_{b a}}{\longrightarrow} 2 \text{i} \text{ Im} \Bigg\{ \sum_{n = 0}^{+ \infty} \text{e}^{\text{i} \omega_n \tau}  \Sigma^{\mathbb{M}}_{a b}(\omega_n) \Bigg\}.
\label{symmetry Sigma tau}
\end{align}

\section{Simplification of the action}

We now consider systematically the effective action term by term, and perform some mathematical manipulations to simplify it. The general strategy consists in applying partial integration in order to transfer the time derivatives from the fields to the Green functions, at the price of receiving boundary values of the fields into the expression of the action, but with an important gain in the fact that time derivatives of the Green functions can be elaborated by means of the Kadanoff-Baym equations \eqref{KB equations}. These are used to remove the single-particle matrix elements and the time derivatives of the Green functions from the effective action, replacing them with self-energy functionals. The final goal is to get expressions for effective exchange parameters in terms of non-equilibrium Green functions and self-energies, which lend themselves naturally to approximation schemes.

\subsection{Elaboration of $S_{QQ}$}

We now elaborate the quantity $\text{i} S_{Q Q}$ [see the first among Eqs.\eqref{RAK functions}]. In this case, we can directly use partial integration with respect to both integration variables $t'$ and $t$, in order to remove the derivatives of the fields $\xi_{a' Q}(t')$ and $\xi^*_{a Q}(t)$. We obtain
\begin{align}
\text{i} S_{Q Q}  & = \sum_{a} \sum_{a'} \int_{t_0}^{\infty} \text{d} t  \int_{t_0}^{\infty} \text{d} t' \xi^*_{a Q}(t) \Big[   \mathcal{A}^{>}_{a a'}(t, t') + \mathcal{A}^{<}_{a a'}(t, t') \Big]  \xi_{a' Q}(t') \nonumber \\
& \quad + \text{i}  \sum_{a} \sum_{a'} \sum_{\eta = >, <}   \int_{t_0}^{\infty} \text{d} t   \Bigg\{ \nonumber \\ 
& \quad \quad \quad \Big[  \xi^*_{a Q}(\infty) \, \Big( X^{\bar{\eta} \downarrow}_{a a'}(\infty, t)   G^{\eta \uparrow}_{a' a}(t, \infty)  - G^{\bar{\eta} \downarrow}_{a a'}(\infty, t)   W^{\eta \uparrow}_{a' a}(t, \infty)   \Big) \nonumber \\
& \quad \quad \quad +  \xi^*_{a Q}(t_0) \, \Big( G^{\bar{\eta} \downarrow}_{a a'}(t_0, t)   W^{\eta \uparrow}_{a' a}(t, t_0) -  X^{\bar{\eta} \downarrow}_{a a'}(t_0, t) G^{\eta \uparrow}_{a' a}(t, t_0) \Big)  \Big] \xi_{a' Q}(t) \nonumber \\
& \quad \quad \quad +    \xi^*_{a Q}(t) \Big[ \Big( G^{\bar{\eta} \downarrow}_{a a'}(t, \infty)   X^{\eta \uparrow}_{a' a}(\infty, t) 
- W^{\bar{\eta} \downarrow}_{a a'}(t, \infty)   G^{\eta \uparrow}_{a' a}(\infty, t) \Big) \, \xi_{a' Q}(\infty) \nonumber \\
& \quad \quad \quad + \Big( W^{\bar{\eta} \downarrow}_{a a'}(t, t_0)   G^{\eta \uparrow}_{a' a}(t_0, t)  - G^{\bar{\eta} \downarrow}_{a a'}(t, t_0)   X^{\eta \uparrow}_{a' a}(t_0, t)   \Big) \, \xi_{a' Q}(t_0)\Big] \Bigg\} \nonumber \\
& \quad - \sum_{a} \sum_{a'} \sum_{\eta = >, <}  \Big( \begin{matrix} \xi^*_{a Q}(t_0) & \xi^*_{a Q}(\infty) \end{matrix} \Big) \nonumber \\
& \quad \quad \cdot \left( \begin{matrix} G^{\bar{\eta} \downarrow}_{a a'}(t_0, t_0)   G^{\eta \uparrow}_{a' a}(t_0, t_0) & - G^{\bar{\eta} \downarrow}_{a a'}(t_0, \infty)   G^{\eta \uparrow}_{a' a}(\infty, t_0) \\ - G^{\bar{\eta} \downarrow}_{a a'}(\infty, t_0)   G^{\eta \uparrow}_{a' a}(t_0, \infty) & G^{\bar{\eta} \downarrow}_{a a'}(\infty, \infty)   G^{\eta \uparrow}_{a' a}(\infty, \infty) \end{matrix}\right) \! \! \left( \begin{matrix} \xi_{a' Q}(t_0) \\ \xi_{a' Q}(\infty) \end{matrix}\right)  ,
\label{S_{Q Q} final}
\end{align}
where we have defined the matrix
\begin{align}
\mathcal{A}^{\lessgtr}_{a a'}(t, t') \equiv & \sum_b  \sum_{b'} \left[ - \delta_{a b} \text{i} \frac{\overrightarrow{\partial}}{\partial t} + \left( 1 - \overrightarrow{P}_{a b} \right)  T_{a b}(t) \right]  G^{\gtrless \downarrow}_{b a'}(t, t')  \, G^{\lessgtr \uparrow}_{b' a}(t', t) \nonumber \\
& \times \left[ - \delta_{a' b'} \text{i} \frac{\overleftarrow{\partial}}{\partial t'} + T_{a' b'}(t') \left( 1 - \overleftarrow{P}_{a' b'} \right)  \right] ;
\label{equation 0}
\end{align}
it can be proved that
\begin{align}
\Big[ \mathcal{A}_{a' a}^{\lessgtr}(t', t) \Big]^* = \mathcal{A}_{a a'}^{\lessgtr}(t, t'),
\label{symmetry A}
\end{align}
which follows from the definition, Eq.\eqref{equation 0}, and Eq.\eqref{cc G mu}. To get an explicit expression for the quantity $\mathcal{A}^{\lessgtr}_{a a'}(t, t')$, which will play an important role in our treatment, we first note that
\begin{align}
& \sum_b \left[ - \delta_{a b} \text{i} \frac{\overrightarrow{\partial}}{\partial t} + \left( 1 - \overrightarrow{P}_{a b} \right)  T_{a b}(t) \right]  G^{\gtrless \downarrow}_{b a'}(t, t') \,  G^{\lessgtr \uparrow}_{b' a}(t', t) \nonumber \\
& =  - G^{\lessgtr \uparrow}_{b' a}(t', t) \,  W^{\gtrless \downarrow}_{a a'}(t, t') + X^{\lessgtr \uparrow}_{b' a}(t', t)  \, G^{\gtrless \downarrow}_{a a'}(t, t')  ;
\label{universal importance}
\end{align}
then, we have
\begin{align}
& \sum_{b'} W^{\gtrless \downarrow}_{a b'}(t, t') \, \overleftarrow{G}^{-1}_{b' a'}(t')  =  \left[ \Sigma^{R \downarrow} \cdot X^{\gtrless \downarrow}  +   \Sigma^{\gtrless \downarrow} \cdot X^{A \downarrow}  +  \Sigma^{\urcorner \downarrow} \star Y^{\ulcorner \downarrow} \right]_{a a'}(t, t'), \nonumber \\
& \sum_{b'} \overrightarrow{G}^{-1}_{a' b'}(t') \, X^{\lessgtr \uparrow}_{b' a}(t', t)  =  \left[ W^{\lessgtr \uparrow} \cdot \Sigma^{A \uparrow}  +  W^{R \uparrow} \cdot \Sigma^{\lessgtr \uparrow}  +  Y^{\urcorner \uparrow} \star \Sigma^{\ulcorner \uparrow} \right]_{a' a}(t', t),
\label{equations 1 and 2}
\end{align}
where we have defined
\begin{align}
& X^{A \sigma}_{a b}(t, t') \equiv \delta(t - t') \, \delta_{a b} - \Theta(t' - t) \left[ X^{> \sigma}_{a b}(t, t') - X^{< \sigma}_{a b}(t, t') \right], \nonumber \\
& W^{R \sigma}_{a b}(t, t') \equiv \delta(t - t') \, \delta_{a b} + \Theta(t - t') \left[ W^{> \sigma}_{a b}(t, t') - W^{< \sigma}_{a b}(t, t') \right].
\end{align}
Inserting Eqs.\eqref{universal importance} into Eq.\eqref{equation 0}, and using Eqs.\eqref{equations 1 and 2}, we obtain:
\begin{align}
\mathcal{A}^{\lessgtr}_{a a'}(t, t')  =  & W^{\gtrless \downarrow}_{a a'}(t, t') \, W^{\lessgtr \uparrow}_{a' a}(t', t) +  X^{\gtrless \downarrow}_{a a'}(t, t') \, X^{\lessgtr \uparrow}_{a' a}(t', t) \nonumber \\
& -  G^{\gtrless \downarrow}_{a a'}(t, t') \, \Big[ W^{\lessgtr \uparrow} \cdot \Sigma^{A \uparrow}  +  W^{R \uparrow} \cdot \Sigma^{\lessgtr \uparrow} +  Y^{\urcorner \uparrow} \star \Sigma^{\ulcorner \uparrow} \Big]_{a' a}(t', t)  \nonumber \\
& - \Big[ \Sigma^{R \downarrow} \cdot X^{\gtrless \downarrow}  +  \Sigma^{\gtrless \downarrow} \cdot X^{A \downarrow}  +  \Sigma^{\urcorner \downarrow}  \star Y^{\ulcorner \downarrow} \Big]_{a a'}(t, t') \, G^{\lessgtr \uparrow}_{a' a}(t', t).  
\label{B(2) mu c-matrix}
\end{align}

\subsection{Elaboration of $S_{CQ}$}

We now elaborate the quantity $\text{i} S_{C Q}$. The procedure is straightforward but a bit long, so we sketch here the main passages. We treat separately the two terms contributing to $\text{i} S_{C Q}$. The first one is
\begin{align}
\text{i} S_{C Q}^{(1)} & \equiv  \int_{t_0}^{\infty} \text{d} t \, \xi^*_{C}(t) \cdot \hat{B}^{(1)}(t) \cdot \xi_{Q}(t) ,
\label{to be compared}
\end{align}
with the operator $\hat{B}^{(1)}_{a a'}(t)$ defined in Eq.\eqref{operator B1}. First, we eliminate all the hopping terms [see Eq.\eqref{operator B1}] by observing that
\begin{align}
& \sum_b G^{< \text{C}}_{a b}(t, t) \, T_{b a}(t) = - X^{< \text{C}}_{a a}(t, t) - \text{i} \left. \frac{ \partial G^{< \text{C}}_{a a}(t, t_1) }{\partial t_1}  \right|_{t_1 = t}, \nonumber \\
& \sum_b T_{a b}(t) \, G^{< \text{C}}_{b a}(t, t) = - W^{< \text{C}}_{a a}(t, t) + \text{i} \left. \frac{ \partial G^{< \text{C}}_{a a}(t_1, t) }{\partial t_1}  \right|_{t_1 = t}, \nonumber \\
& T_{a a'}(t) \, G^{< \uparrow}_{a' a}(t, t) = - \left. \overrightarrow{G}^{-1}_{a a'}(t_1) \, G^{< \uparrow}_{a' a}(t_1, t) \right|_{t_1 = t} + \delta_{a a'} \text{i} \left. \frac{\partial G^{< \uparrow}_{a a}(t_1, t)}{\partial t_1} \right|_{t_1 = t}, \nonumber \\
& G^{< \downarrow}_{a a'}(t, t) \, T_{a' a}(t) = - \left. G^{< \downarrow}_{a a'}(t, t_1) \, \overleftarrow{G}^{-1}_{a' a}(t_1)  \right|_{t_1 = t} - \delta_{a a'} \text{i} \left. \frac{\partial G^{< \downarrow}_{a a}(t, t_1)}{\partial t_1} \right|_{t_1 = t};
\label{useful}
\end{align}
then, we note that
\begin{align}
& \left. \left[ \frac{ \partial G^{< \text{C}}_{a a}(t, t_1) }{\partial t_1} -  \frac{\partial G^{< \downarrow}_{a a}(t, t_1)}{\partial t_1}  -  \frac{ \partial G^{< \text{C}}_{a a}(t_1, t) }{\partial t_1}  +  \frac{\partial G^{< \uparrow}_{a a}(t_1, t)}{\partial t_1}    \right] \right|_{t_1 = t} 
\nonumber \\
& =  \left. \left[ \frac{ \partial G^{< \text{S}}_{a a}(t, t_1) }{\partial t_1} +    \frac{ \partial G^{< \text{S}}_{a a}(t_1, t) }{\partial t_1}  \right] \right|_{t_1 = t} = \frac{ \text{d} G^{< \text{S}}_{a a}(t, t) }{\text{d} t} = \text{i}  \frac{ \text{d} \rho^{\text{S}}_{a}(t) }{\text{d} t} . 
\end{align}
Then, the integrand contains time derivatives of both fields $\xi^*_{a C}(t)$ and $\xi_{a' Q}(t)$. It is possible to eliminate by partial integration one of such field derivatives, but not both. We choose to eliminate the derivative of $\xi_{a Q}(t)$. The result is:
\begin{align}
& \text{i} S_{C Q}^{(1)}  = \sum_{a} \int_{t_0}^{\infty} \text{d} t  \xi^*_{a C}(t) \Bigg\{ 2 \frac{\overleftarrow{\partial}}{\partial t}  \rho^{\text{S}}_{a}(t)  - X^{< \text{C}}_{a a}(t, t)  - W^{< \text{C}}_{a a}(t, t) + 2 \frac{ \text{d} \rho^{\text{S}}_{a}(t)  }{ \text{d} t} \Bigg\} \xi_{a Q}(t) \nonumber \\
& + \sum_{a} \sum_{a'} \int_{t_0}^{\infty} \text{d} t  \xi^*_{a C}(t) \left. \left[   \overrightarrow{G}^{-1}_{a a'}(t_1) \, G^{< \uparrow}_{a' a}(t_1, t)   +  G^{< \downarrow}_{a a'}(t, t_1) \, \overleftarrow{G}^{-1}_{a' a}(t_1)    \right] \right|_{t_1 = t} \xi_{a' Q}(t) \nonumber \\
&  - \sum_a \left[ \xi^*_{a C}(\infty) \, \rho^{\text{S}}_{a}(\infty) \, \xi_{a Q}(\infty) - \xi^*_{a C}(t_0) \, \rho^{\text{S}}_{a}(t_0) \, \xi_{a Q}(t_0)  \right].
\label{S_{C Q}^{(1)} final}
\end{align}

The second term contributing to $\text{i} S_{C Q}$ is given by
\begin{align}
\text{i} S_{C Q}^{(2)} & \equiv  \int_{t_0}^{\infty} \text{d} t  \int_{t_0}^{\infty} \text{d} t'  \xi^*_{C}(t) \cdot \hat{B}^{(2)}(t, t') \cdot \xi_{Q}(t')  ,
\end{align}
with the operator $\hat{B}^{(2)}_{a a'}(t)$ defined in Eq.\eqref{operator B2}. Recalling the second among Eqs.\eqref{RAK functions}, and doing partial integration over $\text{d} t'$ in order to remove the derivative of $\xi_{a' Q}(t')$, we can write
\begin{align}
& \text{i} S_{C Q}^{(2)}  =  \sum_{a} \sum_{a'} \sum_{\eta} \eta \int_{t_0}^{\infty} \text{d} t   \xi^*_{a C}(t) \sum_{b} \sum_{b'} \left[ \delta_{a b} \text{i} \frac{\overleftarrow{\partial}}{\partial t} + \left( 1 - \overrightarrow{P}_{a b} \right)  T_{a b}(t) \right] \nonumber \\
& \quad \times \Bigg\{ \text{i} \delta_{a' b'} \left[ G^{\bar{\eta} \downarrow}_{b a'}(t, \infty) \,  G^{\eta \uparrow}_{b' a}(\infty, t) \, \xi_{a' Q}(\infty) - G^{\bar{\eta} \downarrow}_{b a'}(t, t) \,  G^{\eta \uparrow}_{b' a}(t, t) \, \xi_{a' Q}(t) \right]   \nonumber \\
& \quad  + \int_{t}^{\infty} \text{d} t' G^{\bar{\eta} \downarrow}_{b a'}(t, t') \,  G^{\eta \uparrow}_{b' a}(t', t)  \left[ - \delta_{a' b'} \text{i} \frac{\overleftarrow{\partial}}{\partial t'} + T_{a' b'}(t') \left( 1 - \overleftarrow{P}_{a' b'} \right)  \right]  \xi_{a' Q}(t') \Bigg\} ,
\end{align}
where the index $\eta \in \lbrace >, < \rbrace \equiv \lbrace +, - \rbrace$. The elaboration of this term consists in performing partial integration, and using the sum rule
\begin{align}
\sum_{\eta} \eta G^{\bar{\eta} \downarrow}_{a b}(t, t) \, G^{\eta \uparrow}_{c d}(t, t)  = - \text{i} \delta_{c d} G^{< \downarrow}_{a b}(t, t) + \text{i} \delta_{a b} G^{< \uparrow}_{c d}(t, t) 
=  \delta_{c d}  \rho^{\downarrow}_{a b}(t) -  \delta_{a b}  \rho^{\uparrow}_{c d}(t)   ,
\label{sum rule mu}
\end{align}
which follows from the discontinuity of the equal-times Green function, see Eq.\eqref{discontinuity > <}; as a particular case, when $d = a$ and $b = c = a'$, this quantity equals $- 2 \delta_{a a'} \rho^{\text{S}}_{a}(t)$. Then, one uses Eqs.\eqref{useful}, with C replaced by $\uparrow$ or $\downarrow$, to remove the hopping terms, as well as Eq.\eqref{B(2) mu c-matrix}, and Eq.\eqref{universal importance} for treating the boundary terms.

After the simplifications, the total term $\text{i} S_{C Q} = \text{i} S_{C Q}^{(1)} + \text{i} S_{C Q}^{(2)}$ is given by
\begin{align}
\text{i} S_{C Q}  = & \sum_{a} \sum_{a'} \int_{t_0}^{\infty} \text{d} t \int_{t_0}^{\infty} \text{d} t' \, \xi^*_{a C}(t)  \left[ \sum_{\eta} \eta \, \Theta(t' - t) \, \mathcal{A}^{\eta}_{a a'}(t, t') \right] \xi_{a' Q}(t') \nonumber \\ 
& + \text{i} \sum_a \sum_{a'} \int_{t_0}^{\infty} \text{d}t \xi_{a C}^*(t) \, \mathcal{F}_{a a'}(t) \, \xi_{a' Q}(t) \nonumber \\
& + \text{i} \sum_{a} \sum_{a'} \sum_{\eta} \eta \int_{t_0}^{\infty} \text{d} t  \, \xi^*_{a C}(t) \Big[ G^{\bar{\eta} \downarrow}_{a a'}(t, \infty) \,  X^{\eta \uparrow}_{a' a}(\infty, t)  \nonumber \\
& \quad \quad - W^{\bar{\eta} \downarrow}_{a a'}(t, \infty)  \, G^{\eta \uparrow}_{a' a}(\infty, t)  \Big] \xi_{a' Q}(\infty)  \nonumber \\
& - \text{i}  \sum_{a} \sum_{a'} \sum_{\eta} \eta   \int_{t_0}^{\infty} \text{d} t   \, \xi^*_{a C}(t_0)  \Big[ X^{\bar{\eta} \downarrow}_{a a'}(t_0, t) \,  G^{\eta \uparrow}_{a' a}(t, t_0) \nonumber \\
& \quad \quad - G^{\bar{\eta} \downarrow}_{a a'}(t_0, t) \,  W^{\eta \uparrow}_{a' a}(t, t_0) \Big]   \xi_{a' Q}(t) \nonumber \\
& + \sum_{a} \sum_{a'} \xi^*_{a C}(t_0)  \sum_{\eta} \eta  G^{\bar{\eta} \downarrow}_{a a'}(t_0, \infty) \,  G^{\eta \uparrow}_{a' a}(\infty, t_0) \,  \xi_{a' Q}(\infty) \nonumber \\ 
&  + \sum_a \Big[ \xi^*_{a C}(t_0) \,  \rho^{\text{S}}_{a}(t_0) \, \xi_{a Q}(t_0) + \xi^*_{a C}(\infty) \,  \rho^{\text{S}}_{a}(\infty) \, \xi_{a Q}(\infty) \Big] .
\label{iS_CQ}
\end{align}

In Eq.\eqref{iS_CQ} we have defined
\begin{align}
& \text{i} \mathcal{F}_{a a'}(t) \equiv \delta_{a a'}  \left[  X^{< \text{S}}_{a a}(t, t) - W^{< \text{S}}_{a a}(t, t) \right]  \nonumber \\
& \quad + \text{i}  \sum_{\eta}  \eta    \left[ X^{\bar{\eta} \downarrow}_{a a'}(t, t) \,  G^{\eta \uparrow}_{a' a}(t, t) - G^{\bar{\eta} \downarrow}_{a a'}(t, t) \,  W^{\eta \uparrow}_{a' a}(t, t) \right] \nonumber \\
& =  \, \delta_{a a'} \left[  X^{< \text{C}}_{a a}(t, t) + W^{< \text{C}}_{a a}(t, t) \right]  \nonumber \\
& \quad + \text{i} \left\{ \Big[ X^{< \downarrow}_{a a'}(t, t) - X^{> \downarrow}_{a a'}(t, t) \Big] \, G^{< \uparrow}_{a' a}(t, t) 
        +  G^{< \downarrow}_{a a'}(t, t) \, \Big[  W^{< \uparrow}_{a' a}(t, t) -   W^{> \uparrow}_{a' a}(t, t) \Big]   \right\} ,
\end{align}
where the passage has been done with the help of Eq.\eqref{discontinuity > <}. From Eqs.\eqref{KB equations} it can be shown directly that
\begin{align}
X_{a a'}^<(t, t) - X_{a a'}^>(t, t) = W_{a a'}^<(t, t) - W_{a a'}^>(t, t) = \text{i} \, \overline{\Sigma}_{a a'}(t),
\label{HF self energy identity}
\end{align}
while from Eq.\eqref{symmetry X W} we see that
\begin{align}
X^{< \text{C}}_{a a}(t, t) + W^{< \text{C}}_{a a}(t, t) = - \Big[ W^{< \text{C}}_{a a}(t, t) \Big]^* + W^{< \text{C}}_{a a}(t, t) = 2 \text{i Im}\Big[ W^{< \text{C}}_{a a}(t, t) \Big].
\end{align}
Therefore, we obtain the following expression:
\begin{align}
\mathcal{F}_{a a'}(t) \equiv & 2 \delta_{a a'} \text{Im}\Big[ W^{< \text{C}}_{a a}(t, t) \Big]  +    \text{i} \,  \overline{\Sigma}^{\downarrow}_{a a'}(t) \, G^{< \uparrow}_{a' a}(t, t) + \text{i} \, G^{< \downarrow}_{a a'}(t, t) \,  \overline{\Sigma}^{\uparrow}_{a' a}(t) . 
\label{cal F}
\end{align}
It must be noted that, from the identity \eqref{HF self energy identity} and the property \eqref{symmetry X W}, it follows that the Hartree-Fock self-energy is hermitean:
\begin{align}
\left[ \overline{\Sigma}_{a a'}(t) \right]^* = \overline{\Sigma}_{a' a}(t).
\label{hermiticity HF sigma}
\end{align}
Using Eqs.\eqref{cal F}, \eqref{cc G mu} and \eqref{hermiticity HF sigma}, we obtain that also $\mathcal{F}(t)$ is hermitean:
\begin{align}
\Big[ \mathcal{F}_{a a'}(t) \Big]^* =  \mathcal{F}_{a' a}(t).
\label{symmetry cal F}
\end{align}

\subsection{Elaboration of $S_{QC}$}

The term $\text{i} S_{QC}$ is closely related to the term $\text{i} S_{CQ}$, which we have already treated. In fact, by using the properties $T_{a b}(t) = T^*_{b a}(t)$, as well as Eq.\eqref{cc G mu} and
\begin{align}
\chi^{R}_{b' a b a'}(t, t') = - \left[ \chi^{A}_{a b' a' b}(t', t) \right]^*,
\end{align}
it can be straightforwardly proved, starting from the definitions \eqref{operator C1} and \eqref{operator C2}, that 
\begin{align}
\text{i} S_{QC} = - \left[ \text{i} S_{CQ} \right]^* \Rightarrow S_{QC} =  S_{CQ}^*.
\label{hermiticity QC CQ}
\end{align} 
Given Eq.\eqref{hermiticity QC CQ}, the quantity $\text{i}S_{QC}$ can then be computed directly by taking the complex conjugate of $\text{i}S_{CQ}$ [see Eq.\eqref{iS_CQ}] with the help of Eqs.\eqref{cc G mu}, \eqref{symmetry X W}, \eqref{symmetry A} and \eqref{symmetry cal F}.

\subsection{Elaboration of $S_{vv}$}

We now elaborate the term $\text{i} S_{vv}$, which is given by the sum of two terms, characterized by the kernel operators $\hat{D}^{(1)}$ and $\hat{D}^{(2)}$, defined, respectively, in Eq.\eqref{operator D1} and Eq.\eqref{operator D2}. The first term is equal to
\begin{align}
\text{i} S^{(1)}_{v v} \equiv & \sum_a \sum_{a'} \int_{0}^{\beta} \text{d} \tau  \xi^*_{a v}(\tau) \hat{D}^{(1)}_{a a'}(\tau) \xi_{a' v}(\tau)  \nonumber \\
 = &    \sum_{a}   \int_{0}^{\beta} \text{d} \tau  \xi^*_{a v}(\tau) \Bigg[ 2 \frac{\overleftarrow{\partial}}{\partial \tau} \rho^{\text{S}}_{a a} + \sum_b \Big( \rho^{\text{C}}_{a b} \, T_{b a} + T_{a b} \, \rho^{\text{C}}_{b a}  \Big)  \Bigg] \xi_{a v}(\tau)     \nonumber \\
&  -  \sum_a \sum_{a'} \left( T_{a a'} \, \rho^{\uparrow}_{a' a}  + \rho^{\downarrow}_{a a'} \, T_{a' a} \right) \int_{0}^{\beta} \text{d} \tau  \xi^*_{a v}(\tau) \, \xi_{a' v}(\tau) \nonumber \\
& -   \sum_{a}  \rho^{\text{S}}_{a a} \Big[  \xi^*_{a v}(\beta) \xi_{a v}(\beta) - \xi^*_{a v}(0) \xi_{a v}(0) \Big],
\label{S^1_vv final}
\end{align}
where we have just applied partial integration to remove the derivative of $\xi_{a v}(\tau)$.

In order to manipulate the second term contributing to $\text{i} S_{vv}$, it is convenient to represent the operator $\hat{D}^{(2)}$ [see Eq.\eqref{operator D2}] in terms of the continuous functions $G^{\mathbb{U}}$ and $G^{\mathbb{D}}$, rather then in terms of the Matsubara Green functions $G^{\mathbb{M}}$. To this end, we note that
\begin{align}
G^{\mathbb{M} \downarrow}_{b a'}(\tau - \tau') \, G^{\mathbb{M} \uparrow}_{b' a}(\tau' - \tau) = & \Theta(\tau - \tau') \, G^{\mathbb{D} \downarrow}_{b a'}(\tau - \tau') \, G^{\mathbb{U} \uparrow}_{b' a}(\tau' - \tau) \nonumber \\
& + \Theta(\tau' - \tau) \, G^{\mathbb{U} \downarrow}_{b a'}(\tau - \tau') \, G^{\mathbb{D} \uparrow}_{b' a}(\tau' - \tau).
\end{align} 
We can then write 
\begin{align}
& \text{i} S^{(2)}_{v v} \equiv  - \sum_a \sum_{a'} \int_{0}^{\beta} \text{d} \tau   \int_{0}^{\beta} \text{d} \tau'  \xi^*_{a v}(\tau) \, \hat{D}^{(2)}_{a a'}(\tau, \tau') \, \xi_{a' v}(\tau') \nonumber \\
& =  - \sum_a \sum_{a'} \int_{0}^{\beta} \text{d} \tau   \int_{0}^{\tau} \text{d} \tau'  \xi^*_{a v}(\tau) \sum_b \sum_{b'}  \left[ \delta_{a b}  \frac{\overleftarrow{\partial}}{\partial \tau} - \left( 1 - \overrightarrow{P}_{a b} \right) T_{a b} \right]     \nonumber \\
& \quad \quad \quad  \times G^{\mathbb{D} \downarrow}_{b a'}(\tau - \tau') \, G^{\mathbb{U} \uparrow}_{b' a}(\tau' - \tau) \left[ \delta_{a' b'}  \frac{\overrightarrow{\partial}}{\partial \tau'} -  T_{a' b'}\left( 1 - \overleftarrow{P}_{a' b'} \right)  \right] \xi_{a' v}(\tau') \nonumber \\
& \quad  - \sum_a \sum_{a'} \int_{0}^{\beta} \text{d} \tau   \int_{\tau}^{\beta} \text{d} \tau'  \xi^*_{a v}(\tau) \sum_b \sum_{b'}  \left[ \delta_{a b}  \frac{\overleftarrow{\partial}}{\partial \tau} - \left( 1 - \overrightarrow{P}_{a b} \right) T_{a b} \right]    \nonumber \\
& \quad \quad \quad \times G^{\mathbb{U} \downarrow}_{b a'}(\tau - \tau') \, G^{\mathbb{D} \uparrow}_{b' a}(\tau' - \tau) \left[ \delta_{a' b'}  \frac{\overrightarrow{\partial}}{\partial \tau'} -  T_{a' b'}\left( 1 - \overleftarrow{P}_{a' b'} \right)  \right] \xi_{a' v}(\tau') .
\label{S^2_vv first}
\end{align}
After performing partial integration over $\text{d} \tau'$, we can write $\text{i} S^{(2)}_{v v} \equiv \text{i} S^{(21)}_{v v} + \text{i} S^{(22)}_{v v}$, where the first part requires one integration over the vertical branch, while the second part requires a double integration. The first part reads
\begin{align}
\text{i} S^{(2 1)}_{v v} =  & - \sum_a \sum_{a'} \int_{0}^{\beta} \text{d} \tau  \xi^*_{a v}(\tau) \sum_{b}  \left[ \delta_{a b}  \frac{\overleftarrow{\partial}}{\partial \tau} - \left( 1 - \overrightarrow{P}_{a b} \right) T_{a b} \right]  \nonumber \\
& \quad \quad \times \Bigg\{ \Big[ G^{\mathbb{D} \downarrow}_{b a'}(0) \, G^{\mathbb{U} \uparrow}_{a' a}(0)   - G^{\mathbb{U} \downarrow}_{b a'}(0) \, G^{\mathbb{D} \uparrow}_{a' a}(0) \Big]    \xi_{a' v}(\tau)   \nonumber \\
& \quad \quad + G^{\mathbb{U} \downarrow}_{b a'}(\tau - \beta) \, G^{\mathbb{D} \uparrow}_{a' a}(\beta - \tau) \xi_{a' v}(\beta) - G^{\mathbb{D} \downarrow}_{b a'}(\tau) \, G^{\mathbb{U} \uparrow}_{a' a}(- \tau) \, \xi_{a' v}(0)\Bigg\}.
\label{Svv^{2 1} first passage}
\end{align}
Using Eqs.\eqref{boundary 0 Matsubara}, we see that
\begin{align}
G^{\mathbb{D} \downarrow}_{b a'}(0) \, G^{\mathbb{U} \uparrow}_{a' a}(0)   - G^{\mathbb{U} \downarrow}_{b a'}(0) \, G^{\mathbb{D} \uparrow}_{a' a}(0)  =  \delta_{b a'} \rho^{\uparrow}_{b a} - \delta_{a a'} \rho^{\downarrow}_{b a} ,
\label{to be used 1}
\end{align}
and we further observe that
\begin{align}
& G^{\mathbb{U} \sigma}_{b a}(\tau - \beta) = -  G^{\mathbb{D} \sigma}_{b a}(\tau), \nonumber \\
& G^{\mathbb{D} \sigma}_{b a}(\beta - \tau) = -  G^{\mathbb{U} \sigma}_{b a}(- \tau).
\label{beta translation}
\end{align}
We substitute the relations in Eq.\eqref{to be used 1} and \eqref{beta translation} into Eq.\eqref{Svv^{2 1} first passage}, then we perform partial integration and we use the following relations, which proceed from Eqs.\eqref{beta translation} and \eqref{boundary 0 Matsubara}:
\begin{align}
& G^{\mathbb{D} \downarrow}_{a a'}(\beta) \,   G^{\mathbb{U} \uparrow}_{a' a}(- \beta) = G^{\mathbb{U} \downarrow}_{a a'}(0) \,   G^{\mathbb{D} \uparrow}_{a' a}(0) = - \rho^{\downarrow}_{a a'}   \Big( \rho^{\uparrow}_{a' a} - \delta_{a' a} \Big)  , \nonumber \\
& G^{\mathbb{D} \downarrow}_{a a'}(0) \,   G^{\mathbb{U} \uparrow}_{a' a}(0) = - \Big( \rho^{\downarrow}_{a a'} - \delta_{a a'} \Big) \rho^{\uparrow}_{a' a} ;
\end{align}
after doing this, one needs to evaluate the quantity
\begin{align}
& \sum_{b}  \left[ - \delta_{a b}  \frac{\overrightarrow{\partial}}{\partial \tau} - \left( 1 - \overrightarrow{P}_{a b} \right) T_{a b} \right]  G^{\mathbb{D} \downarrow}_{b a'}(\tau - \tau') \,   G^{\mathbb{U} \uparrow}_{b' a}(\tau' - \tau) \nonumber \\
& =   G^{\mathbb{U} \uparrow}_{b' a}(\tau' - \tau) \, I^{\mathbb{M} \downarrow}_{a a'}(\tau, \tau') -  J^{\mathbb{M} \uparrow}_{b' a}(\tau', \tau) \, G^{\mathbb{D} \downarrow}_{a a'}(\tau - \tau') ,
\label{useful Matsubara}
\end{align}
in the particular case of $\tau' = 0$ and $b' = a'$. The passages leading to Eq.\eqref{useful Matsubara} follow from the equations of motion, Eqs.\eqref{Matsubara final}, under the condition $\tau > \tau'$ [which is satisfied in the term of interest in Eq.\eqref{Svv^{2 1} first passage} because $\tau' = 0$]. We have used the fact that, in the above expression, $T_{a b}$ can be safely replaced by $(T_{a b} - \mu \delta_{a b})$, which can be immediately appreciated by looking at the first line of Eq.\eqref{useful Matsubara} and observing that 
\begin{align}
\left( 1 - \overrightarrow{P}_{a b} \right) T_{a b} f_{a b} = \left( 1 - \overrightarrow{P}_{a b} \right) ( T_{a b} - \mu \delta_{a b} ) f_{a b}
\end{align}
because $\left( 1 - \overrightarrow{P}_{a b} \right) \delta_{a b} f_{a b} = 0$. We obtain
\begin{align}
\text{i} S^{(2 1)}_{v v} = & - \sum_{a} \left[  2 \rho^{\text{S}}_{a} \int_{0}^{\beta} \! \text{d} \tau  \dot{\xi}^*_{a v}(\tau) \, \xi_{a v}(\tau)  +  \sum_{b} \! \Big(  T_{a b}  \rho^{\downarrow}_{b a}  + \rho^{\uparrow}_{a b} T_{b a}    \Big)  \!   \int_{0}^{\beta} \! \text{d} \tau  \left| \xi_{a v}(\tau) \right|^2 \right] \nonumber \\
& + \sum_a \sum_{a'} \Big(  T_{a a'}  \rho^{\uparrow}_{a' a} + \rho^{\downarrow}_{a a'}  T_{a' a}     \Big)  \int_{0}^{\beta} \text{d} \tau  \xi^*_{a v}(\tau) \, \xi_{a' v}(\tau)  \nonumber \\
& - \sum_a \sum_{a'} \Big[ \xi_{a' v}(\beta) - \xi_{a' v}(0) \Big] \nonumber \\
& \quad \quad \times \int_{0}^{\beta} \text{d} \tau  \xi^*_{a v}(\tau) \Big[ G^{\mathbb{U} \uparrow}_{a' a}( - \tau) \, I^{\mathbb{M} \downarrow}_{a a'}(\tau, 0) -  J^{\mathbb{M} \uparrow}_{a' a}(0, \tau) \, G^{\mathbb{D} \downarrow}_{a a'}(\tau) \Big]  \nonumber \\
& + \sum_a \sum_{a'} \rho^{\downarrow}_{a a'}  \rho^{\uparrow}_{a' a} \Big[   \xi^*_{a v}(\beta)  -  \xi^*_{a v}(0)  \Big]     \Big[ \xi_{a' v}(\beta) - \xi_{a' v}(0) \Big]   \nonumber \\
& - \sum_{a}  \Big[   \xi^*_{a v}(\beta)  \, \rho^{\downarrow}_{a a}  -  \xi^*_{a v}(0) \, \rho^{\uparrow}_{a a} \Big]     \Big[ \xi_{a v}(\beta) - \xi_{a v}(0) \Big]     . 
\label{Svv^{2 1} 4th passage}
\end{align}
It is now convenient to sum $\text{i} S^{(2 1)}_{v v}$ with the previously computed quantity $\text{i} S^{(1)}_{v v}$, using the following consequence of Eqs.\eqref{Matsubara final}:
\begin{align}
& \sum_b \Big( T_{a b} \, \rho^{\text{S}}_{b a}    - \rho^{\text{S}}_{a b} \, T_{b a} \Big) = - \text{i} \sum_{b} \Big[ T_{a b} G^{\mathbb{U} \text{S}}_{b a}(0) - G^{\mathbb{U} \text{S}}_{a b}(0) T_{b a} \Big] \nonumber \\
& = \text{i} \, \Bigg\{ I_{a a}^{\mathbb{M} \text{S}}(0, 0) - J_{a a}^{\mathbb{M} \text{S}}(0, 0) + \left. \frac{\partial}{\partial \tau_1}   \Big[ G^{\mathbb{U} \text{S}}_{a a}(\tau_1 - \tau) +   G^{\mathbb{U} \text{S}}_{a a}(\tau - \tau_1) \Big] \right|_{\tau_1 = \tau} \Bigg\} \nonumber \\
& = \text{i} \, \Big[ I_{a a}^{\mathbb{M} \text{S}}(0, 0) - J_{a a}^{\mathbb{M} \text{S}}(0, 0)  \Big]. 
\end{align}
We obtain:
\begin{align}
\text{i} S^{(1)}_{v v} + \text{i} S^{(2 1)}_{v v} =  & \, \text{i} \sum_{a}  \Big[ I_{a a}^{\mathbb{M} \text{S}}(0, 0) - J_{a a}^{\mathbb{M} \text{S}}(0, 0)  \Big]  \int_{0}^{\beta} \text{d} \tau  \xi^*_{a v}(\tau) \xi_{a v}(\tau)   \nonumber \\
& -   \sum_{a}  \rho^{\text{S}}_{a} \Big[  \left| \xi_{a v}(\beta) \right|^2 - \left| \xi_{a v}(0) \right|^2 \Big] \nonumber \\
& - \sum_a \sum_{a'} \Big[ \xi_{a' v}(\beta) - \xi_{a' v}(0) \Big] \nonumber \\
& \quad \quad \times \int_{0}^{\beta} \text{d} \tau  \xi^*_{a v}(\tau) \Big[ G^{\mathbb{U} \uparrow}_{a' a}( - \tau) \, I^{\mathbb{M} \downarrow}_{a a'}(\tau, 0) -  J^{\mathbb{M} \uparrow}_{a' a}(0, \tau) \, G^{\mathbb{D} \downarrow}_{a a'}(\tau) \Big] \nonumber \\
& + \sum_a \sum_{a'}  \rho^{\downarrow}_{a a'} \, \rho^{\uparrow}_{a' a} \, \Big[  \xi^*_{a v}(\beta) -  \xi^*_{a v}(0) \Big]     \Big[ \xi_{a' v}(\beta) - \xi_{a' v}(0) \Big]  \nonumber \\
& - \sum_{a}  \Big[ \xi^*_{a v}(\beta) \, \rho^{\downarrow}_{a a}  - \xi^*_{a v}(0) \, \rho^{\uparrow}_{a a}   \Big]     \Big[ \xi_{a v}(\beta) - \xi_{a v}(0) \Big].
\label{vv 1 + vv 2 final}
\end{align}

The second part from Eq.\eqref{S^2_vv first}, requiring a double time integration, reads
\begin{align}
& \text{i} S^{(2 2)}_{v v} =  - \sum_a \sum_{a'} \int_{0}^{\beta} \text{d} \tau   \int_{0}^{\beta} \text{d} \tau'  \xi^*_{a v}(\tau) \sum_{b b'}  \left[ \delta_{a b}  \frac{\overleftarrow{\partial}}{\partial \tau} - \left( 1 - \overrightarrow{P}_{a b} \right) T_{a b} \right] \nonumber \\
& \times G^{\mathbb{D} \downarrow}_{b a'}(\tau - \tau') \, G^{\mathbb{U} \uparrow}_{b' a}(\tau' - \tau)    \left[ - \delta_{a' b'}  \frac{\overleftarrow{\partial}}{\partial \tau'} -  T_{a' b'}\left( 1 - \overleftarrow{P}_{a' b'} \right)  \right] \xi_{a' v}(\tau') \Theta(\tau - \tau') \nonumber \\
& - \sum_a \sum_{a'} \int_{0}^{\beta} \text{d} \tau   \int_{0}^{\beta} \text{d} \tau'  \xi^*_{a v}(\tau) \sum_{b b'}  \left[ \delta_{a b}  \frac{\overleftarrow{\partial}}{\partial \tau} - \left( 1 - \overrightarrow{P}_{a b} \right) T_{a b} \right] \nonumber \\
& \times G^{\mathbb{U} \downarrow}_{b a'}(\tau - \tau') \, G^{\mathbb{D} \uparrow}_{b' a}(\tau' - \tau)   \left[ - \delta_{a' b'}  \frac{\overleftarrow{\partial}}{\partial \tau'} -  T_{a' b'}\left( 1 - \overleftarrow{P}_{a' b'} \right)  \right] \xi_{a' v}(\tau') \Theta(\tau' - \tau). 
\label{i S^(22) initial}
\end{align}
We see that, analogously to Eq.\eqref{useful Matsubara}, the following holds for $\tau > \tau'$:
\begin{align}
& \sum_{b'} G^{\mathbb{D} \downarrow}_{b a'}(\tau - \tau') \, G^{\mathbb{U} \uparrow}_{b' a}(\tau' - \tau)  \, \left[ - \delta_{a' b'}  \frac{\overleftarrow{\partial}}{\partial \tau'} -  T_{a' b'}\left( 1 - \overleftarrow{P}_{a' b'} \right)  \right] \nonumber \\
& = G^{\mathbb{D} \downarrow}_{b a'}(\tau - \tau') \, I^{\mathbb{M} \uparrow}_{a' a}(\tau' , \tau) - J^{\mathbb{M} \downarrow}_{b a'}(\tau,  \tau') \, G^{\mathbb{U} \uparrow}_{a' a}(\tau' - \tau)
\end{align}
and the following holds for $\tau < \tau'$:
\begin{align}
& \sum_{b'} G^{\mathbb{U} \downarrow}_{b a'}(\tau - \tau') \, G^{\mathbb{D} \uparrow}_{b' a}(\tau' - \tau)  \, \left[ - \delta_{a' b'}  \frac{\overleftarrow{\partial}}{\partial \tau'} -  T_{a' b'}\left( 1 - \overleftarrow{P}_{a' b'} \right)  \right] \nonumber \\
& = G^{\mathbb{U} \downarrow}_{b a'}(\tau - \tau') \, I^{\mathbb{M} \uparrow}_{a' a}(\tau' , \tau) - J^{\mathbb{M} \downarrow}_{b a'}(\tau , \tau') \, G^{\mathbb{D} \uparrow}_{a' a}(\tau' - \tau).
\end{align}
After using these relations and integrating by parts in $\text{d} \tau$, we obtain:
\begin{align}
& \text{i} S^{(2 2)}_{v v} =   - \sum_a \sum_{a'} \int_{0}^{\beta} \text{d} \tau   \int_{0}^{\beta} \text{d} \tau'  \xi^*_{a v}(\tau) \sum_{b}  \left[ - \delta_{a b}  \frac{\overrightarrow{\partial}}{\partial \tau} - \left( 1 - \overrightarrow{P}_{a b} \right) T_{a b} \right] \nonumber \\
& \quad \quad \quad \times  \Big[ G^{\mathbb{M} \downarrow}_{b a'}(\tau - \tau') \, I^{\mathbb{M} \uparrow}_{a' a}(\tau' , \tau) - J^{\mathbb{M} \downarrow}_{b a'}(\tau , \tau') \, G^{\mathbb{M} \uparrow}_{a' a}(\tau' - \tau) \Big]   \xi_{a' v}(\tau')     \nonumber \\
& - \! \sum_a \sum_{a'}   \! \int_{0}^{\beta} \! \text{d} \tau'  \Big[ \xi^*_{a v}(\beta) \, G^{\mathbb{M} \downarrow}_{a a'}(\beta - \tau') \, I^{\mathbb{M} \uparrow}_{a' a}(\tau' , \beta) - \xi^*_{a v}(\beta) \, J^{\mathbb{M} \downarrow}_{a a'}(\beta , \tau') \, G^{\mathbb{M} \uparrow}_{a' a}(\tau' - \beta) \nonumber \\
& \quad \quad \quad - \xi^*_{a v}(0) \, G^{\mathbb{M} \downarrow}_{a a'}( - \tau') \, I^{\mathbb{M} \uparrow}_{a' a}(\tau', 0) + \xi^*_{a v}(0) \, J^{\mathbb{M} \downarrow}_{a a'}( 0, \tau') \, G^{\mathbb{M} \uparrow}_{a' a}(\tau' )\Big]   \xi_{a' v}(\tau')  .
\label{i S^(22) third}
\end{align}
We now use the equations of motion, Eqs.\eqref{Matsubara final}, to eliminate the hopping parameters. We first note that
\begin{align}
&  \overrightarrow{g}^{-1}(\tau)   \cdot  J^{\mathbb{M} \downarrow}(\tau, \tau')  = \left[ I^{\mathbb{M} \downarrow} \star  \Sigma^{\mathbb{M} \downarrow}  \right](\tau, \tau') + \Sigma^{\mathbb{M} \downarrow}(\tau, \tau'), \nonumber \\
&  I^{\mathbb{M} \uparrow}(\tau', \tau) \cdot  \overleftarrow{g}^{-1}(\tau) =  \left[ \Sigma^{\mathbb{M} \uparrow} \star J^{\mathbb{M} \uparrow} \right](\tau', \tau) + \Sigma^{\mathbb{M} \uparrow}(\tau', \tau),
\end{align}
which follow from the definitions of the quantities $I^{\mathbb{M}}$ and $J^{\mathbb{M}}$ [see among Eqs.\eqref{KB equations}]. These give the following simplification:
\begin{align}
& \sum_{b}  \left[ - \delta_{a b}  \frac{\overrightarrow{\partial}}{\partial \tau} - \left( 1 - \overrightarrow{P}_{a b} \right) T_{a b} \right]  \Big[ G^{\mathbb{M} \downarrow}_{b a'}(\tau - \tau') \, I^{\mathbb{M} \uparrow}_{a' a}(\tau' , \tau) \nonumber \\
& \quad \quad \quad - J^{\mathbb{M} \downarrow}_{b a'}(\tau , \tau') \, G^{\mathbb{M} \uparrow}_{a' a}(\tau' - \tau) \Big]  \nonumber \\
& = \sum_{b} \Big[ I^{\mathbb{M} \uparrow}_{a' a}(\tau' , \tau) \, \overrightarrow{g}^{-1}_{a b}(\tau) \, G^{\mathbb{M} \downarrow}_{b a'}(\tau - \tau')  
+  G^{\mathbb{M} \uparrow}_{a' b}(\tau' - \tau) \, \overleftarrow{g}^{-1}_{b a}(\tau) \, J^{\mathbb{M} \downarrow}_{a a'}(\tau , \tau')  \nonumber \\
& \quad \quad \quad - I^{\mathbb{M} \uparrow}_{a' b}(\tau' , \tau) \, \overleftarrow{g}^{-1}_{b a}(\tau) \, G^{\mathbb{M} \downarrow}_{a a'}(\tau - \tau') 
- G^{\mathbb{M} \uparrow}_{a' a}(\tau' - \tau) \, \overrightarrow{g}^{-1}_{a b}(\tau) \, J^{\mathbb{M} \downarrow}_{b a'}(\tau , \tau')  \Big] \nonumber \\
& = I^{\mathbb{M} \downarrow}_{a a'}(\tau , \tau')  \,  I^{\mathbb{M} \uparrow}_{a' a}(\tau' , \tau)  +  J^{\mathbb{M} \downarrow}_{a a'}(\tau , \tau') \, J^{\mathbb{M} \uparrow}_{a' a}(\tau' , \tau) \nonumber \\
& \quad  +  \text{i} \, \delta(\tau - \tau') \delta_{a a'} \Big[ I^{\mathbb{M} \uparrow}_{a a}(\tau, \tau) +  J^{\mathbb{M} \downarrow}_{a a}(\tau, \tau) \Big]  \nonumber \\
&  \quad - G^{\mathbb{M} \downarrow}_{a a'}(\tau - \tau') \, \left[ \Sigma^{\mathbb{M} \uparrow} \star J^{\mathbb{M} \uparrow} \right]_{a' a}(\tau', \tau)   - G^{\mathbb{M} \downarrow}_{a a'}(\tau - \tau')  \, \Sigma^{\mathbb{M} \uparrow}_{a' a}(\tau', \tau)  \nonumber \\
& \quad -  \left[ I^{\mathbb{M} \downarrow} \star  \Sigma^{\mathbb{M} \downarrow}  \right]_{a a'}(\tau, \tau')  \, G^{\mathbb{M} \uparrow}_{a' a}(\tau' - \tau) -  \Sigma^{\mathbb{M} \downarrow}_{a a'}(\tau, \tau') \, G^{\mathbb{M} \uparrow}_{a' a}(\tau' - \tau)   ,
\label{Matsubara double sum}
\end{align}
where we have again implicitly replaced $T_{a b}$ with $( T_{a b} - \mu \delta_{a b} )$, which is correct as can be seen from the first line of Eq.\eqref{Matsubara double sum}. Then, we observe that the boundary condition on Matsubara Green functions, Eq.\eqref{Matsubara boundary}, combined with the equations of motion for Matsubara Green functions [among Eqs.\eqref{KB equations}], implies that
\begin{align}
& J_{b a}^{\mathbb{M}}(\beta , \tau) = -  J_{b a}^{\mathbb{M}}(0 , \tau) \quad \quad \quad \text{for } 0 < \tau < \beta, \nonumber \\
& I_{b a}^{\mathbb{M}}(\tau , \beta) = -  I_{b a}^{\mathbb{M}}(\tau , 0) \quad \quad \quad \text{ for } 0 < \tau < \beta.
\label{I J boundary}
\end{align}
Using Eqs.\eqref{Matsubara double sum}, \eqref{Matsubara boundary} and \eqref{I J boundary} to elaborate Eq.\eqref{i S^(22) third}, we obtain
\begin{align}
& \text{i} S^{(2 2)}_{v v} \equiv   - \sum_a \sum_{a'} \int_{0}^{\beta} \text{d} \tau   \int_{0}^{\beta} \text{d} \tau'  \xi^*_{a v}(\tau) \, \mathcal{K}_{a a'}(\tau, \tau') \,  \xi_{a' v}(\tau')     \nonumber \\
&  - \text{i} \sum_{a}    \int_{0}^{\beta} \text{d} \tau \xi^*_{a v}(\tau)  \Big[   I^{\mathbb{M} \uparrow}_{a a}(\tau, \tau) + J^{\mathbb{M} \downarrow}_{a a}(\tau, \tau)   \Big] \xi_{a v}(\tau) \nonumber \\
& - \sum_a \! \sum_{a'}    \int_{0}^{\beta} \! \text{d} \tau  \Big[ \xi^*_{a v}(\beta) - \xi^*_{a v}(0) \Big] \! \Big[ G^{\mathbb{M} \downarrow}_{a a'}( - \tau) \, I^{\mathbb{M} \uparrow}_{a' a}(\tau) - J^{\mathbb{M} \downarrow}_{a a'}(- \tau) \, G^{\mathbb{M} \uparrow}_{a' a}(\tau) \Big] \xi_{a' v}(\tau)     ,
\label{Svv^{2 2} final}
\end{align}
where we have defined 
\begin{align}
& \mathcal{K}_{a a'}(\tau, \tau') \equiv  I^{\mathbb{M} \downarrow}_{a a'}(\tau , \tau')  \,  I^{\mathbb{M} \uparrow}_{a' a}(\tau' , \tau)  +  J^{\mathbb{M} \downarrow}_{a a'}(\tau , \tau') \, J^{\mathbb{M} \uparrow}_{a' a}(\tau' , \tau)  \nonumber \\
&  - G^{\mathbb{M} \downarrow}_{a a'}(\tau - \tau') \, \left[ \Sigma^{\mathbb{M} \uparrow} \star J^{\mathbb{M} \uparrow} \right]_{a' a}(\tau', \tau)    -  \left[ I^{\mathbb{M} \downarrow} \star  \Sigma^{\mathbb{M} \downarrow}  \right]_{a a'}(\tau, \tau')  \, G^{\mathbb{M} \uparrow}_{a' a}(\tau' - \tau)     \nonumber \\
&   - G^{\mathbb{M} \downarrow}_{a a'}(\tau - \tau')  \, \Sigma^{\mathbb{M} \uparrow}_{a' a}(\tau', \tau)       -  \Sigma^{\mathbb{M} \downarrow}_{a a'}(\tau, \tau') \, G^{\mathbb{M} \uparrow}_{a' a}(\tau' - \tau) .
\label{tilde K}
\end{align}
Finally, we take into account the fact that $I^{\mathbb{M}}(\tau, \tau') = I^{\mathbb{M}}(\tau - \tau')$ and $J^{\mathbb{M}}(\tau, \tau') = J^{\mathbb{M}}(\tau - \tau')$. This implies that
\begin{align}
& I_{a a}^{\mathbb{M} \text{S}}(0, 0) - J_{a a}^{\mathbb{M} \text{S}}(0, 0)  -  I^{\mathbb{M} \uparrow}_{a a}(\tau, \tau) - J^{\mathbb{M} \downarrow}_{a a}(\tau, \tau)  \nonumber \\
&  = I_{a a}^{\mathbb{M} \text{S}}(0) - J_{a a}^{\mathbb{M} \text{S}}(0)  -  I^{\mathbb{M} \uparrow}_{a a}(0) - J^{\mathbb{M} \downarrow}_{a a}(0) = - I^{\mathbb{M} \text{C}}_{a a}(0) - J^{\mathbb{M} \text{C}}_{a a}(0) \equiv \text{i} \beta \mathcal{I}_a,
\label{def cal I}
\end{align}
where, from Eqs.\eqref{I Fourier} and \eqref{J Fourier}, 
\begin{align}
\mathcal{I}_{a} \equiv \frac{1}{2} \sum_{\sigma = \uparrow, \downarrow} \sum_b \sum_{n = -\infty}^{+ \infty} \Big[ \Sigma_{a b}^{\mathbb{M} \sigma}(\omega_n) \, G_{b a}^{\mathbb{M} \sigma}(\omega_{n}) + G_{a b}^{\mathbb{M} \sigma}(\omega_n) \, \Sigma_{b a}^{\mathbb{M} \sigma}(\omega_{n}) \Big] =  \mathcal{I}_a^* ;
\label{cal I}
\end{align}
the reality of this quantity is proved by using Eqs.\eqref{symmetry Matsubara omega} and \eqref{symmetry Sigma omega}.

The total term $\text{i} S_{vv} = \text{i} S^{(1)}_{vv} + \text{i} S^{(21)}_{vv} + \text{i} S^{(22)}_{vv}$ is then written as
\begin{align}
\text{i} S_{v v} = 
& - \sum_a \sum_{a'} \int_{0}^{\beta} \text{d} \tau   \int_{0}^{\beta} \text{d} \tau'  \xi^*_{a v}(\tau) \, \Big[ \mathcal{K}_{a a'}(\tau , \tau') + \beta \delta_{a a'} \delta(\tau - \tau') \, \mathcal{I}_a \Big] \, \xi_{a' v}(\tau')   \nonumber \\
& - \sum_a \sum_{a'}    \int_{0}^{\beta} \text{d} \tau  \Bigg\{ \nonumber \\
& \quad \Big[ \xi^*_{a v}(\beta) - \xi^*_{a v}(0) \Big] \Big[ G^{\mathbb{M} \downarrow}_{a a'}( - \tau) \, I^{\mathbb{M} \uparrow}_{a' a}(\tau) - J^{\mathbb{M} \downarrow}_{a a'}(-\tau) \, G^{\mathbb{M} \uparrow}_{a' a}(\tau) \Big] \xi_{a' v}(\tau) \nonumber \\
& \quad + \Big[ \xi_{a' v}(\beta) - \xi_{a' v}(0) \Big]   \xi^*_{a v}(\tau) \Big[ G^{\mathbb{M} \uparrow}_{a' a}( - \tau) \, I^{\mathbb{M} \downarrow}_{a a'}(\tau) -  J^{\mathbb{M} \uparrow}_{a' a}(- \tau) \, G^{\mathbb{M} \downarrow}_{a a'}(\tau) \Big] \Bigg\} \nonumber \\
& -   \sum_{a}  \rho^{\text{S}}_{a} \Big[  \left| \xi_{a v}(\beta) \right|^2  - \left| \xi_{a v}(0) \right|^2 \Big] \nonumber \\
& - \sum_{a}  \Big[ \xi^*_{a v}(\beta) \, \rho^{\downarrow}_{a}  - \xi^*_{a v}(0) \, \rho^{\uparrow}_{a}   \Big]     \Big[ \xi_{a v}(\beta) - \xi_{a v}(0) \Big]   \nonumber \\
&  + \sum_a \sum_{a'}  \rho^{\downarrow}_{a a'} \, \rho^{\uparrow}_{a' a} \, \Big[  \xi^*_{a v}(\beta) -  \xi^*_{a v}(0) \Big]     \Big[ \xi_{a' v}(\beta) - \xi_{a' v}(0) \Big]  .
\label{Svv rewritten}
\end{align}

It is convenient to write $\mathcal{K}$ in frequency space. The convolutions appearing in Eq.\eqref{tilde K} are given by:
\begin{align}
\left[ I^{\mathbb{M} \downarrow} \star \Sigma^{\mathbb{M} \downarrow} \right](\tau, \tau') & = -\beta^2 \sum_{n = -\infty}^{+ \infty}  \text{e}^{\text{i} \omega_n (\tau - \tau') } \Sigma^{\mathbb{M} \downarrow}(\omega_{n}) \cdot G^{\mathbb{M} \downarrow}(\omega_{n}) \cdot  \Sigma^{\mathbb{M} \downarrow}(\omega_{n}), \nonumber \\
\left[ \Sigma^{\mathbb{M} \uparrow} \star J^{\mathbb{M} \uparrow} \right](\tau', \tau) & = -\beta^2   \sum_{n = -\infty}^{+ \infty}  \text{e}^{\text{i} \omega_n (\tau' - \tau) } \Sigma^{\mathbb{M} \uparrow}(\omega_{n}) \cdot G^{\mathbb{M} \uparrow}(\omega_{n}) \cdot  \Sigma^{\mathbb{M} \uparrow}(\omega_{n}) \nonumber \\
&  = \left[ I^{\mathbb{M} \uparrow} \star \Sigma^{\mathbb{M} \uparrow} \right](\tau', \tau).
\end{align}
Therefore these convolutions, just as $I_{a a'}^{\mathbb{M}}(\tau, \tau')$ and $J_{a a'}^{\mathbb{M}}(\tau, \tau')$, depend on $\tau$ and $\tau'$ only via the difference $(\tau - \tau')$. We then put [cfr. Eq.\eqref{tilde K}]:
\begin{align}
\mathcal{K}_{a a'}(\tau - \tau') \equiv \, \beta^2  \sum_{n = -\infty}^{+ \infty} \sum_{m = -\infty}^{+ \infty} \text{e}^{\text{i} ( \omega_n - \omega_m ) (\tau - \tau') }  \mathcal{K}_{a a'}(\omega_n, \omega_m) ,
\label{def Fourier K}
\end{align}
where
\begin{align}
\mathcal{K}_{a a'}(\omega_n, \omega_m) = &  - \Big[ \Sigma^{\mathbb{M} \downarrow}(\omega_{n}) \cdot G^{\mathbb{M} \downarrow}(\omega_{n}) \Big]_{a a'} \Big[ \Sigma^{\mathbb{M} \uparrow}(\omega_{m}) \cdot G^{\mathbb{M} \uparrow}(\omega_{m}) \Big]_{a' a} \nonumber \\
& -  \Big[ G^{\mathbb{M} \downarrow}(\omega_{n}) \cdot \Sigma^{\mathbb{M} \downarrow}(\omega_{n}) \Big]_{a a'}  \Big[ G^{\mathbb{M} \uparrow}(\omega_{m}) \cdot \Sigma^{\mathbb{M} \uparrow}(\omega_{m}) \Big]_{a' a}   \nonumber \\
&  +  G_{a a'}^{\mathbb{M} \downarrow}(\omega_{n}) \Big[ \Sigma^{\mathbb{M} \uparrow}(\omega_{m}) \cdot G^{\mathbb{M} \uparrow}(\omega_{m})  \cdot \Sigma^{\mathbb{M} \uparrow}(\omega_{m}) \Big]_{a' a} \nonumber \\
& +  \Big[ \Sigma^{\mathbb{M} \downarrow}(\omega_{n}) \cdot G^{\mathbb{M} \downarrow}(\omega_{n}) \cdot  \Sigma^{\mathbb{M} \downarrow}(\omega_{n}) \Big]_{a a'} G_{a' a}^{\mathbb{M} \uparrow}(\omega_{m})  \nonumber \\
& - \frac{1}{\beta^2} \Big[ G_{a a'}^{\mathbb{M} \downarrow}(\omega_{n}) \,  \Sigma_{a' a}^{\mathbb{M} \uparrow}(\omega_{m}) + \Sigma_{a a'}^{\mathbb{M} \downarrow}(\omega_{n}) \, G_{a' a}^{\mathbb{M} \uparrow}(\omega_{m}) \Big] .
\label{K frequency space}
\end{align}
From Eqs.\eqref{symmetry Matsubara omega} and \eqref{symmetry Sigma omega}, it is seen that
\begin{align}
\Big[ \mathcal{K}_{a a'}(- \omega_n, - \omega_m) \Big]^* = \mathcal{K}_{a' a}(\omega_n, \omega_m),
\label{to prove hermiticity J}
\end{align}
which in turn implies 
\begin{align}
\mathcal{K}_{a a'}(\tau) = \Big[ \mathcal{K}_{a' a}(\tau) \Big]^*.
\end{align}

\subsection{Elaboration of $S_{Qv}$}

We now elaborate the term $\text{i} S_{Qv}$, which is given by
\begin{align}
\text{i} S_{Q v} & = \sqrt{2} \sum_{a} \sum_{a'}   \int_{t_0}^{\infty} \text{d}t \int_{0}^{\beta} \text{d} \tau  \xi_{a Q}^*(t)  \sum_{b} \sum_{b'} \left[ - \delta_{ab} \text{i} \overrightarrow{ \frac{\partial}{\partial t} } +  \left( 1 - \overrightarrow{P}_{ab} \right) T_{ab}(t) \right] \nonumber \\
& \quad \quad \quad \quad \quad \times G^{\urcorner \downarrow}_{b a'}(t, \tau) \, G^{\ulcorner \uparrow}_{b' a}(\tau, t) 
 \left[ \delta_{a' b'} \text{i} \overrightarrow{ \frac{\partial}{\partial \tau} } - \text{i} T_{a' b'} \left( 1 - \overleftarrow{P}_{a' b'} \right) \right] \xi_{a' v}(\tau) \nonumber \\
& \quad + \text{i} \sqrt{2} \sum_{a} \sum_{a'} \sum_{b'}  \int_{0}^{\beta} \text{d} \tau  \Big[ \xi_{a Q}^*(\infty) G^{\urcorner \downarrow}_{a a'}(\infty, \tau) \, G^{\ulcorner \uparrow}_{b' a}(\tau, \infty) \nonumber \\
& \quad - \xi_{a Q}^*(t_0) G^{\urcorner \downarrow}_{a a'}(t_0, \tau) \, G^{\ulcorner \uparrow}_{b' a}(\tau, t_0) \Big] \left[ \delta_{a' b'} \text{i} \overrightarrow{ \frac{\partial}{\partial \tau} } - \text{i} T_{a' b'} \left( 1 - \overleftarrow{P}_{a' b'} \right) \right] \xi_{a' v}(\tau) \nonumber \\
& \equiv \text{i} S^{(1)}_{Q v} + \text{i} S^{(2)}_{Q v} ,
\label{S_Qv to be elaborated}
\end{align}
where we have performed partial integration over $\text{d}t$. In the following, we will implicitly replace $T_{a' b'}$ with $(T_{a' b'} - \mu \delta_{a' b'})$, analogously to what we have done in the elaboration of $\text{i} S_{vv}$, which is correct because $T_{a' b'}$ is multiplied by $\left( 1 - \overleftarrow{P}_{a' b'} \right)$ in Eq.\eqref{S_Qv to be elaborated}.  

Equation \eqref{S_Qv to be elaborated} is written as the sum of two terms; the first term requires a double time integration (lines 1 and 2), while the second term requires one integration (lines 3 and 4). The next step is the execution of partial integration over $\text{d} \tau$ in the first term, which then requires the evaluation of the following quantity:
\begin{align}
\mathcal{G}^{\sigma \bar{\sigma}}_{a a'}(t, \tau) \equiv  \sum_{b} \sum_{b'} & \left[ - \delta_{ab} \text{i} \overrightarrow{ \frac{\partial}{\partial t} } +  \left( 1 - \overrightarrow{P}_{ab} \right) T_{ab}(t) \right] G^{\urcorner \sigma}_{b a'}(t, \tau) \, G^{\ulcorner \bar{\sigma}}_{b' a}(\tau, t) \nonumber \\
& \times \left[ \delta_{a' b'}  \overleftarrow{ \frac{\partial}{\partial \tau} } + T_{a' b'} \left( 1 - \overleftarrow{P}_{a' b'} \right) \right].
\label{def cal G}
\end{align}
To simplify this object, we need the Kadanoff-Baym equations for $G^{\urcorner}$ and $G^{\ulcorner}$, see Eqs.\eqref{KB equations}. A consequence of these equations is that Eq.\eqref{universal importance} is valid also with the replacements $\bar{\mu} \rightarrow \urcorner$, $\mu \rightarrow \ulcorner$, $W \rightarrow Y$, $X \rightarrow Y$. Then, we get:
\begin{align}
& \mathcal{G}^{\sigma \bar{\sigma}}_{a a'}(t, \tau) =   \sum_{b'}  \Big[ G^{\urcorner \sigma}_{a a'}(t, \tau) \, Y^{\ulcorner \bar{\sigma}}_{b' a}(\tau, t)   -   
Y^{\urcorner \sigma}_{a a'}(t, \tau) \,  G^{\ulcorner \bar{\sigma}}_{b' a}(\tau, t)   \Big] \nonumber \\
& \quad \quad \times \left[ \delta_{a' b'}  \overleftarrow{ \frac{\partial}{\partial \tau} } + T_{a' b'} \left( 1 - \overleftarrow{P}_{a' b'} \right) \right] \nonumber \\
& = Y^{\urcorner \sigma}_{a a'}(t, \tau)  Z^{\ulcorner \bar{\sigma}}_{a' a}(\tau, t)  +  Z^{\urcorner \sigma}_{a a'}(t, \tau)  Y^{\ulcorner \bar{\sigma}}_{a' a}(\tau, t) \nonumber \\
& \quad - \Bigg[ \Sigma^{\urcorner \sigma}_{a a'}(t, \tau) + \left[ \Sigma^{R \sigma} \cdot Z^{\urcorner \sigma}\right]_{a a'}(t, \tau) + \left[ \Sigma^{\urcorner \sigma} \star J^{\mathbb{M} \sigma} \right]_{a a'}(t, \tau) \Bigg] G^{\ulcorner \bar{\sigma}}_{a' a}(\tau, t)    \nonumber \\
& \quad  -  G^{\urcorner \sigma}_{a a'}(t, \tau) \Bigg[  \left[ Z^{\ulcorner \bar{\sigma}} \cdot \Sigma^{A \bar{\sigma}} \right]_{a' a}(\tau, t) + \left[ I^{\mathbb{M} \bar{\sigma}} \star \Sigma^{\ulcorner \bar{\sigma}} \right]_{a' a}(\tau, t) + \Sigma^{\ulcorner \bar{\sigma}}_{a' a}(\tau, t)  \Bigg] .
\label{cal G}
\end{align}
Including the boundary contributions coming from the partial integrations, the first term from Eq.\eqref{S_Qv to be elaborated} is equal to
\begin{align}
& \text{i} S^{(1)}_{Q v}  =  - \sqrt{2} \text{i} \sum_{a} \sum_{a'}   \int_{t_0}^{\infty} \text{d}t \int_{0}^{\beta} \text{d} \tau  \, \xi_{a Q}^*(t)  \,  \mathcal{G}^{\downarrow \uparrow}_{a a'}(t, \tau) \, \xi_{a' v}(\tau) \nonumber \\
& - \! \sqrt{2} \text{i} \sum_{a} \sum_{a'}   \int_{t_0}^{\infty} \! \text{d}t \,  \xi_{a Q}^*(t)   \Bigg\{ \! \left[ G^{\ulcorner \uparrow}_{a' a}(\beta, t) \, 
Y^{\urcorner \downarrow}_{a a'}(t, \beta) - Y^{\ulcorner \uparrow}_{a' a}(\beta, t) \, G^{\urcorner \downarrow}_{a a'}(t, \beta)   \right] \xi_{a' v}(\beta) 
\nonumber \\
& \quad \quad - \left[ G^{\ulcorner \uparrow}_{a' a}(0, t) \, Y^{\urcorner \downarrow}_{a a'}(t, 0) - Y^{\ulcorner \uparrow}_{a' a}(0, t) \, G^{\urcorner \downarrow}_{a a'}(t, 0)   \right] \xi_{a' v}(0) \Bigg\} .
\label{SQv1}
\end{align}
The second term from Eq.\eqref{S_Qv to be elaborated} is equal to
\begin{align}
\text{i} S^{(2)}_{Q v}  = & - \sqrt{2} \sum_{a} \sum_{a'}  \int_{0}^{\beta} \text{d} \tau  \Bigg\{ \nonumber \\
& \quad  \xi_{a Q}^*(\infty) \, \Big[ G^{\urcorner \downarrow}_{a a'} (\infty, \tau) \, Z^{\ulcorner \uparrow}_{a' a}(\tau, \infty)  - Z^{\urcorner \downarrow}_{a a'} (\infty, \tau) \, G^{\ulcorner \uparrow}_{a' a}(\tau, \infty) \Big]  \nonumber \\
& \quad  - \xi_{a Q}^*(t_0) \, \Big[ G^{\urcorner \downarrow}_{a a'}(t_0, \tau) \, Z^{\ulcorner \uparrow}_{a' a}(\tau, t_0) - Z^{\urcorner \downarrow}_{a a'}(t_0, \tau) \, G^{\ulcorner \uparrow}_{a' a}(\tau, t_0) \Big] \Bigg\}    \xi_{a' v}(\tau) \nonumber \\
&  - \sqrt{2} \sum_{a} \sum_{a'} \Bigg\{  \Big( \begin{matrix} \xi_{a Q}^*(\infty) & \xi_{a Q}^*(t_0) \end{matrix} \Big)  \nonumber \\
& \quad  \cdot \left( \begin{matrix}  - G^{\urcorner \downarrow}_{a a'}(\infty, 0) \, G^{\ulcorner \uparrow}_{a' a}(0, \infty) &  G^{\urcorner \downarrow}_{a a'}(\infty, \beta) \, G^{\ulcorner \uparrow}_{a' a}(\beta, \infty) \\  G^{\urcorner \downarrow}_{a a'}(t_0, 0) \, G^{\ulcorner \uparrow}_{a' a}(0, t_0) & - G^{\urcorner \downarrow}_{a a'}(t_0, \beta) \, G^{\ulcorner \uparrow}_{a' a}(\beta, t_0) \end{matrix}\right) \left( \begin{matrix} \xi_{a' v}(0) \\ \xi_{a' v}(\beta) \end{matrix} \right) \Bigg\} , 
\label{SQv2}
\end{align}
which follows from the application of partial integration and the use of Eqs.\eqref{KB equations}. We then sum Eqs.\eqref{SQv1} and \eqref{SQv2}, and we use the correspondence properties between non-equilibrium Green functions listed in \ref{correspondences}. We obtain
\begin{align}
& \text{i} S_{Q v} =   - \sqrt{2} \text{i} \sum_a \sum_{a'}   \int_{t_0}^{\infty} \text{d}t \int_{0}^{\beta} \text{d} \tau \, \xi_{a Q}^*(t) \, \mathcal{G}_{a a'}^{\downarrow \uparrow}(t, \tau) \, \xi_{a' v}(\tau)   \nonumber \\
& - \sqrt{2} \text{i} \sum_a \sum_{a'}   \int_{t_0}^{\infty} \text{d}t  \, \xi_{a Q}^*(t)   \Bigg\{ \nonumber \\
& \quad \quad \left[ G^{< \uparrow}_{a' a}(t_0, t) \, W^{> \downarrow}_{a a'}(t, t_0) - X^{< \uparrow}_{a' a}(t_0, t) \, G^{> \downarrow}_{a a'}(t, t_0)   \right] \xi_{a' v}(\beta) \nonumber \\
& \quad \quad - \left[ G^{> \uparrow}_{a' a}(t_0, t) \, W^{< \downarrow}_{a a'}(t, t_0) - X^{> \uparrow}_{a' a}(t_0, t) \, G^{< \downarrow}_{a a'}(t, t_0)   \right] \xi_{a' v}(0) \Bigg\} \nonumber \\
&  - \sqrt{2} \sum_a \sum_{a'}  \int_{0}^{\beta} \text{d} \tau  \Bigg\{ \xi_{a Q}^*(\infty) \, \Big[ G^{\urcorner \downarrow}_{a a'} (\infty, \tau) \, Z^{\ulcorner \uparrow}_{a' a}(\tau, \infty)  - Z^{\urcorner \downarrow}_{a a'} (\infty, \tau) \, G^{\ulcorner \uparrow}_{a' a}(\tau, \infty) \Big]  \nonumber \\
& \quad \quad - \xi_{a Q}^*(t_0) \, \Big[ G^{\mathbb{M} \downarrow}_{a a'}(-\tau) \, I^{\mathbb{M} \uparrow}_{a' a}(\tau) - J^{\mathbb{M} \downarrow}_{a a'}(-\tau) \, G^{\mathbb{M} \uparrow}_{a' a}(\tau) \Big] \Bigg\}    \xi_{a' v}(\tau) \nonumber \\
&  - \sqrt{2} \sum_{a} \sum_{a'} \Bigg\{  \Big( \begin{matrix} \xi_{a Q}^*(\infty) & \xi_{a Q}^*(t_0) \end{matrix} \Big)  \nonumber \\
& \quad \quad \cdot \left( \begin{matrix}  - G^{< \downarrow}_{a a'}(\infty, t_0) \, G^{> \uparrow}_{a' a}(t_0, \infty) &  G^{> \downarrow}_{a a'}(\infty, t_0) \, G^{< \uparrow}_{a' a}(t_0, \infty) \\  \Big( - \rho^{\downarrow}_{a a'} \, \rho^{\uparrow}_{a' a} + \delta_{a a'} \rho_a^{\downarrow}\Big)  & \Big( \rho^{\downarrow}_{a a'} \, \rho^{\uparrow}_{a' a} - \delta_{a a'} \rho_a^{\uparrow} \Big)    \end{matrix}\right) \left( \begin{matrix} \xi_{a' v}(0) \\ \xi_{a' v}(\beta) \end{matrix} \right) \Bigg\} .
\label{i S_Qv final}
\end{align}

\subsection{Elaboration of $S_{vQ}$}

The form of the term $\text{i} S_{vQ}$ is analogous to that of $\text{i} S_{Qv}$, except for some substitutions. Comparing Eqs.\eqref{operator E} and \eqref{operator F}, we see that, in order to derive the expression for $\text{i} S_{vQ}$ from that for $\text{i} S_{Qv}$, we must substitute in the latter: $\xi^{*}_{a Q}(t) \rightarrow \xi_{a Q}(t)$, $\xi_{a' v}(\tau) \rightarrow \xi^{*}_{a' v}(\tau)$, $(\uparrow, \downarrow) \rightarrow (\downarrow, \uparrow)$.

\section{Assumption of continuous field}

By looking at the final expressions derived in the previous Section, one can see that the total action on the Kadanoff-Baym contour, $S = S_{QQ} + S_{CQ} + S_{QC} + S_{vv} + S_{Qv} + S_{vQ}$ can be re-written as the sum of four contributions:
\begin{align}
S \equiv S_{\text{cc}} + S_{\text{bc}} + S_{\text{cb}} + S_{\text{bb}},
\label{action in four}
\end{align}
where $S_{\text{cc}}$ includes all the contributions requiring a double integration on the contour, $S_{\text{cb}}$ and $S_{\text{bc}}$ include the contributions requiring one integration on the contour and depending on the values of the fields at the boundaries, while $S_{\text{bb}}$ depends only on the values of fields at the boundaries. More precisely, we can parameterize the whole Kadanoff-Baym contour by means of a single parameter $z \in \left[ 0, z_f \right]$, with $z_f = 2 (t_{\infty} - t_0) + \beta$ such that
\begin{align}
& t = t_0 + z   \quad \quad \quad \quad \quad  \text{for } z \in \left[ 0, (t_{\infty} - t_0) \right] \quad \quad \quad \quad \quad \quad \, \Rightarrow t \in \left[ t_0, t_{\infty} \right] \equiv \gamma_+ , \nonumber \\
& t = 2 t_{\infty} - t_0 - z   \quad \quad  \text{for } z \in \left[ (t_{\infty} - t_0), 2 (t_{\infty} - t_0) \right] \quad \quad \,\, \Rightarrow t \in \left[ t_{\infty}, t_0 \right] \equiv \gamma_- ,\nonumber \\
& \tau = z - 2 (t_{\infty} - t_0) \quad  \text{for } z \in \left[ 2 (t_{\infty} - t_0), 2 (t_{\infty} - t_0) + \beta \right] \Rightarrow \tau \in \left[ 0, \beta \right] \equiv \gamma_v ,
\label{correspondence z t}
\end{align}
and write the action as a functional of the single field $\xi_a(z)$, which, with respect to the fields living on the various branches introduced above, is given by:
\begin{align}
\xi_{a}(z)  = & \Theta[(t_{\infty} - t_0) - z] \, \xi_{a +}(t_0 + z) \nonumber \\
& + \Theta[z - (t_{\infty} - t_0)] \, \Theta[2 (t_{\infty} - t_0) - z] \, \xi_{a -}(2 t_{\infty} - t_0 - z) \nonumber \\
& + \Theta[z - 2 (t_{\infty} - t_0)] \, \xi_{a v}(z - 2 t_{\infty} + 2 t_0).
\label{correspondence fields z t}
\end{align}
In the above equations, we have denoted as $t_{\infty}$ the time value corresponding to the end of the branch $\gamma_+$. As mentioned in Section 2, this can be taken as an arbitrary value, and it is usually sent to $\infty$. In addition to the initial point $z = 0$ and the final point $z = z_f$, there are two other special points: $z = z_{f+} \equiv t_{\infty} - t_0$ and $z = z_{f-} \equiv 2 (t_{\infty} - t_0)$, corresponding, respectively, to the end points of the branches $\gamma_+$ and $\gamma_-$. 

To proceed, we \emph{assume} that the field $\xi_a(z)$ is a \emph{continuous} function of $z$ on the whole contour, which means, in particular, that
\begin{align}
& \lim_{\epsilon \rightarrow 0^+} \Big[ \xi_a(z_{f+} - \epsilon) - \xi_a(z_{f+} + \epsilon) \Big] = 0 \Rightarrow \xi_{a+}(t_{\infty}) = \xi_{a-}(t_{\infty}), \nonumber \\
& \lim_{\epsilon \rightarrow 0^+} \Big[ \xi_a(z_{f-} - \epsilon) - \xi_a(z_{f-} + \epsilon) \Big] = 0 \Rightarrow \xi_{a-}(t_0) = \xi_{a v}(0),
\label{continuity}
\end{align}
so that the quantities $\xi_a(z_{f+})$ and $\xi_a(z_{f-})$ are uniquely defined. This assumption provides a significant simplification of the total action. It must be noted that, since it affects only the terms of the action labelled as $S_{\text{bc}}$, $S_{\text{cb}}$ and $S_{\text{bb}}$, while the magnetic interactions appear in $S_{\text{cc}}$, this simplification does not affect the discussion of time-dependent exchange interactions, which is the focus of this Article, so it must not be regarded as an additional approximation.

To exploit it, we take the total action and we separate it into the four parts introduced in Eq.\eqref{action in four}. Then, the first among Eqs.\eqref{continuity} implies $\xi_{Q}(\infty) = 0$, which sends to zero all the terms which would require the evaluation of Green functions at $t \rightarrow \infty$. To exploit the second among Eqs.\eqref{continuity}, we transform the fields $\xi_C(t_0)$ and $\xi_Q(t_0)$ back to $[\xi_+(t_0) + \xi_-(t_0)]/\sqrt{2}$ and $[\xi_+(t_0) - \xi_-(t_0)]/\sqrt{2}$, respectively, then we substitute $\xi_+(t_0) \equiv \xi(0)$, $\xi_-(t_0) = \xi_{v}(0)$, and $\xi_{v}(\beta) \equiv \xi(z_f)$. We obtain that all the boundary terms evaluated at the special points $z = z_{f+}$ and $z = z_{f-}$ vanish, consistently with the fact that they are not boundary points anymore for the field $\xi$. The resulting expressions for the four terms contributing to the action are:
\begin{align}
\text{i} S_{\text{cc}} = & \sum_{a} \sum_{a'} \int_{t_0}^{\infty} \text{d} t  \int_{t_0}^{\infty} \text{d} t' \Bigg\{  \Big[  \mathcal{A}^{>}_{a a'}(t, t') + \mathcal{A}^{<}_{a a'}(t, t') \Big] \xi^*_{a Q}(t) \,  \xi_{a' Q}(t') \nonumber \\
& \quad \quad \quad + \Big[ \mathcal{A}^{>}_{a a'}(t, t') - \mathcal{A}^{<}_{a a'}(t, t')   \Big] \Big[ \Theta(t' - t)  \, \xi^*_{a C}(t) \,   \xi_{a' Q}(t') \nonumber \\
& \quad \quad \quad \quad \quad \quad \quad \quad   -  \Theta(t - t') \,  \xi^*_{a Q}(t)   \,  \xi_{a' C}(t') \Big] \nonumber \\
&  \quad \quad \quad + \text{i} \, \delta(t - t')  \Big[ \xi^*_{a C}(t) \, \xi_{a' Q}(t) + \xi^*_{a Q}(t) \, \xi_{a' C}(t) \Big]  \mathcal{F}_{a a'}(t) \Bigg\} \nonumber \\
& - \sqrt{2} \text{i} \sum_a \sum_{a'}   \int_{t_0}^{\infty} \text{d}t \int_{0}^{\beta} \text{d} \tau \Big[  \xi_{a Q}^*(t) \, \xi_{a' v}(\tau) \, \mathcal{G}_{a a'}^{\downarrow \uparrow}(t, \tau)  \nonumber \\
& \quad \quad \quad \quad \quad \quad \quad \quad +  \xi_{a' Q}(t) \, \xi^*_{a v}(\tau)  \, \mathcal{G}_{a' a}^{\uparrow \downarrow}(t, \tau) \Big] \nonumber \\
& - \sum_{a} \sum_{a'}   \int_{0}^{\beta} \text{d} \tau  \int_{0}^{\beta} \text{d} \tau'     \xi^*_{a v}(\tau) \Big[ \mathcal{K}_{a a'}(\tau - \tau') + \beta \delta_{a a'} \delta(\tau- \tau') \mathcal{I}_a \Big] \xi_{a' v}(\tau') ,
\label{Scc}
\end{align}
\begin{align}
\text{i} S_{\text{bc}} = &  \sum_{a} \sum_{a'}  \Big[ \xi^*_{a}(0)     - \xi^*_{a}(z_f)   \Big]   \nonumber \\
& \quad \cdot \Bigg\{ \int_{0}^{\beta} \text{d} \tau    \left[ G^{\mathbb{M} \downarrow}_{a a'}( - \tau) \, I^{\mathbb{M} \uparrow}_{a' a}(\tau) - J^{\mathbb{M} \downarrow}_{a a'}(- \tau) \, G^{\mathbb{M} \uparrow}_{a' a}(\tau)  \right] \xi_{a' v}(\tau)   \nonumber \\
&  \quad \quad  + \sqrt{2}  \text{i}  \int_{t_0}^{\infty} \! \text{d} t    \Big[  G^{< \downarrow}_{a a'}(t_0, t)   W^{> \uparrow}_{a' a}(t, t_0) -  X^{< \downarrow}_{a a'}(t_0, t) G^{> \uparrow}_{a' a}(t, t_0)  \Big] \xi_{a' Q}(t) \Bigg\} ,
\label{Sbc}
\end{align}
\begin{align}
\text{i} S_{\text{cb}} = &  \sum_a \sum_{a'} \Bigg\{  \sqrt{2} \text{i}  \int_{t_0}^{\infty} \text{d} t  \, \xi^*_{a Q}(t) \Big[ W^{> \downarrow}_{a a'}(t, t_0)   G^{< \uparrow}_{a' a}(t_0, t)  - G^{> \downarrow}_{a a'}(t, t_0)   X^{< \uparrow}_{a' a}(t_0, t) \Big]      \nonumber \\ 
& \quad \quad +  \int_{0}^{\beta} \text{d} \tau  \xi^*_{a v}(\tau) \Big[ I^{\mathbb{M} \downarrow}_{a a'}(\tau) \, G^{\mathbb{M} \uparrow}_{a' a}( - \tau)  - G^{\mathbb{M} \downarrow}_{a a'}(\tau) \, J^{\mathbb{M} \uparrow}_{a' a}(- \tau)  \Big]  \Bigg\} \nonumber \\
& \quad \quad \cdot \Big[ \xi_{a'}(0)  - \xi_{a'}(z_f)  \Big] ,
\label{Scb}
\end{align}
\begin{align}
\text{i} S_{\text{bb}} = & \sum_a \sum_{a'} \Big[ \begin{matrix}  \xi^*_a(0)  & \xi^*_a(z_f)  \end{matrix} \Big] \nonumber \\ 
& \quad \quad \quad \quad \cdot \left\{  \rho^{\downarrow}_{a a'}  \rho^{\uparrow}_{a' a} \left[ \begin{matrix} 1 & -1 \\  -1  & 1 \end{matrix}  \right]    
+  \delta_{a a'} \left[ \begin{matrix} - \rho^{\text{C}}_{a a}    & \rho^{\uparrow}_{a a}  \\ \rho^{\downarrow}_{a a}   & - \rho^{\text{C}}_{a a}  \end{matrix}  \right]  \right\}  
\cdot \left[ \begin{matrix}  \xi_{a'}(0)   \\ \xi_{a'}(z_f)  \end{matrix} \right] .
\label{Sbb}
\end{align}
In the remainder of this Article, we focus on the analysis of the term $S_{\text{cc}}$, given in Eq.\eqref{Scc}, which includes the magnetic interactions between the time-dependent spin fields. If one assumes that $\xi_{a}(0) = \xi_{a}(z_f)$, then the terms $S_{\text{bc}}$, $S_{\text{cb}}$ and $S_{\text{bb}}$ vanish [as it can be seen by looking at Eqs.\eqref{Sbc}, \eqref{Scb} and \eqref{Sbb}], and the action then reduces to $S_{\text{cc}}$.

\section{Exchange and twist-exchange interactions}

We now prove that the effective action for the small deviations from collinear spin configurations for the Hubbard model, that we have just derived, is equivalent to that of a time-dependent spin model, which means that the action can be written as the sum of interactions between spins. To show this, we notice that it is possible to transform the $\xi$ fields to unit spin vectors via the relation
\begin{align}
\xi_{A}^* \xi_{B} \cong \frac{1}{4} \left[ \boldsymbol{e}_{A} \cdot \boldsymbol{e}_{B} - 1 + \frac{\theta_{A}^2 + \theta_{B}^2}{2} + \text{i} \, \boldsymbol{u}_z \cdot \Big( \boldsymbol{e}_{A} \times \boldsymbol{e}_{B} \Big) \right] ,
\label{xi to vectors}
\end{align} 
where $A$ and $B$ are sets of indexes, including Hubbard site-orbital, branch $(+, -, v)$ and time labels. The transformation Eq.\eqref{xi to vectors} is correct up to the quadratic order in the azimuthal angles $\theta$'s, which is consistent with the degree of accuracy of our formulation. The products of $C$ and $Q$ fields appearing in in Eq.\eqref{Scc} become:
\begin{align}
& \xi^*_{a Q}(t) \, \xi_{a' Q}(t') \cong \frac{1}{4} \Big\{ \boldsymbol{e}_{a Q}(t) \cdot \boldsymbol{e}_{a' Q}(t') + \text{i} \boldsymbol{u}_z \cdot \Big[ \boldsymbol{e}_{a Q}(t) \times \boldsymbol{e}_{a' Q}(t')\Big] \Big\}, \nonumber \\
& \xi^*_{a C}(t) \, \xi_{a' Q}(t') \cong \frac{1}{4} \Big\{ \boldsymbol{e}_{a C}(t) \cdot \boldsymbol{e}_{a' Q}(t') + \text{i} \boldsymbol{u}_z \cdot \Big[ \boldsymbol{e}_{a C}(t) \times \boldsymbol{e}_{a' Q}(t')\Big] \Big\} \nonumber \\
& \quad \quad \quad \quad \quad \quad \quad + \frac{\theta_{a' +}^2(t') - \theta_{a' -}^2(t')}{8}, \nonumber \\
& \xi^*_{a Q}(t) \, \xi_{a' C}(t') \cong \frac{1}{4} \Big\{ \boldsymbol{e}_{a Q}(t) \cdot \boldsymbol{e}_{a' C}(t') + \text{i} \boldsymbol{u}_z \cdot \Big[ \boldsymbol{e}_{a Q}(t) \times \boldsymbol{e}_{a' C}(t')\Big] \Big\} \nonumber \\
& \quad \quad \quad \quad \quad \quad \quad + \frac{\theta_{a +}^2(t) - \theta_{a -}^2(t)}{8},
\end{align}
where we have first undone the Keldysh transform from $(\xi_Q, \xi_C)$ to $(\xi_+, \xi_-)$ in order to apply Eq.\eqref{xi to vectors}, and then we have defined
\begin{align}
& \boldsymbol{e}_{a C}(t) \equiv \left[ \boldsymbol{e}_{a +}(t) + \boldsymbol{e}_{a -}(t) \right] / \sqrt{2} ,   \nonumber \\
& \boldsymbol{e}_{a Q}(t) \equiv \left[ \boldsymbol{e}_{a +}(t) - \boldsymbol{e}_{a -}(t) \right] / \sqrt{2} .
\end{align}
After this transformation, we can write $S_{\text{cc}}$ as the sum of two contributions:
\begin{align}
S_{\text{cc}} \equiv  S_{\text{exchange}} +  S_{\text{twist}},
\end{align}
where
\begin{align}
& \text{i} S_{\text{exchange}}  \equiv \frac{1}{4} \sum_{a} \sum_{a'} \int_{t_0}^{\infty} \text{d} t  \int_{t_0}^{\infty} \text{d} t' \Bigg\{ 2 \text{i} \, \boldsymbol{e}_{a C}(t) \cdot \boldsymbol{e}_{a' Q}(t')  \nonumber \\
& \quad \quad \cdot  \Big[  \delta(t - t') \,  \text{Re}\left[ \mathcal{F}_{a a'}(t) \right] + \Theta(t' - t) \, \text{Im}\left[\mathcal{A}^{>}_{a a'}(t, t') - \mathcal{A}^{<}_{a a'}(t, t') \right]    \Big] \nonumber \\
&  \quad  +  \boldsymbol{e}_{a Q}(t)  \cdot  \boldsymbol{e}_{a' Q}(t')  \,  \text{Re}\left[ \mathcal{A}^{>}_{a a'}(t, t') + \mathcal{A}^{<}_{a a'}(t, t')  \right]   \Bigg\} \nonumber \\
&  \quad - \frac{\text{i}}{2 \sqrt{2}} \sum_a \sum_{a'}   \int_{t_0}^{\infty} \text{d}t \int_{0}^{\beta} \text{d} \tau   \, \boldsymbol{e}_{a Q}(t) \cdot \boldsymbol{e}_{a' v}(\tau) \,   \Big[ \mathcal{G}_{a a'}^{\downarrow \uparrow}(t, \tau) + \mathcal{G}_{a a'}^{\uparrow \downarrow}(t, \tau) \Big] \nonumber \\
&  \quad - \frac{1}{4}  \sum_{a} \sum_{a'}   \int_{0}^{\beta} \text{d} \tau  \int_{0}^{\beta} \text{d} \tau' \Big[ \boldsymbol{e}_{a v}(\tau) \cdot \boldsymbol{e}_{a' v}(\tau') - 1  \Big]  \mathcal{K}_{a a'}(\tau - \tau')
\label{symmetrized exchange action}
\end{align}
is the exchange part (including all the terms depending on scalar products $\boldsymbol{e}_{A} \cdot \boldsymbol{e}_{B}$), and
\begin{align}
& \text{i} S_{\text{twist}}  \equiv  \frac{1}{4}  \boldsymbol{u}_z \cdot \sum_{a} \sum_{a'} \int_{t_0}^{\infty} \text{d} t  \int_{t_0}^{\infty} \text{d} t' \Bigg\{ 2 \text{i} \, \boldsymbol{e}_{a C}(t) \times \boldsymbol{e}_{a' Q}(t')     \nonumber \\
& \quad \quad  \cdot  \Big[  -  \, \delta(t - t') \, \text{Im}\left[ \mathcal{F}_{a a'}(t) \right] + \Theta(t' - t)  \text{Re} \left[ \mathcal{A}^{>}_{a a'}(t, t')  -   \mathcal{A}^{<}_{a a'}(t, t') \right]  \Big] \nonumber \\
& \quad -    \boldsymbol{e}_{a Q}(t)    \times  \boldsymbol{e}_{a' Q}(t')  \,  \text{Im}\left[ \mathcal{A}^{>}_{a a'}(t, t') + \mathcal{A}^{<}_{a a'}(t, t') \right]   \Bigg\} \nonumber \\
& \quad +   \frac{1}{2 \sqrt{2}} \boldsymbol{u}_z \cdot \sum_a \sum_{a'}   \int_{t_0}^{\infty}  \text{d}t \int_{0}^{\beta}  \text{d} \tau   \, \boldsymbol{e}_{a Q}(t)   \times \boldsymbol{e}_{a' v}(\tau)  \Big[ \mathcal{G}_{a a'}^{\downarrow \uparrow}(t, \tau) - \mathcal{G}_{a a'}^{\uparrow \downarrow}(t, \tau) \Big]   \nonumber \\
& \quad  - \frac{1}{4} \text{i} \, \boldsymbol{u}_z \cdot \sum_{a} \sum_{a'}   \int_{0}^{\beta} \text{d} \tau  \int_{0}^{\beta} \text{d} \tau' \, \boldsymbol{e}_{a v}(\tau) \times \boldsymbol{e}_{a' v}(\tau') \, \mathcal{K}_{a a'}(\tau - \tau') 
\label{symmetrized DM action}
\end{align}
is an additional part, including all the terms depending on vector products $\boldsymbol{e}_{A} \times \boldsymbol{e}_{B}$), that corresponds to a new kind of magnetic interaction, which we call \emph{twist exchange}. In Eq.\eqref{symmetrized exchange action} we have already removed an obviously null term equal to $\Big[ \boldsymbol{e}_{a v}(\tau) \cdot \boldsymbol{e}_{a' v}(\tau') - 1  \Big] \delta_{a a'} \delta(\tau - \tau') \beta \mathcal{I}_a$, and we have exploited the symmetry of $\boldsymbol{e}_{a \pm}(t) \cdot \boldsymbol{e}_{a' \pm}(t')$ under the interchange of both $a \leftrightarrow a'$ and $t \leftrightarrow t'$, combined with the symmetry properties of the quantities $\mathcal{A}^{\lessgtr}_{a a'}(t, t')$ and $\mathcal{F}_{a a'}(t)$ under the same transformations. Similarly, in Eq.\eqref{symmetrized DM action} we have removed an obviously null term equal to $ \boldsymbol{e}_{a v}(\tau) \times \boldsymbol{e}_{a' v}(\tau') \, \delta_{a a'} \delta(\tau - \tau') \beta \mathcal{I}_a$, and we have exploited the antisymmetry of $\boldsymbol{e}_{a \pm}(t) \times \boldsymbol{e}_{a' \pm}(t')$ under the interchange of both $a \leftrightarrow a'$ and $t \leftrightarrow t'$.

From the transformation Eq.\eqref{xi to vectors}, it is seen that also an action term $S_{\theta^2}$ appears, depending on the squares of the azimuthal angles $\theta_{A}^2$. This \emph{anisotropy} term, however, is identically zero as a consequence of the symmetry properties of the kernel matrices. We give the detailed proof of this fact in \ref{nullStheta}. Therefore, we correctly find that the system does not exhibit any anisotropy, consistently with the fact that we are not including spin-orbit, and the action is entirely written in terms of interactions between spins.

The term $S_{\text{twist}}$ is particularly interesting, and deserves special attention. While at a first sight it seems to represent an effective Dzyaloshinskii-Moriya interaction, we know that this cannot be the right interpretation, since our system does not include spin-orbit. On the other hand, the action of a non-relativistic system must be invariant under rotation of all spins of the same angle about an arbitrary axis. Therefore, the allowed effective interactions between spins must be of the forms $\propto \boldsymbol{\sigma}_1 \cdot \boldsymbol{\sigma}_2$, or $\propto  \left( \boldsymbol{\sigma}_1 \times \boldsymbol{\sigma}_2 \right) \cdot \boldsymbol{\sigma}_3$, or products of these combinations. While the first form is commonly identified as the exchange interaction, the second one is less known. To the best of our knowledge, the first discussion of a similar interaction was given by Bogoliubov \cite{BogoliubovBook} in 1949, in the context of the \emph{theory of the polar model} \cite{Schubin34} (a precursor of the Hubbard model). His treatment is based on what we would call today a half-filled single-band Hubbard model (in equilibrium), and a three-spin interaction of the form $\propto  \left( \boldsymbol{\sigma}_1 \times \boldsymbol{\sigma}_2 \right) \cdot \boldsymbol{\sigma}_3$ appears at the third order in the expansion of the Hamiltonian in powers of $T / U$. However, in that case the corresponding term vanished by symmetry. Here, we are not making a perturbation expansion in terms of $T / U$, since in our expansion the quantities which are assumed to be small are the angles of deviation of the spins from the equilibrium configuration. Therefore, we receive all terms allowed by symmetries, irrespectively of their dependence on the Hubbard parameters. Since our expansion is up to the second order in the deviation angles, it corresponds to interactions $\propto \delta\boldsymbol{\sigma}_1 \cdot \delta \boldsymbol{\sigma}_2$ and, without loss of generality, $\propto \left( \delta \boldsymbol{\sigma}_1 \times \delta \boldsymbol{\sigma}_2 \right) \cdot \boldsymbol{\sigma}_3$. In the second form, only $\boldsymbol{\sigma}_1$ and $\boldsymbol{\sigma}_2$ are rotated, while the third spin vector is pointing along the original quantization axis, which is $\boldsymbol{u}_z$ in our model. Therefore, since the third vector is fixed, the interaction has the form $\propto \left( \boldsymbol{e}_1 \times \boldsymbol{e}_2 \right) \cdot \boldsymbol{u}_z$, which is precisely what we find. Therefore, despite its resemblance with the Dzyaloshinskii-Moriya interaction, we conclude that the contribution $S_{\text{twist}}$ describes a three-spin non-relativistic interaction; the unit vector $\boldsymbol{u}_z$ is to be interpreted as the quantization axis of the third spin, which is not rotated in the expansion. Note that the Dzyaloshinskii-Moriya vector is \emph{not} parallel to $\boldsymbol{u}_z$ \cite{Katsnelson10}. Differently from the case studied by Bogoliubov, we are here treating a multiband Hubbard model in non-equilibrium, and in our case this interaction does not vanish in general.

\section{Equilibrium}

Before treating the general non-equilibrium case, which is the main goal of this Article, we here specialize the discussion to the equilibrium regime. This had already been studied, e.g., in Ref.\cite{Katsnelson00, Katsnelson02} with different methods, and we now show that we are able to reproduce the known results. 

If the system is in equilibrium, the general treatment pursued so far is overkill. In fact, consider the expression of the time-dependent expectation value of an operator $\hat{O}$, as given in Eq.\eqref{exp value operator}. In the equilibrium limit, i.e., $\hat{H}(t) = \hat{H}_0$, one has $\hat{U}(t, t') = \exp[- \text{i} (t - t') \hat{H}_0]$. Exploiting the cyclic property of the trace, as well as $[\hat{N}, \hat{H}_0] = 0$, one obtains the equilibrium expression:
\begin{align}
O(t) =  \frac{\text{Tr} \left[ \text{e}^{- \beta \left( \hat{H}_0 - \mu \hat{N} \right)   }  \hat{O}   \right]}{\text{Tr} \left[  \text{e}^{- \beta \left( \hat{H}_0 - \mu \hat{N} \right)   } \right]} = \frac{\text{Tr} \left[  \hat{U}_v(t_0 - \text{i} \beta, t_0)  \hat{O}   \right]}{\text{Tr} \left[  \text{e}^{- \beta \left( \hat{H}_0 - \mu \hat{N} \right)   } \right]},
\label{exp value equilibrium}
\end{align}
which shows that the equilibrium expectation value of any operator can be evaluated by just considering the vertical branch of the Kadanoff-Baym contour. In the path-integral formulation, this results in an effective partition function given by
\begin{align}
\mathcal{Z}^{\text{EQ}} \equiv \int \mathcal{D}\left[ \bar{c}, c \right] \text{e}^{\text{i} S_{vv}\left[ \bar{c}, c\right]} \equiv \int \mathcal{D}\left[ \bar{c}, c \right] \text{e}^{- S^{\text{EQ}}\left[ \bar{c}, c\right]},
\label{Z_eq}
\end{align}
where $S^{\text{EQ}} = - \text{i} S_{vv}$. Only $\text{i} S_{vv}$ is therefore needed to compute the equilibrium properties, and in this Section we study this portion of the action in order to recover the known equilibrium results.

\subsection{Equilibrium action}

The equilibrium action (excluding the boundary terms) is given by:
\begin{align}
S^{\text{EQ}}[\boldsymbol{e}(\tau)] = & \frac{1}{4} \sum_a \sum_{a'} \int_0^{\beta} \text{d} \tau \int_0^{\beta} \text{d} \tau' \mathcal{K}_{a a'}(\tau - \tau') \Big\{ \boldsymbol{e}_{a v}(\tau) \cdot \boldsymbol{e}_{a' v}(\tau') - 1 \nonumber \\
& + \text{i} \boldsymbol{u}_z \cdot \Big[ \boldsymbol{e}_{a v}(\tau) \times \boldsymbol{e}_{a' v}(\tau') \Big]\Big\}.
\end{align}
This action is a functional of the $\tau$-dependent fields $\boldsymbol{e}_{a v}(\tau)$. For the particular choice of \emph{constant} fields, i.e.,
\begin{align}
\xi_{a v}(\tau) \equiv \xi_{a v} \quad \forall \tau \in [ 0, \beta ],
\label{assumption xi_v constant}
\end{align}
which is equivalent to neglecting the so-called \emph{quantum fluctuations} (see, e.g., the discussion in Chapter 6 of Ref.\cite{AltlandBook}), after integrating over $\text{d} \tau$ and $\text{d} \tau'$ the quantity $\mathcal{K}_{a a'}(\tau - \tau')$ we obtain the action
\begin{align}
S^{\text{EQ}}[\boldsymbol{e}] \equiv & \frac{1}{4} \beta^4 \sum_a \sum_{a'} \sum_{n = - \infty}^{+ \infty} \mathcal{K}_{a a'}(\omega_n, \omega_n) \Big[ \boldsymbol{e}_{a v} \cdot \boldsymbol{e}_{a' v} - 1 + \text{i} \boldsymbol{u}_z \cdot \Big( \boldsymbol{e}_{a v} \times \boldsymbol{e}_{a' v} \Big) \Big] .
\end{align}
From Eq.\eqref{to prove hermiticity J}, one can show that
\begin{align}
\sum_{n = - \infty}^{+ \infty} \mathcal{K}_{a a'}(\omega_n, \omega_n) = \left[ \sum_{n = - \infty}^{+ \infty} \mathcal{K}_{a' a}(\omega_n, \omega_n) \right]^*,
\label{symmetry trace K}
\end{align}
and by using this relation together with the symmetry of $\Big( \boldsymbol{e}_{a v} \cdot \boldsymbol{e}_{a' v} - 1 \Big)$ and the antisymmetry of $\boldsymbol{e}_{a v} \times \boldsymbol{e}_{a' v}$ under $a \leftrightarrow a'$, we obtain
\begin{align}
S^{\text{EQ}}[\boldsymbol{e}] & \equiv \beta^4  \sum_a \sum_{a'} \Big[ - \left( \boldsymbol{e}_{a v} \cdot \boldsymbol{e}_{a' v} - 1 \right) \mathcal{J}_{a a'}  +  \left( \boldsymbol{e}_{a v} \times \boldsymbol{e}_{a' v} \right) \cdot \boldsymbol{u}_z \mathcal{Y}_{a a'} \Big] ,
\label{equilibrium action 3}
\end{align}
where we have introduced the \emph{equilibrium exchange parameters}
\begin{align}
\mathcal{J}_{a a'} \equiv - \frac{1}{4} \text{Re} \left[ \sum_{n = -\infty}^{+ \infty} \mathcal{K}_{a a'}(\omega_n, \omega_n) \right] 
\label{Jaa'}
\end{align}
and the \emph{equilibrium twist-exchange parameters}
\begin{align}
\mathcal{Y}_{a a'} & \equiv  -  \frac{1}{4} \text{Im} \left[ \sum_{n = -\infty}^{+ \infty} \mathcal{K}_{a a'}(\omega_n, \omega_n) \right] .
\label{D-M vector}
\end{align}
Evidently, from Eq.\eqref{symmetry trace K} it follows that $\mathcal{J}_{a a'} = \mathcal{J}_{a' a}$ and $\mathcal{Y}_{a a'} = - \mathcal{Y}_{a' a}$.

We can make a further step by re-writing Eqs.\eqref{Jaa'} and \eqref{D-M vector} in the $\tau$-representation, which is achieved by inverting the Fourier transforms:
\begin{align}
& \sum_{n = -\infty}^{+ \infty}  \mathcal{K}_{a a'}(\omega_n, \omega_n)   =  \, \frac{1}{\beta^4} \int_{0}^{\beta} \! \text{d} \tau_1 \! \int_{0}^{\beta} \! \text{d} \tau_2 \! \int_{0}^{\beta} \! \text{d} \tau_3 \! \int_{0}^{\beta} \! \text{d} \tau_4  \sum_{n = -\infty}^{+ \infty} \text{e}^{-\text{i} \omega_n (\tau_1 + \tau_2 + \tau_3 + \tau_4)} \nonumber \\
& \quad \quad \quad \quad \quad \quad  \times  \Bigg\{ - \Big[ \Sigma^{\mathbb{M} \downarrow}(\tau_1) \cdot G^{\mathbb{M} \downarrow}(\tau_2) \Big]_{a a'} \Big[ \Sigma^{\mathbb{M} \uparrow}(\tau_3) \cdot G^{\mathbb{M} \uparrow}(\tau_4) \Big]_{a' a} \nonumber \\
& \quad \quad \quad \quad \quad \quad  -  \Big[ G^{\mathbb{M} \downarrow}(\tau_1) \cdot \Sigma^{\mathbb{M} \downarrow}(\tau_2) \Big]_{a a'}  \Big[ G^{\mathbb{M} \uparrow}(\tau_3) \cdot \Sigma^{\mathbb{M} \uparrow}(\tau_4) \Big]_{a' a}   \nonumber \\
& \quad \quad \quad \quad \quad \quad  +  G_{a a'}^{\mathbb{M} \downarrow}(\tau_1) \Big[ \Sigma^{\mathbb{M} \uparrow}(\tau_2) \cdot G^{\mathbb{M} \uparrow}(\tau_3)  \cdot \Sigma^{\mathbb{M} \uparrow}(\tau_4) \Big]_{a' a} \nonumber \\
& \quad \quad \quad \quad \quad \quad  +  \Big[ \Sigma^{\mathbb{M} \downarrow}(\tau_1) \cdot G^{\mathbb{M} \downarrow}(\tau_2) \cdot  \Sigma^{\mathbb{M} \downarrow}(\tau_3) \Big]_{a a'} G_{a' a}^{\mathbb{M} \uparrow}(\tau_4) \Bigg\}  \nonumber \\ 
& - \frac{1}{\beta^4} \int_{0}^{\beta}  \text{d} \tau_1  \int_{0}^{\beta}  \text{d} \tau_2  \sum_{n = -\infty}^{+ \infty} \text{e}^{-\text{i} \omega_n (\tau_1 + \tau_2 )}   \Big[ G_{a a'}^{\mathbb{M} \downarrow}(\tau_1) \,  \Sigma_{a' a}^{\mathbb{M} \uparrow}(\tau_2) + \Sigma_{a a'}^{\mathbb{M} \downarrow}(\tau_1) \, G_{a' a}^{\mathbb{M} \uparrow}(\tau_2) \Big] .
\label{trace K tau general}
\end{align}
Now, the sums of the form
\begin{align}
\sum_{n = -\infty}^{+ \infty} \text{e}^{-\text{i} \omega_n \tau} = \sum_{n = -\infty}^{+ \infty} \text{cos}( \omega_n \tau )
\end{align}
are real. Moreover, since the system that we are studying is non-relativistic (NR), we can choose a basis set of real single-particle wave functions, which implies that $T_{ab}$ is real, hence the driving operator on the LHS of the Matsubara equations [the last two among Eqs.\eqref{KB equations}] is real, as well as the equilibrium pair correlation function $\rho_{a b}$. From Eqs.\eqref{boundary 0 Matsubara}, it then follows that the Matsubara Green function $G_{a b}^{\mathbb{M}}(\tau)$ is purely imaginary close to the discontinuity point $\tau = 0$, and will therefore stay imaginary on the whole domain $- \beta < \tau < \beta$. This also implies that the self-energy $\Sigma_{a b}^{\mathbb{M}}(\tau)$ be imaginary. With this choice of the basis set, it must be noted that from Eq.\eqref{symmetry Matsubara tau} it follows that $G_{a b}^{\mathbb{M}}(\tau) =  G_{b a}^{\mathbb{M}}(\tau)$, and from Eq.\eqref{symmetry Sigma tau} it follows that $\Sigma_{a b}^{\mathbb{M}}(\tau) = \Sigma_{b a}^{\mathbb{M}}(\tau)$. Therefore, the quantity appearing in Eq.\eqref{trace K tau general} is real, and from Eq.\eqref{D-M vector} we have $\mathcal{Y}_{a a'} = 0$, consistently with the discussion by Bogoliubov \cite{BogoliubovBook}.

\subsection{Local single-band case in equilibrium}

To establish the correspondence with the results of Ref.\cite{Katsnelson00}, we now study the equilibrium action \eqref{equilibrium action 3} in the \emph{local single-band} (LSB) approximation. Here \emph{local} means that the self-energy is assumed to be diagonal in site index, and because of the fact that each site carries a single orbital (\emph{single band}), we put
\begin{align}
\Sigma_{a b}^{\mathbb{M} \sigma}(\omega_{n})  \stackrel{ \text{LSB} }{\longrightarrow} \delta_{a b} \Sigma_{a}^{\mathbb{M} \sigma}(\omega_{n}).
\end{align}
Under this assumption, 
from \eqref{K frequency space} we obtain
\begin{align}
\sum_{n = -\infty}^{+ \infty}  \mathcal{K}_{a a'}(\omega_n, \omega_n)   \stackrel{ \text{LSB} }{=} & \, 4 \sum_{n = -\infty}^{+ \infty}  \Sigma_{a}^{\mathbb{M} \text{S}}(\omega_{n}) \, G_{a a'}^{\mathbb{M} \downarrow}(\omega_{n}) \, \Sigma_{a'}^{\mathbb{M} \text{S}}(\omega_{n}) \, G_{a' a}^{\mathbb{M} \uparrow}(\omega_{n}) \nonumber \\
&  \, -  \frac{\delta_{a a'}}{\beta^2}  \sum_{n = - \infty}^{+ \infty}  \Big[ G_{a a}^{\mathbb{M} \downarrow}(\omega_{n}) \,  \Sigma_{a}^{\mathbb{M} \uparrow}(\omega_{n}) + \Sigma_{a}^{\mathbb{M} \downarrow}(\omega_{n}) \, G_{a a}^{\mathbb{M} \uparrow}(\omega_{n}) \Big] ,
\label{trace K LNR}
\end{align}
where $\Sigma_{a}^{\mathbb{M} \text{S}} \equiv \left( \Sigma_{a}^{\mathbb{M} \uparrow} - \Sigma_{a}^{\mathbb{M} \downarrow} \right) / 2$. The term proportional to $\delta_{a a'}$ appearing in Eq.\eqref{trace K LNR} does not contribute to the action, as it can be seen by looking at Eq.\eqref{equilibrium action 3}, hence it can be immediately suppressed. Eliminating this term, in the LSB case we can write the effective exchange parameters just as
\begin{align}
\mathcal{J}_{a a'} \stackrel{ \text{LSB} }{\equiv} - \sum_{n = -\infty}^{+ \infty}  \Sigma_{a}^{\mathbb{M} \text{S}}(\omega_{n}) \,  G_{a a'}^{\mathbb{M} \downarrow}(\omega_{n}) \, \Sigma_{a'}^{\mathbb{M} \text{S}}(\omega_{n}) \, G_{a' a}^{\mathbb{M} \uparrow}(\omega_{n}) ,
\label{J_aa' diag without delta}
\end{align}
which is precisely the result obtained in Ref.\cite{Katsnelson00} with a different method. Our analysis provides the more general expression Eq.\eqref{Jaa'}, which can be applied in the case of a non-local (non-diagonal in site index) self-energy, as well as the interpretation of the results of Ref.\cite{Katsnelson00} in the more general framework of the path-integral formulation. Our action includes also terms depending on $\tau$-dependent spin vectors (quantum fluctuations), which contribute to the complete treatment of the equilibrium properties.

\section{Non-equilibrium}

We now consider the full non-equilibrium action, given by Eqs.\eqref{symmetrized exchange action} and \eqref{symmetrized DM action}. We notice that both the exchange and the twist-exchange contributions are written as sums of several different terms: 1) a term $S^{RR}$ requiring a double integration on the real-time branch, which is related to the real-time evolution of the system, 2) a term $S^{RI}$ requiring one integration on the real-time branch and one integration on the imaginary-time branch, which describes the influence of the initial correlations on the evolution of the system, and 3) a term $S^{II}$ requiring a double integration on the imaginary-time branch, which describes the system in equilibrium. 

In turn, the first term can be written as the sum of three different contributions. The term involving products of two $Q$ spin vectors is the source of noise \cite{KamenevBook, Fransson08}, which has contributions of both exchange and twist-exchange forms. The pertinent parameters are: 
\begin{align}
& \mathcal{J}^{\text{noise}}_{a a'}(t, t') \equiv   \frac{1}{4} \text{Re}\left[ \mathcal{A}^{>}_{a a'}(t, t') + \mathcal{A}^{<}_{a a'}(t, t')  \right], \nonumber \\
& \mathcal{Y}^{\text{noise}}_{a a'}(t, t') \equiv  - \frac{1}{4} \text{Im}\left[ \mathcal{A}^{>}_{a a'}(t, t') + \mathcal{A}^{<}_{a a'}(t, t')  \right] .
\end{align}

The term involving the products of a $C$ and a $Q$ spin vectors can be interpreted by first going back to the representation in terms of branch vectors, $\boldsymbol{e}_+$ and $\boldsymbol{e}_-$. One has $ 2 \boldsymbol{e}_{a C}(t) \cdot \boldsymbol{e}_{a' Q}(t') =  \Big[ \boldsymbol{e}_{a +}(t) \cdot \boldsymbol{e}_{a' +}(t') - \boldsymbol{e}_{a -}(t) \cdot \boldsymbol{e}_{a' -}(t') \Big] + \Big[ \boldsymbol{e}_{a -}(t) \cdot \boldsymbol{e}_{a' +}(t') - \boldsymbol{e}_{a +}(t) \cdot \boldsymbol{e}_{a' -}(t')  \Big]$, where the first (second) pair of square brackets enclose a term which is symmetric (antisymmetric) with respect to the tranformation $(a, t) \leftrightarrow (a', t')$. Analogously, $ 2 \boldsymbol{e}_{a C}(t) \times \boldsymbol{e}_{a' Q}(t') =  \Big[ \boldsymbol{e}_{a +}(t) \times \boldsymbol{e}_{a' +}(t') - \boldsymbol{e}_{a -}(t) \times \boldsymbol{e}_{a' -}(t') \Big] + \Big[ \boldsymbol{e}_{a -}(t) \times \boldsymbol{e}_{a' +}(t') - \boldsymbol{e}_{a +}(t) \times \boldsymbol{e}_{a' -}(t')  \Big]$, where the term enclosed by the first (second) pair of square brackets is antisymmetric (symmetric) under $(a, t) \leftrightarrow (a', t')$. Then, exploiting the symmetries provided by Eqs.\eqref{symmetry A} and \eqref{symmetry cal F}, we can write the resulting coefficient of $\Big[ \boldsymbol{e}_{a +}(t) \cdot \boldsymbol{e}_{a' +}(t')  - \boldsymbol{e}_{a -}(t) \cdot \boldsymbol{e}_{a' -}(t') \Big]$ in Eq.\eqref{symmetrized exchange action}, which we call the \emph{non-equilibrium exchange parameter}, as 
\begin{align}
\mathcal{J}_{a a'}(t, t') \! \equiv \! - \frac{1}{4} \! \left\{ \delta(t - t') \,  \text{Re} \! \left[ \mathcal{F}_{a a'}(t) \right] - \! \frac{\text{sign}(t - t') }{2} \text{Im}\!\left[\mathcal{A}^{>}_{a a'}(t, t') - \!  \mathcal{A}^{<}_{a a'}(t, t') \right] \right\} \! ,
\label{noneq exchange}
\end{align}
and we can write the coefficient of the term $\Big[ \boldsymbol{e}_{a +}(t) \times \boldsymbol{e}_{a' +}(t')  - \boldsymbol{e}_{a -}(t) \times \boldsymbol{e}_{a' -}(t') \Big]$ in Eq.\eqref{symmetrized DM action}, which we call the \emph{non-equilibrium twist-exchange parameter}, as 
\begin{align}
\mathcal{Y}_{a a'}(t, t') \! \equiv \! \frac{1}{4}  \left\{  \delta(t - t') \,  \text{Im}\left[ \mathcal{F}_{a a'}(t) \right]  +   \frac{\text{sign}(t - t')}{2} \text{Re}\!\left[\mathcal{A}^{>}_{a a'}(t, t') -  \mathcal{A}^{<}_{a a'}(t, t') \right]  \right\}.
\label{noneq DM}
\end{align}
It can be immediately seen that $\mathcal{J}_{a a'}(t, t')  = \mathcal{J}_{a' a}(t', t)$, while $\mathcal{Y}_{a a'}(t, t')  = - \mathcal{Y}_{a' a}(t', t)$, analogously to the properties of the equilibrium parameters $\mathcal{J}_{a a'}$ and $\mathcal{Y}_{a a'}$. The structures of the non-equilibrium parameters, as defined in Eqs.\eqref{noneq exchange} and \eqref{noneq DM}, are analogous to those of the equilibrium quantities, but the Matsubara Green functions appearing in the latter are replaced here by real-time non-equilibrium Green functions. The last contribution to $S^{RR}$ is represented by additional terms which are related to the causality structure of the action: we define
\begin{align}
\Delta \mathcal{J}_{a a'}(t, t') \equiv - \frac{1}{8} \text{Im}\left[ \mathcal{A}^{>}_{a a'}(t, t') - \mathcal{A}^{<}_{a a'}(t, t')\right], 
\end{align}
and 
\begin{align}
\Delta \mathcal{Y}_{a a'}(t, t') \equiv - \frac{1}{8} \text{Re}\left[ \mathcal{A}^{>}_{a a'}(t, t') - \mathcal{A}^{<}_{a a'}(t, t')\right].
\end{align}
These two parameters have \emph{anomalous} symmetry properties: $\Delta \mathcal{J}_{a a'}(t, t') = - \Delta \mathcal{J}_{a' a}(t', t)$ and $\Delta \mathcal{Y}_{a a'}(t, t') = \Delta \mathcal{Y}_{a' a}(t', t)$. Hence, they can be called, respectively, \emph{anomalous exchange} and \emph{anomalous twist-exchange parameters}. In terms of the quantities introduced above, the exchange and twist-exchange parts of $S^{RR}$ can be written as
\begin{align}
\text{i} S^{RR}_{\text{exchange}}  = & \sum_{a} \sum_{a'} \int_{t_0}^{\infty} \text{d} t  \! \int_{t_0}^{\infty} \text{d} t' \Bigg\{  \boldsymbol{e}_{a Q}(t)  \cdot  \boldsymbol{e}_{a' Q}(t')  \,  \mathcal{J}^{\text{noise}}_{a a'}(t, t') \nonumber \\
& - 2 \text{i} \, \boldsymbol{e}_{a C}(t) \cdot \boldsymbol{e}_{a' Q}(t')  \,  \Big[ \mathcal{J}_{a a'}(t, t')  + \Delta  \mathcal{J}_{a a'}(t, t')   \Big]   \Bigg\} 
\label{exchange RR action}
\end{align}
and
\begin{align}
\text{i} S^{RR}_{\text{twist}}  = &  \sum_{a} \sum_{a'} \int_{t_0}^{\infty} \text{d} t  \int_{t_0}^{\infty} \text{d} t' \Bigg\{ \left[ \boldsymbol{e}_{a Q}(t)    \times  \boldsymbol{e}_{a' Q}(t')  \right] \cdot \boldsymbol{u}_z \mathcal{Y}^{\text{noise}}_{a a'}(t, t')  \nonumber \\
& - 2 \text{i} \, \left[ \boldsymbol{e}_{a C}(t) \times \boldsymbol{e}_{a' Q}(t')   \right]  \cdot  \boldsymbol{u}_z \Big[ \mathcal{Y}_{a a'}(t, t') + \Delta \mathcal{Y}_{a a'}(t, t') \Big]  \Bigg\} .
\label{DM RR action}
\end{align}

It is apparent that the computation of the parameters entering Eqs.\eqref{exchange RR action} and \eqref{DM RR action}, as well as of the remaining ones from Eqs.\eqref{symmetrized exchange action} and \eqref{symmetrized DM action}, would be cumbersome in the most general case. In order to obtain a tractable problem, as well as to gain insight into the physical processes contributing to non-equilibrium magnetic interactions, we shall discuss two approximations. The first one is the \emph{time-dependent Hartree-Fock} (HF) approximation, which consists in neglecting memory effects in the electronic self-energies. The second one is the assumption of \emph{slow spin dynamics} (SSD), i.e., that the dynamics of the spin degrees of freedom be much slower than the dynamics of the electronic degrees of freedom. The two approximations are independent because they involve different quantities (the spin vectors in the former case, the self-energies in the latter case), therefore it is possible to apply either just one of them, or both. In the following, we discuss them separately.

\subsection{Time-dependent Hartree-Fock approximation}

We now consider, for the general non-equilibrium case, the time-dependent Hartree-Fock (HF) approximation \cite{Leeuwen06}. This amounts to neglecting all self-energies contributions having time arguments corresponding to different points on the Kadanoff-Baym contour. In other words, from Eqs.\eqref{KB equations} we neglect $\Sigma^>$, $\Sigma^<$, $\Sigma^{\urcorner}$ and $\Sigma^{\ulcorner}$, and we keep only the equal-time singular parts of the self-energies, i.e., from Eqs.\eqref{ret adv G} we obtain
\begin{align}
\Sigma^R(t, t') \stackrel{\text{HF}}{=} \Sigma^A(t, t') \stackrel{\text{HF}}{=} \delta(t - t') \, \overline{\Sigma}(t),
\label{noneq HF}
\end{align}
and, analogously, the Matsubara self-energy becomes
\begin{align}
\Sigma^{\mathbb{M}}(\tau) \stackrel{\text{HF}}{=} \beta \delta(\tau) \, \overline{\Sigma}^{\mathbb{M}},
\label{Matsubara HF}
\end{align}
where we have introduced the factor $\beta$ just for convenience. This approximation allows to compute trivially all the integral convolutions between Green functions and self-energies. Physically, it corresponds to neglecting memory effects in the self-energies. The technicalities implied by this approximation are discussed in \ref{HF LNR details}, while here we will present the results for the main quantities of interest. In the case in which the self-energy can be assumed to be diagonal in Hubbard indexes, the action can be further simplified as shown in \ref{HFD approximation}.

The expression for the exchange and twist-exchange parameters can be simplified by means of the representation of Green functions in terms of time-dependent occupation numbers and of the spectral function, which is defined as
\begin{align}
A(t, t') \equiv \text{i} \left[ G^R(t, t') - G^A(t, t') \right] = \text{i} \left[ G^>(t, t') - G^<(t, t') \right]. 
\label{spectral function}
\end{align}
In equilibrium conditions the Hamiltonian is time-independent, therefore the Green functions $G^{\lessgtr}(t, t')$ depend only on $(t - t')$, hence it is possible to take their Fourier transform with respect to the relative time, 
\begin{align}
{G}_{b a}^{\lessgtr \text{ EQ}}(\omega) \equiv \int_{-\infty}^{+\infty} \text{d} t \, \text{e}^{\text{i} \omega t} {G}_{b a}^{\lessgtr \text{ EQ}}(t). 
\end{align}
It can be shown (see the proof in \ref{equilibrium occupation numbers}) that the following relation holds:
\begin{align}
{G}_{b a}^{< \text{ EQ}}(\omega) = \text{i} f(\omega) \, {A}_{b a}^{\text{ EQ}}(\omega),
\label{equilibrium occ number relation}
\end{align}
where
\begin{align}
f(\omega) \equiv \frac{1}{1 + \text{e}^{\beta (\hbar \omega - \mu)}}
\label{Fermi}
\end{align}
is the Fermi distribution, and ${A}_{b a}^{\text{ EQ}}(\omega)$ is the Fourier transform of the equilibrium spectral function.

In non-equilibrium the Green functions $G^{\lessgtr}_{b a}(t, t')$, as well as the spectral function, can be written as functions of $t_r = t - t'$ and $T = (t + t')/2$. Taking the Fourier transform with respect to $t_r$ leads to the following definitions:
\begin{align}
{G}_{b a}^{<}(\omega, T) \equiv \text{i} f(\omega, T) \, {A}_{b a}(\omega, T) ,
\label{nonequilibrium occ number relation}
\end{align}
where
\begin{align}
{G}_{b a}^{<}(\omega, T) = \int_{- \infty}^{+ \infty} \text{d} (t - t') \, \text{e}^{\text{i} \omega (t - t')} G^<_{b a}(t, t'),
\end{align}
and $f(\omega, T)$ and ${A}_{b a}(\omega, T)$ are, respectively, $T$-dependent occupation number and spectral function. Equation \eqref{nonequilibrium occ number relation} is effectively the definition of $f(\omega, T)$. The definition of the spectral function, Eq.\eqref{spectral function}, then implies ${G}_{b a}^{>}(\omega, T) = \text{i} \left[ f(\omega, T) - 1 \right]  {A}_{b a}(\omega, T)$. 

The transformation from the Green functions $G^{\lessgtr}$ to the spectral function and the time-dependent occupation number is particularly useful in the HF regime. Using Eqs.\eqref{cal A HF} and \eqref{cal F HF}, we obtain the following expressions:
\begin{align}
\mathcal{A}^{<}_{a a'}(t, t') & \mp \mathcal{A}^{>}_{a a'}(t, t')  \stackrel{\text{HF}}{=} - \int_{-\infty}^{+\infty} \frac{\text{d} \omega}{2 \pi} \int_{-\infty}^{+\infty} \frac{\text{d} \omega'}{2 \pi}  \text{e}^{- \text{i} (\omega - \omega') t_r} f_{\mp}(\omega, T) \nonumber \\
& \quad \times \Bigg\{\left[ \overline{\Sigma}^{\downarrow}\left(T + \frac{t_r}{2}\right) \cdot {A}^{\downarrow}(\omega, T) \right]_{a a'} \, \left[ \overline{\Sigma}^{\uparrow}\left( T - \frac{t_r}{2} \right) \cdot {A}^{\uparrow}(\omega', T)\right]_{a' a}  \nonumber \\
& \quad \quad -  {A}^{\downarrow}_{a a'}(\omega, T) \, \left[ \overline{\Sigma}^{\uparrow}\left( T - \frac{t_r}{2} \right) \cdot {A}^{\uparrow}(\omega', T) \cdot \overline{\Sigma}^{\uparrow}\left( T + \frac{t_r}{2} \right)  \right]_{a' a} \nonumber \\
& \quad \quad + \left[ {A}^{\downarrow}(\omega, T) \cdot \overline{\Sigma}^{\downarrow}\left( T - \frac{t_r}{2} \right) \right]_{a a'} \, \left[ {A}^{\uparrow}(\omega', T) \cdot \overline{\Sigma}^{\uparrow}\left( T + \frac{t_r}{2} \right)\right]_{a' a} \nonumber\\ 
& \quad \quad - \left[ \overline{\Sigma}^{\downarrow}\left( T + \frac{t_r}{2} \right) \cdot {A}^{\downarrow}(\omega, T) \cdot \overline{\Sigma}^{\downarrow}\left( T - \frac{t_r}{2} \right) \right]_{a a'}  \, {A}^{\uparrow}_{a' a}(\omega', T) \Bigg\} ,
\label{occ simpl}
\end{align}
where $t_r = t - t'$ and $T = (t + t')/2$, and
\begin{align}
& f_-(\omega, T) = f(\omega, T) - f(\omega', T), \nonumber \\
& f_+(\omega, T) = 2  f(\omega, T) \, f(\omega', T) - f(\omega, T) - f(\omega', T);
\end{align}
\begin{align}
\mathcal{F}_{a a'}(t) \stackrel{\text{HF}}{=} - \int_{-\infty}^{+\infty} \frac{\text{d} \omega}{2 \pi} f(\omega, t) \Big[ \overline{\Sigma}_{a a'}^{\downarrow}(t) \, {A}_{a' a}^{\uparrow}(\omega, t) + {A}_{a a'}^{\downarrow}(\omega, t) \, \overline{\Sigma}_{a' a}^{\uparrow}(t) \Big]
\label{F occ}
\end{align}
where we have removed the term proportional to $\delta_{a a'}$ coming from $\mathcal{F}_{a a'}$ [see Eq.\eqref{cal F HF}], because it gives a null contribution to the action. In fact, the quantity $\mathcal{F}_{a a'}(t)$ enters the action in the product with $\delta(t - t')$ and either $\Big[ \boldsymbol{e}_{a +}(t) \cdot \boldsymbol{e}_{a' +}(t')  - \boldsymbol{e}_{a -}(t) \cdot \boldsymbol{e}_{a' -}(t') \Big]$ or $\Big[ \boldsymbol{e}_{a +}(t) \times \boldsymbol{e}_{a' +}(t')  - \boldsymbol{e}_{a -}(t) \times \boldsymbol{e}_{a' -}(t') \Big]$. Both combinations of spin vectors vanish identically for $a = a'$ and $t = t'$, therefore the components of $\mathcal{F}_{a a'}$ proportional to $\delta_{a a'}$ can be suppressed.

The quantities of interest can be written down using Eqs.\eqref{occ simpl} and \eqref{F occ}. For example the exchange parameters, from Eq.\eqref{noneq exchange}, become
\begin{align}
& \mathcal{J}_{a a'}(t, t') \stackrel{\text{HF}}{=}    \frac{1}{4} \delta(t_r) \,  \text{Re}  \Bigg\{ \int_{-\infty}^{+\infty} \frac{\text{d} \omega}{2 \pi} f(\omega, t) \Big[ \overline{\Sigma}^{\downarrow}_{a a'}(t) \, {A}^{\uparrow}_{a' a}(\omega, t) \nonumber \\
& \quad \quad \quad + {A}^{\downarrow}_{a a'}(\omega, t) \, \overline{\Sigma}^{\uparrow}_{a' a}(t)  \Big] \Bigg\} \nonumber \\
& + \frac{\text{sign}(t_r)}{8}  \text{Im} \Bigg\{ \int_{-\infty}^{+\infty} \frac{\text{d} \omega}{2 \pi} \int_{-\infty}^{+\infty} \frac{\text{d} \omega'}{2 \pi}  \text{e}^{- \text{i} (\omega - \omega') t_r} \Big[ f(\omega, T) - f(\omega', T) \Big] \nonumber \\
& \quad \times \Bigg\{\left[ \overline{\Sigma}^{\downarrow}\left( T + \frac{t_r}{2} \right) \cdot {A}^{\downarrow}(\omega, T) \right]_{a a'} \, \left[ \overline{\Sigma}^{\uparrow}\left( T - \frac{t_r}{2} \right) \cdot {A}^{\uparrow}(\omega', T)\right]_{a' a}  \nonumber \\
& \quad \quad - {A}^{\downarrow}_{a a'}(\omega, T) \, \left[ \overline{\Sigma}^{\uparrow}\left( T - \frac{t_r}{2} \right) \cdot {A}^{\uparrow}(\omega', T) \cdot \overline{\Sigma}^{\uparrow}\left( T + \frac{t_r}{2} \right)  \right]_{a' a} \nonumber \\
& \quad \quad + \left[ {A}^{\downarrow}(\omega, T) \cdot \overline{\Sigma}^{\downarrow}\left( T - \frac{t_r}{2} \right) \right]_{a a'} \, \left[ {A}^{\uparrow}(\omega', T) \cdot \overline{\Sigma}^{\uparrow}\left( T + \frac{t_r}{2} \right) \right]_{a' a} \nonumber\\ 
& \quad \quad - \left[ \overline{\Sigma}^{\downarrow}\left( T + \frac{t_r}{2} \right) \cdot {A}^{\downarrow}(\omega, T) \cdot \overline{\Sigma}^{\downarrow}\left( T - \frac{t_r}{2} \right) \right]_{a a'}  \, {A}^{\uparrow}_{a' a}(\omega', T) \Bigg\} \Bigg\}  .
\end{align}

\subsection{Slow spin dynamics}

To implement the condition of slow spin dynamics (SSD), we look back to the term $\text{i} S^{RR}$ of the action, which is given by Eqs.\eqref{exchange RR action} and \eqref{DM RR action}. While the spin vector $\boldsymbol{e}_{a Q}$ is associated with the fast quantum fluctuations of the local magnetization, we can recognize the \emph{classical} magnetization in the spin vector $\boldsymbol{e}_{a C}$, see e.g. Ref.\cite{Bhattacharjee12}. It is then possible to assume that the $\boldsymbol{e}_C$ vectors be slowly-varying functions of time with respect to the rapid electronic processes, which are embedded in the exchange and twist-exchange parameters. Therefore, one can expand
\begin{align}
g(t, t') \, \boldsymbol{e}_{a C}(t) \approx  g(t, t') \left[ \boldsymbol{e}_{a C}(t') + (t - t') \dot{\boldsymbol{e}}_{a C}(t') + \frac{(t - t)^2}{2} \ddot{\boldsymbol{e}}_{a C}(t') + \ldots \right],
\label{SSD approximation}
\end{align}
where $g(t, t')$ is an electronic-related function (such as the exchange parameters). No approximation can be done, instead, on the quantum vectors $\boldsymbol{e}_{Q}$, since their dependence on time is fluctuatory and not under control. When only the term of zero order in $(t - t')$ is kept in the expansion \eqref{SSD approximation}, one obtains
\begin{align}
& - 2 \text{i} \, \sum_{a} \sum_{a'} \int_{t_0}^{\infty} \text{d} t   \int_{t_0}^{\infty} \text{d} t'    \boldsymbol{e}_{a C}(t) \cdot \boldsymbol{e}_{a' Q}(t')  \,  \Big[  \mathcal{J}_{a a'}(t, t')  + \Delta  \mathcal{J}_{a a'}(t, t')   \Big]  \nonumber \\
& \approx - 2 \text{i} \, \sum_{a} \sum_{a'} \int_{t_0}^{\infty} \text{d} t'    \boldsymbol{e}_{a C}(t') \cdot \boldsymbol{e}_{a' Q}(t')  \,  \Big[  \mathcal{J}_{a a'}(t')  + \Delta  \mathcal{J}_{a a'}(t')   \Big]  ,
\label{exchange SSD 0}
\end{align}
and
\begin{align}
&  - 2 \text{i} \,  \sum_{a} \sum_{a'} \int_{t_0}^{\infty} \text{d} t  \int_{t_0}^{\infty} \text{d} t'  \left[ \boldsymbol{e}_{a C}(t) \times \boldsymbol{e}_{a' Q}(t')   \right]  \cdot \boldsymbol{u}_z \Big[ \mathcal{Y}_{a a'}(t, t') + \Delta \mathcal{Y}_{a a'}(t, t') \Big]  \nonumber \\
& \approx - 2 \text{i} \,  \sum_{a} \sum_{a'} \int_{t_0}^{\infty} \text{d} t'  \left[ \boldsymbol{e}_{a C}(t') \times \boldsymbol{e}_{a' Q}(t')   \right]  \cdot \boldsymbol{u}_z \Big[ \mathcal{Y}_{a a'}(t') + \Delta \mathcal{Y}_{a a'}(t') \Big] ,
\label{DM SSD 0}
\end{align}
where 
\begin{align}
& \mathcal{J}_{a a'}(t') \equiv \int_{t_0}^{\infty} \text{d} t \, \mathcal{J}_{a a'}(t, t') , \quad \quad \Delta \mathcal{J}_{a a'}(t') \equiv \int_{t_0}^{\infty} \text{d} t \, \Delta\mathcal{J}_{a a'}(t, t'), \nonumber \\
& \mathcal{Y}_{a a'}(t') \equiv \int_{t_0}^{\infty} \text{d} t \, \mathcal{Y}_{a a'}(t, t') , \quad \quad \Delta \mathcal{Y}_{a a'}(t') \equiv \int_{t_0}^{\infty} \text{d} t \, \Delta \mathcal{Y}_{a a'}(t, t')
\label{SSD parameters}
\end{align}
are the parameters that account for spin precession due to magnetic interactions within the Hubbard system. The terms in the expansion \eqref{exchange SSD 0} depending linearly and quadratically on $(t - t')$ generate, respectively, spin damping and spin nutation \cite{Bhattacharjee12}.

The SSD approximation may be useful for computational purposes because, if the expansion \eqref{SSD approximation} is truncated at any order in $(t - t')$, it gives an effective action where the classical spin vectors $\boldsymbol{e}_{a C}$ (and their derivatives) are evaluated at the same time $t'$ as the quantum spin vectors $\boldsymbol{e}_{a' Q}$. The price to pay is that the parameters of the magnetic interactions, depending on two times $(t, t')$, must be integrated over $\text{d} t$ on the whole real axis. A technical simplification can be achieved by using the sum rule Eq.\eqref{t dependent sum rule}, combined with the symmetry properties \eqref{symmetry A} and \eqref{symmetry cal F}: for example, the exchange parameters can be written as:
\begin{align}
\mathcal{J}_{a a'}(t') & = - \frac{1}{4}  \left\{  \text{Re} \! \left[ \mathcal{F}_{a a'}(t') \right] - \frac{1}{2} \int_{t_0}^{\infty} \! \text{d} t  \, \text{sign}(t - t')  \, \text{Im}\!\left[\mathcal{A}^{>}_{a a'}(t, t') - \!  \mathcal{A}^{<}_{a a'}(t, t') \right] \right\} \nonumber \\
& =  - \frac{1}{8}  \left\{  \text{Re} \! \left[ \mathcal{F}_{a a'}(t') \right] + \frac{1}{2} \int_{0}^{\beta} \text{d} \tau \Big[ \mathcal{G}^{\downarrow \uparrow}_{a' a}(t', \tau) + \mathcal{G}^{\uparrow \downarrow}_{a' a}(t', \tau)  \Big] \right. \nonumber \\
& \left. \quad - \int_{t'}^{\infty} \! \text{d} t  \,  \text{Im}\!\left[\mathcal{A}^{>}_{a a'}(t, t') - \!  \mathcal{A}^{<}_{a a'}(t, t') \right] \right\} .
\end{align}
For large values of $t'$, this representation avoids the need of computing an integral over $\text{d}t$ from $t_0$ to $t'$, replacing it with an integral over the imaginary branch. 

It is then clear that the further level of simplification consists in just applying both HF and SSD. This can be done by computing Eqs.\eqref{SSD parameters} with the approximate expressions for $\mathcal{A}$, $\mathcal{F}$ and $\mathcal{G}$ given in \ref{HF LNR details} [Eqs.\eqref{cal A HF}, \eqref{cal F HF}, \eqref{cal G HF}].

\section{Dynamical spin stiffness}

The spectrum of spin waves \cite{SpinWavesBook} in ferromagnets, in the non-relativistic case and in the absence of external fields, can be computed directly from the spatial Fourier transform $\mathcal{J}_{\boldsymbol{q}}$ of $\mathcal{J}_{a a'}$, as
\begin{align}
\omega_{\boldsymbol{q}} = \frac{4}{M} \Big( \mathcal{J}_{\boldsymbol{0}} - \mathcal{J}_{\boldsymbol{q}}\Big) ,
\label{SW dispersion}
\end{align}
where $M$ is the magnetic moment per ion. In the long-wavelength limit, the dispersion $\omega_{\boldsymbol{q}}$ is quadratic, 
\begin{align}
\omega_{\boldsymbol{q}} \stackrel{\boldsymbol{q} \rightarrow \boldsymbol{0}}{\equiv} \sum_{\alpha} \sum_{\beta} D_{\alpha \beta} q_{\alpha} q_{\beta},
\end{align}
where $\alpha, \beta \in \lbrace x, y, z \rbrace$ and the tensor $D_{\alpha \beta}$ is known as the spin stiffness. This tensor is a relevant quantity for comparison with experiments, since it can be determined from measurements of the spin-wave spectrum. Furthermore, for the equilibrium case it has been proved \cite{Lichtenstein01} that, if the self-energy is local, then the expression for the spin stiffness computed by applying Eq.\eqref{J_aa' diag without delta} and Eq.\eqref{SW dispersion} is exact, despite the approximation of neglecting the vertex in the Dyson equation for the two-particle Green functions, Eq.\eqref{novertex}.

However, our analysis shows that when an external field is applied, the situation is more complicated. First, there is a non-zero time-dependent twist-exchange interaction, which may in principle alter the spin-wave spectrum. Moreover, even in regimes where this effect is negligible, the presence of a time-dependent external field leads to a modification of the concept of spin-wave itself. Only in the SSD case, corresponding to adiabatic spin dynamics, we can think of generalizing Eq.\eqref{SW dispersion} as
\begin{align}
\omega_{\boldsymbol{q}}(t) \stackrel{\text{SSD}}{\approx} \frac{4}{M} \Big[ \mathcal{J}_{\boldsymbol{0}}(t) - \mathcal{J}_{\boldsymbol{q}}(t) \Big] ,
\label{SW dispersion t}
\end{align}
where $\mathcal{J}_{\boldsymbol{q}}(t)$ is the spatial Fourier transform of $\mathcal{J}_{a a'}(t)$ given by Eqs.\eqref{SSD parameters}. Equation \eqref{SW dispersion t} assumes that the magnetic moment per ion $M$ remains constant and that $\mathcal{J}_{a a'}(t)$ depends spatially only on the lattice vector $\boldsymbol{R}_{a a'} = \boldsymbol{R}_{a} - \boldsymbol{R}_{a'}$, as in equilibrium. This is justified if the external time-dependent field can be assumed to be almost uniform over a portion of the sample large with respect to the interatomic distances, a condition which is usually met in experiments on laser-induced spin dynamics, where the typical diameter of uniformity of a laser beam is of the order of some tens of $\mu$m.

We shall therefore consider for simplicity the case of a single-band Hubbard model, and assume that the non-equilibrium twist-exchange interactions be negligible. Under these conditions, we apply the HF approximation, obtaining effectively the HFD approximation discussed in \ref{HFD approximation}. To determine the spin stiffness, we put
\begin{align}
G^{\eta \sigma}_{a a'}(t, t') = \frac{1}{n} \sum_{\boldsymbol{k}} \text{e}^{\text{i} \boldsymbol{k} \cdot \boldsymbol{R}_{a a'}} G_{\boldsymbol{k}}^{\eta \sigma}(t, t'),
\end{align}
where $n$ is the number of $\boldsymbol{k}$ vectors in the first Brillouin zone (1BZ). We use Eq.\eqref{J HFD} for the exchange parameters, imposing also the SSD approximation, and we directly get
\begin{align}
\mathcal{J}_{a a'}(t') = \frac{1}{n}  \sum_{\boldsymbol{q}} \text{e}^{\text{i} \boldsymbol{q} \cdot \boldsymbol{R}_{a a'}} \mathcal{J}_{\boldsymbol{q}}(t'),
\end{align}
where
\begin{align}
\mathcal{J}_{\boldsymbol{q}}(t') \equiv \frac{1}{4 \text{i}} \sum_{\eta} \eta \sum_{\sigma} \int_{t_0}^{\infty} \text{d} t \, \text{sign}(t - t')  \,  \overline{\Sigma}^{\text{S}}(t') \, \overline{\Sigma}^{\text{S}}(t) \, \frac{1}{n} \sum_{\boldsymbol{k}} G^{\eta \sigma}_{\boldsymbol{k}}(t, t') \, G^{\bar{\eta} \bar{\sigma}}_{\boldsymbol{k} - \boldsymbol{q}}(t', t) .
\label{J_q}
\end{align}
In the small $|\boldsymbol{q}|$ limit, we can expand
\begin{align}
G^{\bar{\eta} \bar{\sigma}}_{\boldsymbol{k} - \boldsymbol{q}}(t', t) \approx G^{\bar{\eta} \bar{\sigma}}_{\boldsymbol{k}}(t', t) - \sum_{\alpha} q_{\alpha} \frac{\partial G^{\bar{\eta} \bar{\sigma}}_{\boldsymbol{k}}(t', t)}{\partial k_{\alpha}} + \frac{1}{2} \sum_{\alpha} \sum_{\beta} q_{\alpha} q_{\beta} \frac{\partial^2 G^{\bar{\eta} \bar{\sigma}}_{\boldsymbol{k}}(t', t)}{\partial k_{\alpha} \partial k_{\beta}},
\end{align}
but the term linear in $q_{\alpha}$ does not contribute to Eq.\eqref{J_q} because $G_{\boldsymbol{k}} = G_{- \boldsymbol{k}}$. Computing Eq.\eqref{SW dispersion t} with the help of Eq.\eqref{J_q}, we find indeed that the dispersion is quadratic, with a spin stiffness tensor given by
\begin{align}
D_{\alpha \beta}(t) \equiv & - \frac{\text{i}}{2 M} \sum_{\eta} \eta \sum_{\sigma} \int_{t_0}^{\infty} \text{d} t' \, \text{sign}(t' - t)  \,  \overline{\Sigma}^{\text{S}}(t) \, \overline{\Sigma}^{\text{S}}(t') \nonumber \\
& \times \frac{1}{n} \sum_{\boldsymbol{k}} \frac{\partial G^{\eta \sigma}_{\boldsymbol{k}}(t', t)}{\partial k_{\alpha}} \, \frac{ \partial G^{\bar{\eta} \bar{\sigma}}_{\boldsymbol{k}}(t, t')}{ \partial k_{\beta}} .
\label{spin stiffness result}
\end{align}
Details on this derivation are given in \ref{sum 1BZ}.

\section{Conclusions}

In this Article we have provided for the first time a method to compute the parameters characterizing the evolution of the low-energy magnetic interactions in a general realistic system (a multi-band Hubbard model) under a generic time-dependent electrostatic perturbation. While confining the analysis to the low-energy sector (small spin deviations from the equilibrium configuration) is crucial to obtain a tractable problem, there is no restriction on the time-scale of the perturbing field, which allows us to describe ultrafast excitations. Importantly, our method requires only the computation of Green functions which are diagonal in spin, which is expected to be a major simplification in view of numerical implementation.

Our findings show that the evolution of the system is a highly non-trivial process, involving complicated interactions between the macroscopic magnetization and quantum fluctuations. In particular we find the emergence of a new magnetic interaction, the \emph{twist exchange}, that is formally similar to a Dzyaloshinskii-Moriya interaction, despite the absence of spin-orbit in the system that we have treated. This term can be interpreted, generalizing Bogoliubov's work \cite{BogoliubovBook}, as a three-spin interaction. Moreover, we have derived expressions for noise contributions arising from the interaction between quantum fluctuations. Finally, we have discussed relevant approximations (time-dependent Hartree-Fock and slow spin dynamics), and we have showed how to compute the time-dependent spin stiffness tensor, which may be relevant for comparison with experiments.

Future developments may involve the incorporation of spin-orbit interaction and external magnetic fields, and the derivation of effective equations of motion for the magnetization. The numerical implementation of the formalism described here, which is best done in the framework of time-dependent dynamical mean-field theory \cite{Freericks06, Tsuij11, Eckstein11}, combined with employing appropriate hopping and interaction parameters, should allow to simulate realistic systems under ultrafast (femtosecond) perturbations.

\section*{Acknowledgements}

We acknowledge inspiring discussions with Johan Mentink, Martin Eckstein, Laszlo Szunyogh, Oksana Chubykalo-Fesenko, Ulrich Nowak, Peter Oppeneer, Theo Rasing, and Roy Chantrell. This work is supported by the European Union ERC Grant Agreement No. 281043 (FEMTOSPIN) and SFB925 (Germany).

\appendix

\section{Kadanoff-Baym functions with particular arguments}\label{correspondences}

Some relations hold between different Kadanoff-Baym Green functions, for certain values of their arguments. In this Appendix we list such relations. We first observe that
\begin{align}
& G^{\ulcorner}_{a a'}(0, t) = - \text{i} \left< \hat{\psi}_a \hat{\psi}^{\dagger}_{a'}(t) \right> = G^{>}_{a a'}(t_0, t), \nonumber \\
& G^{\urcorner}_{a' a}(t, 0) =  \text{i} \left< \hat{\psi}^{\dagger}_a \hat{\psi}_{a'}(t) \right> = G^{<}_{a' a}(t, t_0),
\label{0t t0}
\end{align}
\begin{align}
& G^{\ulcorner}_{a a'}(\beta, t) = - \text{i} \left< \hat{\psi}_a(\beta) \, \hat{\psi}^{\dagger}_{a'}(t) \right> = - \text{i} \left< \hat{\psi}^{\dagger}_{a'}(t) \, \hat{\psi}_a \right> = - G_{a a'}^{<}(t_0, t), \nonumber \\
& G^{\urcorner}_{a' a}(t, \beta) =   \text{i} \left< \hat{\psi}^{\dagger}_a(\beta) \, \hat{\psi}_{a'}(t) \right> =   \text{i} \left< \hat{\psi}_{a'}(t) \, \hat{\psi}^{\dagger}_a \right> = - G_{a' a}^{>}(t, t_0),
\label{betat tbeta}
\end{align}
and
\begin{align}
& G^{\ulcorner}_{b a}(\tau, t_0) = - \text{i} \left< \hat{\psi}_b(\tau) \hat{\psi}^{\dagger}_a \right> = - \text{i} \left< \mathcal{T}_{\tau} \left[ \hat{\psi}_b(\tau) \hat{\psi}^{\dagger}_a (0)     \right] \right>   = G^{\mathbb{M}}_{b a}(\tau), \nonumber \\
& G^{\urcorner}_{b a}(t_0, \tau) = \text{i} \left<  \hat{\psi}^{\dagger}_a(\tau) \hat{\psi}_b \right> = - \text{i} \left< \mathcal{T}_{\tau} \left[ \hat{\psi}_b(0) \hat{\psi}^{\dagger}_a (\tau)     \right] \right>   = G^{\mathbb{M}}_{b a}(- \tau).
\label{taut0 t0tau}
\end{align}
As particular cases,
\begin{align}
& G^{\ulcorner}_{b a}(0, t_0) = - \text{i} \left<  \hat{\psi}_{b} \hat{\psi}^{\dagger}_{a}   \right> = G^{>}_{b a}(t_0, t_0) = \text{i} \rho_{b a} -  \text{i} \delta_{b a} , \nonumber \\
& G^{\urcorner}_{b a}(t_0, 0) = \text{i} \left<  \hat{\psi}^{\dagger}_{a}  \hat{ \psi}_{b}  \right> = G^{<}_{b a}(t_0, t_0) = \text{i} \rho_{b a}, \nonumber \\
& G^{\ulcorner}_{b a}(\beta, t_0) = - \text{i} \left<  \hat{\psi}_{b}(\beta)  \, \hat{\psi}^{\dagger}_{a}   \right> = - \text{i} \left<  \hat{\psi}^{\dagger}_{a}  \hat{ \psi}_{b}  \right>  = -  G^{<}_{b a}(t_0, t_0) = - \text{i} \rho_{b a}, \nonumber \\
& G^{\urcorner}_{b a}(t_0, \beta) = \text{i} \left<  \hat{\psi}^{\dagger}_{a}(\beta) \,  \hat{ \psi}_{b}  \right> = \text{i} \left<  \hat{\psi}_{b} \hat{\psi}^{\dagger}_{a}   \right>  = -  G^{>}_{b a}(t_0, t_0) = - \text{i} \rho_{b a} +  \text{i} \delta_{b a} .
\label{first correspondence}
\end{align}
As a consequence of Eq.\eqref{0t t0},
\begin{align}
& Y^{\urcorner}(t, 0) = \Big[ \Sigma^{R} \cdot G^{\urcorner}  + \Sigma^{\urcorner} \star G^{\mathbb{M}}  \Big](t, 0) = \Big[ \Sigma^{R} \cdot G^{<}  \Big](t, t_0) + \Big[ \Sigma^{\urcorner} \star G^{\mathbb{M}}  \Big](t, 0), \nonumber \\
& Y^{\ulcorner}(0, t) = \Big[ G^{\ulcorner} \cdot \Sigma^{A}  + G^{\mathbb{M}} \star \Sigma^{\ulcorner}  \Big](0, t) = \Big[ G^{>} \cdot \Sigma^{A}  \Big](t_0, t) + \Big[ G^{\mathbb{M}} \star \Sigma^{\ulcorner}   \Big](0, t).
\end{align}
As a consequence of Eq.\eqref{betat tbeta} and Eq.\eqref{Matsubara boundary},
\begin{align}
Y^{\urcorner}(t, \beta) & = \Big[ \Sigma^R \cdot G^{\urcorner} + \Sigma^{\urcorner} \star G^{\mathbb{M}}\Big](t, \beta) = - \Big[ \Sigma^R \cdot G^{>}\Big](t, t_0) - \Big[ \Sigma^{\urcorner} \star G^{\mathbb{M}}\Big](t, 0) \nonumber \\
&  = - \Big[ \Sigma^R \cdot G^{>}\Big](t, t_0) - \Big[ \Sigma^{\urcorner} \star G^{\ulcorner}\Big](t, t_0) = - W^{>}(t, t_0), \nonumber \\
Y^{\ulcorner}(\beta, t) & = \Big[ G^{\ulcorner} \cdot \Sigma^A  +  G^{\mathbb{M}} \star \Sigma^{\ulcorner}  \Big](\beta, t) = - \Big[ G^{<} \cdot \Sigma^A \Big](t_0, t) - \Big[ G^{\mathbb{M}} \star \Sigma^{\ulcorner}  \Big](0, t) \nonumber \\
&  = - \Big[ G^{<} \cdot \Sigma^A \Big](t_0, t) - \Big[ G^{\urcorner} \star \Sigma^{\ulcorner}  \Big](t_0, t) = - X^{<}(t_0, t).
\label{cons betat tbeta}
\end{align}
As a consequence of Eq.\eqref{taut0 t0tau},
\begin{align}
W^{<}(t, t_0) & = \Big[ \Sigma^{R} \cdot G^{<}   +  \Sigma^{\urcorner} \star G^{\ulcorner}   +  \Sigma^{<} \cdot G^{A}  \Big](t, t_0) \nonumber \\
& = \Big[ \Sigma^{R} \cdot G^{<}  \Big](t, t_0) + \Big[ \Sigma^{\urcorner} \star G^{\mathbb{M}}  \Big](t, 0) = Y^{\urcorner}(t, 0) , \nonumber \\
X^{>}(t_0, t) & = \Big[ G^{>}  \cdot \Sigma^{A}  +  G^{\urcorner} \star \Sigma^{\ulcorner}   +  G^{R} \cdot \Sigma^{>}  \Big](t_0, t) \nonumber \\
& = \Big[ G^{>}  \cdot \Sigma^{A}  \Big](t_0, t) + \Big[ G^{\mathbb{M}} \star \Sigma^{\ulcorner}  \Big](0, t) = Y^{\ulcorner}(0, t),
\label{cons 0t t0}
\end{align}
where we have used $G^{A}_{b a}(t', t_0) = 0$ as well as $G^{R}_{b a}(t_0, t') = 0$ because $t' > t_0$. Then, it also holds that
\begin{align}
& Z^{\urcorner}(t_0, \tau) = \Big[ G^{R} \cdot \Sigma^{\urcorner}   +  G^{\urcorner} \star \Sigma^{\mathbb{M}}  \Big](t_0, \tau) = \Big[ G^{\mathbb{M}} \star \Sigma^{\mathbb{M}}  \Big](0, \tau) = J^{\mathbb{M}}(-\tau)  , \nonumber \\
& Z^{\ulcorner}(\tau, t_0) = \Big[ \Sigma^{\ulcorner} \cdot G^{A}   +  \Sigma^{\mathbb{M}} \star G^{\ulcorner}   \Big](\tau, t_0) = \Big[ \Sigma^{\mathbb{M}} \star G^{\mathbb{M}}   \Big](\tau, 0) = I^{\mathbb{M}}(\tau) .
\end{align}

\section{Proof of $S_{\theta^2} = 0$}\label{nullStheta}

The action term produced by the transformation Eq.\eqref{xi to vectors} and depending on the squares of the azimuthal angles is equal to:
\begin{align}
& \text{i} S_{\theta^2} = \frac{1}{8} \sum_{a} \int_{t_0}^{\infty} \text{d} t  \Big[ \theta_{a +}^2(t) - \theta_{a -}^2(t) \Big]  \sum_{a'}    \Bigg\{  \text{i} \, \Big[ \mathcal{F}_{a a'}(t)  + \mathcal{F}_{a' a}(t) \Big]  \nonumber \\
& \quad \quad   + \int_{t_0}^{t} \text{d} t' \Big[ \mathcal{A}^{<}_{a a'}(t, t')- \mathcal{A}^{>}_{a a'}(t, t') + \mathcal{A}^{>}_{a' a}(t', t)  -  \mathcal{A}^{<}_{a' a}(t', t) \Big] \nonumber\\
& \quad \quad   - \text{i} \, \int_{0}^{\beta} \text{d} \tau   \Big[ \mathcal{G}_{a a'}^{\downarrow \uparrow}(t, \tau) + \mathcal{G}_{a a'}^{\uparrow \downarrow}(t, \tau) \Big] \Bigg\} \nonumber \\
&  - \frac{1}{8}  \sum_{a} \int_{0}^{\beta} \text{d} \tau \theta_{a v}^2(\tau) \Bigg\{  2 \beta  \mathcal{I}_a  + \sum_{a'} \int_{0}^{\beta} \text{d} \tau'   \Big[ \mathcal{K}_{a a'}(\tau - \tau')      +      \mathcal{K}_{a' a}(\tau' - \tau)   \Big] \Bigg\}  .
\label{Hubbard action vectors theta sq}
\end{align}
We shall now prove that $S_{\theta^2} = 0$.

We start by computing the last line of Eq.\eqref{Hubbard action vectors theta sq}, which contains terms related to the vertical branch of the Kadanoff-Baym contour. We have:
\begin{align}
\int_{0}^{\beta} \text{d} \tau'   \Big[ \mathcal{K}_{a a'}(\tau - \tau')      +  \mathcal{K}_{a' a}(\tau' - \tau)   \Big]  = \beta^3 \! \sum_{n = - \infty}^{+ \infty} \Big[ \mathcal{K}_{a a'}(\omega_n, \omega_n) + \mathcal{K}_{a' a}(\omega_n, \omega_n)\Big] ,
\label{passage 1}
\end{align}
where we have used Eqs.\eqref{def Fourier K} and \eqref{delta Matsubara}. Then, we need to sum the RHS of Eq.\eqref{passage 1} over $a'$: we use \eqref{K frequency space} as well as a sum rule valid for the Matsubara Green functions, which reads 
\begin{align}
\beta^2  G^{\mathbb{M} \sigma}(\omega_n) \cdot  \Sigma^{\mathbb{M} \text{S}}(\omega_n) \cdot G^{\mathbb{M} \bar{\sigma}}(\omega_n)  =  - G^{\mathbb{M} \text{S}}(\omega_n);
\label{sum rule M}
\end{align}
we present the full derivation of this sum rule in \ref{sumRule}. One then obtains:
\begin{align}
\sum_{a'}  \mathcal{K}_{a a'}(\omega_n, \omega_n) & =  - \frac{1}{\beta^2}   \Big[  \Sigma^{\mathbb{M} \downarrow}(\omega_{n}) \cdot G^{\mathbb{M} \downarrow} (\omega_{n})   +   G^{\mathbb{M} \uparrow}(\omega_{n}) \cdot \Sigma^{\mathbb{M} \uparrow}(\omega_{n})    \Big]_{a a},
\label{sum aa'}
\end{align}
and
\begin{align}
\sum_{a'}  \mathcal{K}_{a' a}(\omega_n, \omega_n) & =  - \frac{1}{\beta^2}   \Big[ G^{\mathbb{M} \downarrow}(\omega_{n}) \cdot \Sigma^{\mathbb{M} \downarrow}(\omega_{n})  +  \Sigma^{\mathbb{M} \uparrow}(\omega_{n}) \cdot G^{\mathbb{M} \uparrow} (\omega_{n})  \Big]_{a a} ,
\label{sum a'a}
\end{align}
which imply
\begin{align}
\beta^3 \sum_{n = - \infty}^{+ \infty} \Big[ \mathcal{K}_{a a'}(\omega_n, \omega_n) + \mathcal{K}_{a' a}(\omega_n, \omega_n)\Big] = -2 \beta \mathcal{I}_a,
\label{sum rule cal I}
\end{align}
as can be seen by comparison with Eq.\eqref{cal I}. Therefore, the last line in Eq.\eqref{Hubbard action vectors theta sq} is equal to zero.

To proceed, we need to elaborate the following sum and integral:
\begin{align}
\sum_{a'} \int_{t_0}^t \text{d} t' \Big[ \mathcal{A}^{<}_{a a'}(t, t')- \mathcal{A}^{>}_{a a'}(t, t') + \mathcal{A}^{>}_{a' a}(t', t)  -  \mathcal{A}^{<}_{a' a}(t', t) \Big] .
\label{sum and integral}
\end{align}
To this aim, it is convenient to write the quantities $\mathcal{A}^{\lessgtr}$, from Eq.\eqref{equation 0}, as:
\begin{align}
\mathcal{A}^{\lessgtr}_{a a'}(t, t')  =  \sum_{b'} & \Big[ - W^{\gtrless \downarrow}_{a a'}(t, t') \, G^{\lessgtr \uparrow}_{b' a}(t', t) + G^{\gtrless \downarrow}_{a a'}(t, t') \, X^{\lessgtr \uparrow}_{b' a}(t', t) \Big] \nonumber \\
& \cdot \left[ - \delta_{a' b'} \text{i} \, \frac{\overleftarrow{\partial}}{\partial t'} + T_{a' b'}(t') \, \Big( 1 - \overleftarrow{P}_{a' b'} \Big) \right] , 
\end{align}
and
\begin{align}
\mathcal{A}^{\lessgtr}_{a' a}(t', t)  =  \sum_{b'} &  \left[ - \delta_{a' b'} \text{i} \, \frac{\overrightarrow{\partial}}{\partial t'} +  \Big( 1 - \overrightarrow{P}_{a' b'} \Big) \, T_{a' b'}(t')  \right] \nonumber \\
& \cdot \Big[ - W^{\lessgtr \uparrow}_{a a'}(t, t') \, G^{\gtrless \downarrow}_{b' a}(t', t) + G^{\lessgtr \uparrow}_{a a'}(t, t') \, X^{\gtrless \downarrow}_{b' a}(t', t) \Big] .
\end{align}
In this way, when one computes the sum over $a'$ in Eq.\eqref{sum and integral}, the terms depending on the hopping parameters give zero by symmetry, and the remaining terms are derivatives with respect to $t'$, which can then be integrated trivially. With the help of Eq.\eqref{HF self energy identity} and the definition \eqref{cal F}, we then obtain
\begin{align}
& \sum_{a'} \int_{t_0}^t \text{d} t' \Big[ \mathcal{A}^{<}_{a a'}(t, t')- \mathcal{A}^{>}_{a a'}(t, t') + \mathcal{A}^{>}_{a' a}(t', t)  -  \mathcal{A}^{<}_{a' a}(t', t) \Big] \nonumber \\
&  = - \text{i}  \sum_{a'} \Bigg\{  \mathcal{F}_{a a'}(t) + \mathcal{F}_{a' a}(t)  +  \! \sum_{\sigma} \Big[ W^{> \sigma}_{a a'}(t, t_0)  G^{< \bar{\sigma}}_{a' a}(t_0, t) - G^{> \sigma}_{a a'}(t, t_0)  X^{< \bar{\sigma}}_{a' a}(t_0, t)  \nonumber \\
& \quad \quad -  W^{< \sigma}_{a a'}(t, t_0)  G^{> \bar{\sigma}}_{a' a}(t_0, t) + G^{< \sigma}_{a a'}(t, t_0)  X^{> \bar{\sigma}}_{a' a}(t_0, t)   \Big] \Bigg\}.
\end{align}

We now evaluate the following sum and integral:
\begin{align}
-  \text{i} \,  \sum_{a'} \int_{0}^{\beta} \text{d} \tau   \Big[ \mathcal{G}_{a a'}^{\downarrow \uparrow}(t, \tau) + \mathcal{G}_{a a'}^{\uparrow \downarrow}(t, \tau) \Big] = -  \text{i} \,  \sum_{a'} \sum_{\sigma} \int_{0}^{\beta} \text{d} \tau    \mathcal{G}_{a a'}^{\sigma \bar{\sigma}}(t, \tau) .
\end{align}
To do this, it is convenient to write the quantities $\mathcal{G}_{a a'}^{\sigma \bar{\sigma}}$, from Eq.\eqref{def cal G}, as
\begin{align}
\mathcal{G}_{a a'}^{\sigma \bar{\sigma}}(t, \tau) = - \sum_{b'} & \Big[ - Y^{\urcorner \sigma}_{a a'}(t, \tau) \, G^{\ulcorner \bar{\sigma}}_{b' a}(\tau, t) + G^{\urcorner \sigma}_{a a'}(t, \tau) \, Y^{\ulcorner \bar{\sigma}}_{b' a}(\tau, t) \Big] \nonumber \\
& \cdot \left[ - \delta_{a' b'} \frac{\overleftarrow{\partial}}{\partial \tau} - T_{a' b'} \Big( 1 - \overleftarrow{P}_{a' b'} \Big) \right],
\end{align}
and then proceed analogously to the previous computation of Eq.\eqref{sum and integral}. We obtain
\begin{align}
& -  \text{i} \,  \sum_{a'} \sum_{\sigma} \int_{0}^{\beta} \text{d} \tau    \mathcal{G}_{a a'}^{\sigma \bar{\sigma}}(t, \tau) \nonumber \\
&  \quad = - \text{i} \,  \sum_{a'} \sum_{\sigma}   \Big[ - Y^{\urcorner \sigma}_{a a'}(t, \beta) \, G^{\ulcorner \bar{\sigma}}_{a' a}(\beta, t) + G^{\urcorner \sigma}_{a a'}(t, \beta) \, Y^{\ulcorner \bar{\sigma}}_{a' a}(\beta, t)  \nonumber \\
& \quad \quad \quad \quad \quad \quad \quad + Y^{\urcorner \sigma}_{a a'}(t, 0) \, G^{\ulcorner \bar{\sigma}}_{a' a}(0, t) - G^{\urcorner \sigma}_{a a'}(t, 0) \, Y^{\ulcorner \bar{\sigma}}_{a' a}(0, t) \Big]   \nonumber \\
&  \quad = - \text{i} \,  \sum_{a'} \sum_{\sigma}   \Big[ - W^{> \sigma}_{a a'}(t, t_0) \, G^{< \bar{\sigma}}_{a' a}(t_0, t) + G^{> \sigma}_{a a'}(t, t_0) \, X^{< \bar{\sigma}}_{a' a}(t_0, t)  \nonumber \\
& \quad \quad \quad \quad \quad \quad \quad + W^{< \sigma}_{a a'}(t, t_0) \, G^{> \bar{\sigma}}_{a' a}(t_0, t) - G^{< \sigma}_{a a'}(t, t_0) \, X^{> \bar{\sigma}}_{a' a}(t_0, t) \Big]  ,
\end{align}
where for the last passage we have used Eqs.\eqref{0t t0}, \eqref{cons 0t t0}, \eqref{betat tbeta}, \eqref{cons betat tbeta}. Collecting all the terms, one obtains
\begin{align}
& \text{i} \, \Big[ \mathcal{F}_{a a'}(t)  + \mathcal{F}_{a' a}(t) \Big]  + \int_{t_0}^{t} \text{d} t' \Big[ \mathcal{A}^{<}_{a a'}(t, t')- \mathcal{A}^{>}_{a a'}(t, t') + \mathcal{A}^{>}_{a' a}(t', t)  -  \mathcal{A}^{<}_{a' a}(t', t) \Big]  \nonumber \\
& -  \text{i} \,   \int_{0}^{\beta} \text{d} \tau   \Big[ \mathcal{G}_{a a'}^{\downarrow \uparrow}(t, \tau) + \mathcal{G}_{a a'}^{\uparrow \downarrow}(t, \tau) \Big] = 0,
\label{t dependent sum rule}
\end{align}
which, together with Eq.\eqref{sum rule cal I}, proves that $S_{\theta^2} = 0$. $\blacksquare$

\section{Sum rule for Matsubara Green functions}\label{sumRule}

The proof of $S_{\theta^2} = 0$, presented in \ref{nullStheta}, requires a sum rule that we will now derive. We start from writing the KB equations of motion for the Matsubara Green functions, among Eqs.\eqref{KB equations}, in the frequency domain. To do this, we multiply both sides of the KB equations by $\text{e}^{- \text{i} \omega_n (\tau - \tau')}$ and we apply $\int_{- \tau_1}^{\beta - \tau_1} \text{d} (\tau - \tau')$, where $0 < \tau_1 < \beta$. After performing the integration with the help of Eqs. \eqref{Matsubara boundary}, \eqref{Fourier Matsubara}, \eqref{I Fourier} and \eqref{J Fourier}, we obtain
\begin{align}
&  g^{-1}(\omega_n) \cdot G^{\mathbb{M} \sigma}(\omega_n) = \frac{1}{\beta}  - \beta  \Sigma^{\mathbb{M} \sigma}(\omega_n) \cdot G^{\mathbb{M} \sigma}(\omega_n)  , \nonumber \\
&  G^{\mathbb{M} \sigma}(\omega_n) \cdot g^{-1}(\omega_n) = \frac{1}{\beta}  - \beta  G^{\mathbb{M} \sigma}(\omega_n) \cdot \Sigma^{\mathbb{M} \sigma}(\omega_n)  ,
\label{Matsubara omega space}
\end{align}
where we have put
\begin{align}
\text{i} g^{-1}_{a b}(\omega_n) \equiv - \text{i} \omega_n \delta_{a b} - ( T_{a b} - \mu \delta_{a b} ).
\end{align}
We now multiply both sides of the second among Eqs.\eqref{Matsubara omega space} by $G^{\mathbb{M} \bar{\sigma}}(\omega_n)$ from the right, obtaining
\begin{align}
&  G^{\mathbb{M} \sigma}(\omega_n) \cdot g^{-1}(\omega_n) \cdot G^{\mathbb{M} \bar{\sigma}}(\omega_n) = \frac{1}{\beta} G^{\mathbb{M} \bar{\sigma}}(\omega_n)  - \beta   G^{\mathbb{M} \sigma}(\omega_n) \cdot \Sigma^{\mathbb{M} \sigma}(\omega_n) \cdot G^{\mathbb{M} \bar{\sigma}}(\omega_n) ,
\label{Matsubara omega space 2}
\end{align}
then we use the first among Eqs.\eqref{Matsubara omega space} to eliminate $g^{-1}(\omega_n)$ in the LHS of Eq.\eqref{Matsubara omega space 2}, and we obtain the required sum rule:
\begin{align}
&  G^{\mathbb{M} \sigma}(\omega_n) \cdot \!\! \left[ \frac{1}{\beta} - \! \beta  \Sigma^{\mathbb{M} \bar{\sigma}}(\omega_n) \cdot G^{\mathbb{M} \bar{\sigma}}(\omega_n)  \right] = \left[ \frac{1}{\beta}  - \! \beta G^{\mathbb{M} \sigma}(\omega_n) \cdot \! \Sigma^{\mathbb{M} \sigma}(\omega_n) \right] \! \cdot  G^{\mathbb{M} \bar{\sigma}}(\omega_n) \nonumber \\
& \Rightarrow \frac{G^{\mathbb{M} \sigma}(\omega_n) - G^{\mathbb{M} \bar{\sigma}}(\omega_n)}{\beta}   =   - \beta   G^{\mathbb{M} \sigma}(\omega_n) \cdot \Big[ \Sigma^{\mathbb{M} \sigma}(\omega_n) - \Sigma^{\mathbb{M} \bar{\sigma}}(\omega_n) \Big] \cdot G^{\mathbb{M} \bar{\sigma}}(\omega_n)  \nonumber \\
& \Rightarrow \beta^2  G^{\mathbb{M} \sigma}(\omega_n) \cdot  \Sigma^{\mathbb{M} \text{S}}(\omega_n) \cdot G^{\mathbb{M} \bar{\sigma}}(\omega_n)  =  - G^{\mathbb{M} \text{S}}(\omega_n) .
\label{Matsubara omega space 3}
\end{align}

\section{Technical consequences of the Hartree-Fock approximation}\label{HF LNR details}

We here list the identities which follow from adopting the Hartree-Fock (HF) approximation. From Eqs.\eqref{KB equations}, using Eqs.\eqref{noneq HF} and \eqref{Matsubara HF} we obtain (matrix notation):
\begin{align}
& W^{\lessgtr \sigma}(t, t') \stackrel{\text{HF}}{=} \left[ \Sigma^{R \sigma} \cdot G^{\lessgtr \sigma}\right](t,t') \stackrel{\text{HF}}{=} \overline{\Sigma}^{\sigma}(t) \cdot G^{\lessgtr \sigma}(t, t'), \nonumber \\
& X^{\lessgtr \sigma}(t, t') \stackrel{\text{HF}}{=} \left[ G^{\lessgtr \sigma} \cdot \Sigma^{A \sigma} \right](t,t') \stackrel{\text{HF}}{=} G^{\lessgtr \sigma}(t, t') \cdot  \overline{\Sigma}^{\sigma}(t'), \nonumber \\
& Y^{\urcorner \sigma}(t, \tau) \stackrel{\text{HF}}{=} \left[ \Sigma^{R \sigma} \cdot G^{\urcorner \sigma}\right](t,\tau) \stackrel{\text{HF}}{=}  \overline{\Sigma}^{\sigma}(t) \cdot G^{\urcorner \sigma}(t, \tau), \nonumber \\
& Y^{\ulcorner \sigma}(\tau, t) \stackrel{\text{HF}}{=} \left[ G^{\ulcorner \sigma} \cdot \Sigma^{A \sigma} \right](\tau,t) \stackrel{\text{HF}}{=}  G^{\ulcorner \sigma}(\tau, t) \cdot  \overline{\Sigma}^{\sigma}(t), \nonumber \\
& Z^{\urcorner \sigma}(t, \tau) \stackrel{\text{HF}}{=} \left[ G^{\urcorner \sigma} \star \Sigma^{\mathbb{M} \sigma} \right](t, \tau) \stackrel{\text{HF}}{=} - \text{i} \,  G^{\urcorner \sigma}(t, \tau) \cdot \overline{\Sigma}^{\mathbb{M} \sigma} , \nonumber \\
& Z^{\ulcorner \sigma}(\tau, t) \stackrel{\text{HF}}{=} \left[ \Sigma^{\mathbb{M} \sigma} \star G^{\ulcorner \sigma} \right](\tau, t) \stackrel{\text{HF}}{=} - \text{i} \, \overline{\Sigma}^{\mathbb{M} \sigma} \cdot G^{\ulcorner \sigma}(\tau, t)  .
\end{align}
Inserting these approximate identities into Eqs.\eqref{B(2) mu c-matrix}, \eqref{cal F} and \eqref{cal G}, we obtain:
\begin{align}
\mathcal{A}^{\lessgtr}_{a a'}(t, t') \stackrel{\text{HF}}{=}  & \left[ \overline{\Sigma}^{\downarrow}(t) \cdot G^{\gtrless \downarrow}(t, t') \right]_{a a'} \, \left[ \overline{\Sigma}^{\uparrow}(t') \cdot G^{\lessgtr \uparrow}(t', t)\right]_{a' a} \nonumber \\
& -  G^{\gtrless \downarrow}_{a a'}(t, t') \, \left[ \overline{\Sigma}^{\uparrow}(t') \cdot G^{\lessgtr \uparrow}(t', t) \cdot \overline{\Sigma}^{\uparrow}(t)  \right]_{a' a} \nonumber \\
&  + \left[ G^{\gtrless \downarrow}(t, t') \cdot \overline{\Sigma}^{\downarrow}(t') \right]_{a a'} \, \left[ G^{\lessgtr \uparrow}(t', t) \cdot \overline{\Sigma}^{\uparrow}(t)\right]_{a' a} \nonumber\\ 
& -  \left[ \overline{\Sigma}^{\downarrow}(t) \cdot G^{\gtrless \downarrow}(t, t') \cdot \overline{\Sigma}^{\downarrow}(t') \right]_{a a'}  \, G^{\lessgtr \uparrow}_{a' a}(t', t) ,  
\label{cal A HF}
\end{align}
\begin{align}
\mathcal{F}_{a a'}(t) \stackrel{\text{HF}}{=}  &  -\frac{\text{i}}{2} \delta_{a a'} \Big[ \overline{\Sigma}^{\uparrow}(t) \cdot G^{< \uparrow}(t, t)  +  \overline{\Sigma}^{\downarrow}(t) \cdot G^{< \downarrow}(t, t)  +  G^{< \uparrow}(t, t) \cdot \overline{\Sigma}^{\uparrow}(t)  \nonumber \\
& \quad \quad + G^{< \downarrow}(t, t) \cdot \overline{\Sigma}^{\downarrow}(t) \Big]_{a a}  + \text{i} \,  \overline{\Sigma}_{a a'}^{\downarrow}(t) \, G^{< \uparrow}_{a' a}(t, t) + \text{i} \, G^{< \downarrow}_{a a'}(t, t) \,  \overline{\Sigma}_{a' a}^{\uparrow}(t)  ,
\label{cal F HF}
\end{align}
\begin{align}
\mathcal{G}_{a a'}^{\sigma \bar{\sigma}}(t, \tau)  \stackrel{\text{HF}}{=}  & \, \text{i} \,  \left[ \overline{\Sigma}^{\sigma}(t) \cdot G^{\urcorner \sigma}(t, \tau) \cdot \overline{\Sigma}^{\mathbb{M} \sigma}\right]_{a a'}  \, G^{\ulcorner \bar{\sigma}}_{a' a}(\tau, t)  \nonumber \\
&  + \text{i} \,  G^{\urcorner \sigma}_{a a'}(t, \tau) \,  \left[ \overline{\Sigma}^{\mathbb{M} \bar{\sigma}} \cdot G^{\ulcorner \bar{\sigma}}(\tau, t) \cdot \overline{\Sigma}^{\bar{\sigma}}(t) \right]_{a' a} \nonumber \\
&  - \text{i} \, \left[ \overline{\Sigma}^{\sigma}(t) \cdot G^{\urcorner \sigma}(t, \tau) \right]_{a a'}  \left[ \overline{\Sigma}^{\mathbb{M} \bar{\sigma}} \cdot G^{\ulcorner \bar{\sigma}}(\tau, t) \right]_{a' a}  \nonumber \\
&  - \text{i} \, \left[  G^{\urcorner \sigma}(t, \tau) \cdot \overline{\Sigma}^{\mathbb{M} \sigma}\right]_{a a'}  \left[ G^{\ulcorner \bar{\sigma}}(\tau, t) \cdot \overline{\Sigma}^{\bar{\sigma}}(t) \right]_{a' a}.
\label{cal G HF}
\end{align}

\section{Occupation numbers representation in equilibrium}\label{equilibrium occupation numbers}

For the sake of completeness, we here prove Eq.\eqref{equilibrium occ number relation}. In equilibrium, $\hat{H}(t) \equiv \hat{H}_0$, and $G^{>}(t_1, t_2)$ becomes
\begin{align}
G^{> \text{EQ}}_{b a}(t) = - \frac{\text{i}}{\mathcal{Z}_0} \text{Tr}\left\{ \text{e}^{- \beta \left(\hat{H}_0 - \mu \hat{N} \right)} \text{e}^{\text{i} t \hat{H}_0} \hat{\psi}_b \text{e}^{- \text{i} t \hat{H}_0} \hat{\psi}_a^{\dagger} \right\},
\end{align}
where $\mathcal{Z}_0 \equiv \text{Tr}\left\{ \text{exp}\left[ - \beta \left( \hat{H}_0 - \mu \hat{N}\right)\right]\right\}$, and $t = t_1 - t_2$. We put $\hat{K} \equiv \hat{H}_0 - \mu \hat{N}$, and we observe that 
\begin{align}
\text{e}^{\text{i} t \mu \hat{N}} \hat{\psi}_b = \hat{\psi}_b \text{e}^{\text{i} t \mu \hat{N}} \text{e}^{- \text{i} t \mu},
\end{align} 
hence we can write
\begin{align}
G^{> \text{EQ}}_{b a}(t) = - \frac{\text{i} \, \text{e}^{- \text{i} t \mu}}{\mathcal{Z}_0} \text{Tr}\left\{ \text{e}^{- \text{i} t \hat{K}}  \hat{\psi}_a^{\dagger} \text{e}^{\text{i} t \hat{K}} \text{e}^{- \beta \hat{K}}  \hat{\psi}_b   \right\};
\end{align}
analogously, one obtains
\begin{align}
G^{< \text{EQ}}_{b a}(t) = \frac{\text{i} \, \text{e}^{- \text{i} t \mu}}{\mathcal{Z}_0} \text{Tr}\left\{ \text{e}^{- \text{i} t \hat{K}}  \hat{\psi}_a^{\dagger} \text{e}^{\text{i} t \hat{K}}  \hat{\psi}_b  \text{e}^{- \beta \hat{K}}  \right\}.
\end{align}
In both expressions, we explicitly express the trace by employing a complete set of many-body eigenstates $\left| n \right>$ of $\hat{K}$ (with eigenvalues $K_n$), and we also insert the spectral decomposition of the identity between $\hat{\psi}_a^{\dagger}$ and $\hat{\psi}_b$. We obtain
\begin{align}
G^{> \text{EQ}}_{b a}(t) =  - \frac{\text{i} \, \text{e}^{- \text{i} t \mu}}{\mathcal{Z}_0} \sum_n \sum_m  \text{e}^{- \text{i} t \left( K_n - K_m \right) } \text{e}^{- \beta K_m} \left< n \left| \hat{\psi}_a^{\dagger} \right| m \right> \left< m \left| \hat{\psi}_b \right| n \right>   ,
\end{align}
\begin{align}
G^{< \text{EQ}}_{b a}(t) =  \frac{\text{i} \, \text{e}^{- \text{i} t \mu}}{\mathcal{Z}_0} \sum_n \sum_m  \text{e}^{- \text{i} t \left( K_n - K_m \right) } \text{e}^{- \beta K_n} \left< n \left| \hat{\psi}_a^{\dagger} \right| m \right> \left< m \left| \hat{\psi}_b \right| n \right>   .
\end{align}
The Fourier transforms with respect to $t$ are
\begin{align}
{G}^{< \text{EQ}}_{b a}(\omega) = \frac{2 \pi \text{i} }{\mathcal{Z}_0} \sum_n \sum_m & \delta\left(- \omega + K_n - K_m + \mu\right)  \text{e}^{- \beta K_n} \nonumber \\
& \times \left< n \left| \hat{\psi}_a^{\dagger} \right| m \right> \left< m \left| \hat{\psi}_b \right| n \right>   
\end{align}
and ${G}^{> \text{EQ}}_{b a}(\omega) = - \text{exp}\left[ \beta \left(\omega - \mu \right)\right] \, {G}^{< \text{EQ}}_{b a}(\omega)$. Equations \eqref{equilibrium occ number relation} and \eqref{Fermi} follow immediately. $\blacksquare$

\section{Hartree-Fock approximation with diagonal self-energy}\label{HFD approximation}

In the Hartree-Fock approximation, if it is possible to further assume that the self-energy be diagonal in Hubbard site-orbital indexes (which we denote as HFD approximation), the expressions that we have derived for the dynamical parameters of the magnetic interactions undergo a remarkable simplification. This approximation is somehow analogous to the LSB approximation that we have discussed in the equilibrium case and, analogously, it may be relevant only in the special case of a single-band Hubbard model (with a local Hartree-Fock self-energy). From the identities following from the HF approximation, see \ref{HF LNR details}, with the additional assumption of diagonal self-energy, we get:
\begin{align}
& W^{\lessgtr \sigma}_{a a'}(t, t') \stackrel{\text{HFD}}{=}  \overline{\Sigma}_{a}^{\sigma}(t) \, G^{\lessgtr \sigma}_{a a'}(t, t'), \nonumber \\
& X^{\lessgtr \sigma}_{a a'}(t, t') \stackrel{\text{HFD}}{=}  G^{\lessgtr \sigma}_{a a'}(t, t') \, \overline{\Sigma}_{a'}^{\sigma}(t'), \nonumber \\
& Y^{\urcorner \sigma}_{a a'}(t, \tau) \stackrel{\text{HFD}}{=} \overline{\Sigma}_{a}^{\sigma}(t) \, G^{\urcorner \sigma}_{a a'}(t, \tau), \nonumber \\
& Y^{\ulcorner \sigma}_{a a'}(\tau, t) \stackrel{\text{HFD}}{=} G^{\ulcorner \sigma}_{a a'}(\tau, t) \,\overline{\Sigma}_{a'}^{\sigma}(t), \nonumber \\
& Z^{\urcorner \sigma}_{a a'}(t, \tau) \stackrel{\text{HFD}}{=} - \text{i} \, G^{\urcorner \sigma}_{a a'}(t, \tau) \, \overline{\Sigma}_{a'}^{\mathbb{M} \sigma} , \nonumber \\
& Z^{\ulcorner \sigma}_{a a'}(\tau, t) \stackrel{\text{HFD}}{=} - \text{i} \, \overline{\Sigma}_{a}^{\mathbb{M} \sigma} \, G^{\ulcorner \sigma}_{a a'}(\tau, t)  ;
\end{align}
it follows that
\begin{align}
& \mathcal{A}^{\lessgtr}_{a a'}(t, t') \stackrel{\text{HFD}}{=} - 4 \overline{\Sigma}^{\text{S}}_{a'}(t') \, G^{\lessgtr \uparrow}_{a' a}(t', t) \, \overline{\Sigma}^{\text{S}}_{a}(t) \, G^{\gtrless \downarrow}_{a a'}(t, t'), \nonumber \\
& \mathcal{F}_{a a'}(t) \stackrel{\text{HFD}}{=} - 4 \text{i} \, \delta_{a a'} \overline{\Sigma}^{\text{S}}_a(t) \, G^{< \text{S}}_{a a}(t,t) = 4  \delta_{a a'} \overline{\Sigma}^{\text{S}}_a(t) \,  \rho^{\text{S}}_{a}(t), \nonumber \\
& \mathcal{G}_{a a'}^{\sigma \bar{\sigma}}(t, \tau)  \stackrel{\text{HFD}}{=}  4 \text{i} \,   \overline{\Sigma}_a^{\text{S}}(t) \, G_{a a'}^{\urcorner \sigma}(t, \tau) \, \overline{\Sigma}_{a'}^{\mathbb{M} \text{S}}  \, G^{\ulcorner \bar{\sigma}}_{a' a}(\tau, t) .
\label{cal G HFD}
\end{align}
Since $\mathcal{F}_{a a'}(t)$ is now proportional to $\delta_{a a'}$, it does not contribute to the exchange and twist-exchange actions, hence it can be removed from the expressions of the parameters. After eliminating the irrelevant term, the expression for the exchange parameters becomes very simple:
\begin{align}
\mathcal{J}_{a a'}(t, t') \stackrel{\text{HFD}}{=}  \text{sign}(t - t')  \,  \overline{\Sigma}^{\text{S}}_{a'}(t') \, \overline{\Sigma}^{\text{S}}_{a}(t) \, \frac{1}{4 \text{i}} \sum_{\eta} \eta \sum_{\sigma} G^{\eta \sigma}_{a a'}(t, t') \, G^{\bar{\eta} \bar{\sigma}}_{a' a}(t', t)  .
\label{J HFD}
\end{align}
It must be noted that the part multiplying $\text{sign}(t - t')$ is identically zero for $t = t'$, so one has $\mathcal{J}_{a a'}(t, t) \stackrel{\text{HFD}}{=} 0$, with no ambiguities related to the definition of the function $\text{sign}(x)$. The twist-exchange parameters in HFD conditions are given by
\begin{align}
\mathcal{Y}_{a a'}(t, t') \stackrel{\text{HFD}}{=} - \text{sign}(t - t')   \overline{\Sigma}^{\text{S}}_{a'}(t') \, \overline{\Sigma}^{\text{S}}_{a}(t) \, \frac{1}{4}  \sum_{\eta} \eta  \sum_{\sigma}  \sigma \, G^{\eta \sigma}_{a a'}(t, t') \, G^{\bar{\eta} \bar{\sigma}}_{a' a}(t', t)  . 
\label{D HFD}
\end{align}
In the occupation numbers representation,
\begin{align}
\mathcal{J}_{a a'}(t, t') \stackrel{\text{HFD}}{=} & - \text{sign}(t_r)  \,  \overline{\Sigma}^{\text{S}}_{a}\left( T + \frac{t_r}{2} \right) \, \overline{\Sigma}^{\text{S}}_{a'}\left( T - \frac{t_r}{2} \right) \,   \frac{1}{4 \text{i}} \int_{-\infty}^{+ \infty} \frac{\text{d} \omega}{2 \pi} \int_{-\infty}^{+ \infty} \frac{\text{d} \omega'}{2 \pi}  \nonumber \\
& \times \text{e}^{- \text{i} (\omega - \omega') t_r} \Big[ f(\omega, T) - f(\omega', T) \Big] \sum_{\sigma} {A}^{\sigma}_{a a'}(\omega, T) \, {A}^{\bar{\sigma}}_{a' a}(\omega', T)  ,
\end{align}
and
\begin{align}
\mathcal{Y}_{a a'}(t, t') \stackrel{\text{HFD}}{=} &  \text{sign}(t_r)     \, \overline{\Sigma}^{\text{S}}_{a}\left( T + \frac{t_r}{2} \right)  \, \overline{\Sigma}^{\text{S}}_{a'}\left( T - \frac{t_r}{2} \right) \, \frac{1}{4}  \int_{-\infty}^{+ \infty} \frac{\text{d} \omega}{2 \pi} \int_{-\infty}^{+ \infty} \frac{\text{d} \omega'}{2 \pi} \nonumber \\
& \times \text{e}^{- \text{i} (\omega - \omega') t_r}  \Big[ f(\omega, T) - f(\omega', T) \Big] \sum_{\sigma} \sigma {A}^{\sigma}_{a a'}(\omega, T) \, {A}^{\bar{\sigma}}_{a' a}(\omega', T)  ,
\end{align}
where $t_r = t - t'$ and $T = (t + t')/2$.

\section{Derivation of the spin stiffness}\label{sum 1BZ}

We here add some details on the derivation of Eq.\eqref{spin stiffness result}. Using Eq.\eqref{J_q}, we directly obtain
\begin{align}
D_{\alpha \beta}(t) \equiv & \frac{\text{i}}{2 M} \sum_{\eta} \eta \sum_{\sigma} \int_{t_0}^{\infty} \text{d} t' \, \text{sign}(t' - t)  \,  \overline{\Sigma}^{\text{S}}(t) \, \overline{\Sigma}^{\text{S}}(t') \nonumber \\
& \times \frac{1}{n} \sum_{\boldsymbol{k}} G^{\eta \sigma}_{\boldsymbol{k}}(t', t) \, \frac{ \partial^2 G^{\bar{\eta} \bar{\sigma}}_{\boldsymbol{k}}(t, t')}{\partial k_{\alpha} \partial k_{\beta}} .
\label{spin stiffness first}
\end{align}
To remove the second derivatives of the Green functions, we first put
\begin{align}
G^{\eta \sigma}_{\boldsymbol{k}} \frac{ \partial^2 G^{\bar{\eta} \bar{\sigma}}_{\boldsymbol{k}}}{\partial k_{\alpha} \partial k_{\beta}} = \frac{\partial}{\partial k_{\alpha}} \left[ G^{\eta \sigma}_{\boldsymbol{k}} \frac{\partial}{\partial k_{\beta}} G^{\bar{\eta} \bar{\sigma}}_{\boldsymbol{k}} \right] - \frac{\partial G^{\eta \sigma}_{\boldsymbol{k}}}{\partial k_{\alpha}}   \frac{\partial G^{\bar{\eta} \bar{\sigma}}_{\boldsymbol{k}} }{\partial k_{\beta}}  ,
\label{derivatives}
\end{align}
then we transform the sum over the $\boldsymbol{k}$ vectors in Eq.\eqref{spin stiffness first} into the corresponding integral over the volume $V$ of the 1BZ, and we use 
\begin{align}
\int_{V} \frac{\partial}{\partial k_{\alpha}} \left[ G^{\eta \sigma}_{\boldsymbol{k}} \frac{\partial}{\partial k_{\beta}} G^{\bar{\eta} \bar{\sigma}}_{\boldsymbol{k}} \right] \text{d} V = \oint_{\partial V} \left[ G^{\eta \sigma}_{\boldsymbol{k}} \frac{\partial}{\partial k_{\beta}} G^{\bar{\eta} \bar{\sigma}}_{\boldsymbol{k}}  \right] \text{d} S_{\alpha}, 
\end{align}
where $\text{d} S_{\alpha}$ is the projection along the $\alpha$ direction of the oriented infinitesimal element of surface of the 1BZ at $\boldsymbol{k}$. Separating the contributions with $k_{\alpha} > 0$ and $k_{\alpha} < 0$, we see that they cancel with each other, because $G_{\boldsymbol{k}} = G_{- \boldsymbol{k}}$ while $\text{d} S_{\alpha}$ changes sign under $k_{\alpha} \rightarrow - k_{\alpha}$. Therefore, the integral of the first term of the RHS of Eq.\eqref{derivatives} is zero. The remaining term gives Eq.\eqref{spin stiffness result}.

\end{document}